\documentclass[a4paper,11pt]{article}
\usepackage{jheppub} 
\usepackage{comment}
\usepackage{lineno}
\usepackage[dvipsnames]{xcolor}
\usepackage[utf8]{inputenc}  
\usepackage{fix-cm}  
\usepackage{tikz}    
\usepackage{array}
\usetikzlibrary{arrows}
\usetikzlibrary { decorations.pathmorphing, decorations.pathreplacing, decorations.shapes, }
\usepackage{tikz-cd} 
\usepackage{tkz-euclide}
\usepackage{circuitikz}
\usepgfmodule{shapes}
\usetikzlibrary{decorations.pathmorphing}
\usetikzlibrary{backgrounds, arrows,calc,shapes,decorations.pathreplacing, automata,positioning}
\usepackage{cancel} 
\usepackage{graphicx}
\usepackage{mathdots}
\usetikzlibrary{decorations.markings}

\usetikzlibrary{shapes.misc}

\usepackage{amsmath} 
\usepackage{amssymb}
\usepackage{graphicx}
\graphicspath{ {images/} }
\usepackage{caption}
\usepackage{subcaption}

\usepackage[most]{tcolorbox}
\usepackage{hyperref}

\definecolor{CScolor}{rgb}{1.0, 0.0, 0.0}
\colorlet{BFcolor}{OliveGreen}
\colorlet{FIcolor}{orange}

\def\mon{\mathfrak{M}}

\def\identityN[#1,#2]{
	_{#1}\mathbb{I}_{#2}^{[N]}
}

\def\identityOne[#1,#2]{
	_{#1}\mathbb{I}_{#2}^{[1^N]}
}

\def\DeltaN[#1]{
	\Delta^{[N]}\left( #1 \right)
}

\def\DeltaOne[#1]{
	\Delta^{[1^N]}\left( #1 \right)
}

\tikzstyle{flavor}=[rectangle,draw=black,thick,inner sep = 0pt, minimum size = 6mm]
\tikzstyle{manifest}=[rectangle,draw=blue!50,thick,inner sep = 0pt, minimum size = 6mm]
\tikzstyle{gauge}=[circle,draw=black,thick,inner sep = 0pt, minimum size = 6mm] 
\tikzstyle{gauge2}=[circle,draw=black!50,thick,inner sep = 0pt, minimum size = 4.5mm] 
\tikzstyle{gauge3}=[rounded rectangle, draw=black!100, thick, minimum size=5mm] 
\tikzset{->-/.style={decoration={
  markings,
  mark=at position .5 with {\arrow{>}}},postaction={decorate}}}
\tikzset{-<-/.style={decoration={
  markings,
  mark=at position .5 with {\arrow{<}}},postaction={decorate}}}
\tikzstyle{BFline}=[dashed]
  
\def\nodeCS(#1,#2)(#3,#4,#5){ 	
	\node at (#1,#2) (#3) [gauge,black] {#4};
	\draw[CScolor] (#3) ++(6pt,-7pt) node[anchor=west] {\tiny#5};
}

\def\flavorCS(#1,#2)(#3,#4,#5){ 	
	\node at (#1,#2) (#3) [flavor,black] {#4};
	\draw[CScolor] (#3) ++(6pt,-8pt) node[anchor=west] {\tiny#5};
}

\def\arrowBF(#1,#2)(#3){ 
	\begin{scope}[every node/.style={auto, outer sep=-1pt}]
	\path (#1) edge[->-] node[BFcolor,midway] {\tiny#3} (#2);
	\end{scope}
}

\def\arrowBFlr(#1,#2)(#3,#4){ 
	\begin{scope}[every node/.style={auto=#4, outer sep=-1pt}]
	\path (#1) edge[->-] node[BFcolor,midway] {\tiny#3} (#2);
	\end{scope}
}

\def\dottedBF(#1,#2)(#3){ 
	\begin{scope}[every node/.style={auto, outer sep=-1pt}]
	\path (#1) edge[dashed] node[BFcolor,midway] {\tiny#3} (#2);
	\end{scope}
}

\def\simpFund[#1]{ 
	\begin{tikzpicture}
        \node at (0,0) (g) [flavor, black] {$#1$};
        \node at (0,1.5) (f) [flavor,black] {$1$};
        \draw[CScolor] (g)+(0,-.6) node {$_{( \text{-}\frac{1}{2}, \text{-}\frac{1}{2})}$};
        \draw[->-] (g)--(f);
        \draw[right] (g)+(.4,0) node {$\mathbb{I}^{[#1]}$};
	\end{tikzpicture}
}

\def\simpAFund[#1]{ 
	\begin{tikzpicture}
        \node at (0,0) (g) [flavor, black] {$#1$};
        \node at (0,1.5) (f) [flavor,black] {$1$};
        \draw[CScolor] (g)+(0,-.6) node {$_{(\text{-}\frac{1}{2}, \text{-}\frac{1}{2})}$};
        \draw[->-] (f)--(g);
        \draw[right] (g)+(.4,0) node {$\mathbb{I}^{[#1]}$};
	\end{tikzpicture}
}

\def\simpFundNMirror[#1]{ 
	\begin{tikzpicture}
		\node at (-1.5,0) (f1) [flavor,black] {$#1$};
        \node at (0,0) (g1) [gauge, black] {$1^#1$};
        \node at (1,1) (g2) [gauge,black] {$1^#1$};
        \node at (2.5,1) (f2) [flavor,black] {$#1$};
        \draw[->-] (g1)--(g2);
        \draw[decorate,decoration={coil,segment length=4pt}] (f1)--(g1) node[midway,above] {$+$};
        \draw[decorate,decoration={coil,segment length=4pt}] (g2)--(f2)node[midway,above] {$-$};
	\end{tikzpicture}
}

\def\simpAFundNMirror[#1]{
	\begin{tikzpicture}
		\node at (-1.5,1) (f1) [flavor,black] {$#1$};
        \node at (0,1) (g1) [gauge, black] {$1^#1$};
        \node at (1,0) (g2) [gauge,black] {$1^#1$};
        \node at (2.5,0) (f2) [flavor,black] {$#1$};
        \draw[->-] (g2)--(g1);
        \draw[decorate,decoration={coil,segment length=4pt}] (f1)--(g1) node[midway,above] {$+$};
        \draw[decorate,decoration={coil,segment length=4pt}] (g2)--(f2)node[midway,above] {$-$};
	\end{tikzpicture}
}

\def\simpBifLNN[#1]{
	\begin{tikzpicture}
        \node at (0,0) (g) [flavor, black] {$#1$};
        \node at (1.5,0) (f) [flavor,black] {$#1$};
        \draw[CScolor] (g)+(0,-.6) node {$_{(\text{-}\frac{#1}{2},\text{-}\frac{#1}{2})}$};
        \draw[CScolor] (f)+(0,-.6) node {$_{(\text{-}\frac{#1}{2},\text{-}\frac{#1}{2})}$};
        \draw[BFcolor] (.75,.3) node {$_{+1}$};
        \draw[->-] (f)--(g);
	\end{tikzpicture}
}

\def\simpBifRNN[#1]{
	\begin{tikzpicture}
		\node at (0,0) (g) [flavor, black] {$#1$};
        \node at (1.5,0) (f) [flavor,black] {$#1$};
        \draw[CScolor] (g)+(0,-.6) node {$_{(\text{-}\frac{#1}{2},\text{-}\frac{#1}{2})}$};
        \draw[CScolor] (f)+(0,-.6) node {$_{(\text{-}\frac{#1}{2},\text{-}\frac{#1}{2})}$};
        \draw[BFcolor] (.75,.3) node {$_{+1}$};
		\draw[->-] (g)--(f);
	\end{tikzpicture}
}

\def\simpBifLNNMirror[#1]{
\;
	\begin{tikzpicture}
		\node at (-1.5,0) (f1) [flavor,black] {$#1$};
        \node at (0,0) (g) [gauge, black] {$1^#1$};
        \node at (1.5,0) (f2) [flavor,black] {$#1$};
        \node at (1,-1) (f3) [flavor,black] {$1$};
        \draw[decorate,decoration={coil,segment length=4pt}] (f1)--(g) node[midway,above] {$+$};
        \draw[decorate,decoration={coil,segment length=4pt}] (g)--(f2)node[midway,above] {$-$};
		\draw[->-] (f3)--(g);
	\end{tikzpicture}
\;
}

\def\simpBifRNNMirror[#1]{
\;
	\begin{tikzpicture}
		\node at (-1.5,0) (f1) [flavor,black] {$#1$};
        \node at (0,0) (g) [gauge, black] {$1^#1$};
        \node at (1.5,0) (f2) [flavor,black] {$#1$};
        \node at (-1,1) (f3) [flavor,black] {$1$};
        \draw[decorate,decoration={coil,segment length=4pt}] (f1)--(g) node[midway,above] {$+$};
        \draw[decorate,decoration={coil,segment length=4pt}] (g)--(f2)node[midway,above] {$-$};
		\draw[->-] (g)--(f3);
	\end{tikzpicture}
\;
}

\def\simpGenBifund[#1]{ 
\;
	\begin{tikzpicture}[baseline=0]
		    \node at (0,0) (g1) [flavor, black] {$#1$};
		    \node at (1.3,0) (f1) [flavor,black] {$1^#1$};
		    \draw[decorate,decoration={coil,segment length=4pt}] (f1)--(g1);
	\end{tikzpicture}
}

\def\simpGenBifundA[#1]{ 
\;
	\begin{tikzpicture}[baseline=0]
		    \node at (0,0) (g1) [flavor, black] {$1^#1$};
		    \node at (1.3,0) (f1) [flavor,black] {$#1$};
		    \draw[decorate,decoration={coil,segment length=4pt}] (f1)--(g1);
	\end{tikzpicture}
}

\def\simpBifLOne[#1]{
\;
	\begin{tikzpicture}[baseline=10]
		    \node at (1,0) (g) [flavor, black] {$#1$};
		    \node at (0,1) (f) [flavor,black] {$#1$};
		    \draw[->-] (g)--(f);
	\end{tikzpicture}
\;
}

\def\simpBifROne[#1]{
\;
	\begin{tikzpicture}[baseline=10]
		    \node at (0,0) (g) [flavor, black] {$#1$};
		    \node at (1,1) (f) [flavor,black] {$#1$};
		    \draw[->-] (g)--(f);
	\end{tikzpicture}
\;
}

\def\simpFundOneMirror[#1]{
\;
	\begin{tikzpicture}[baseline=0]
		\node at (-1.5,0) (f1) [flavor,black] {$1^#1$};
		    \node at (0,0) (g1) [gauge, black] {$#1$};
      \draw[red] (-.5,-.6) node {$_{(-\frac{1}{2}, #1-\frac{1}{2})}$};
        \draw[OliveGreen] (.5,.5) node {$_1$};
      \draw[red] (1.5,-.6) node {$_{(-\frac{1}{2}, #1-\frac{1}{2})}$};
		    \node at (1,0) (g2) [gauge,black] {$#1$};
		    \node at (2.5,0) (f2) [flavor,black] {$1^#1$};
		    \draw[->-] (g1)--(g2);
		    \draw[decorate,decoration={coil,segment length=4pt}] (f1)--(g1) node[midway,above] {$+$};
			\draw[decorate,decoration={coil,segment length=4pt}] (g2)--(f2)node[midway,above] {$-$};
	\end{tikzpicture}
\;
}

\def\simpFundOne[#1]{
\;
	\begin{tikzpicture}[baseline=0]
		    \node at (0,0) (g1) [flavor, black] {$1^#1$};
		    \node at (0,1.5) (f2) [flavor,black] {$1$};
		    \draw[->-] (g1)--(f2);
	\end{tikzpicture}
\;
\mathbb{I}^{[1^#1]}
\;
}

\def\simpAFundOneMirror[#1]{
\;
	\begin{tikzpicture}[baseline=0]
		\node at (-1.5,0) (f1) [flavor,black] {$1^#1$};
		    \node at (0,0) (g1) [gauge, black] {$#1$};
      \draw[red] (-.5,-.6) node {$_{(-\frac{1}{2}, #1-\frac{1}{2})}$};
        \draw[OliveGreen] (.5,.5) node {$_1$};
      \draw[red] (1.5,-.6) node {$_{(-\frac{1}{2}, #1-\frac{1}{2})}$};
		    \node at (1,0) (g2) [gauge,black] {$#1$};
		    \node at (2.5,0) (f2) [flavor,black] {$1^#1$};
		    \draw[->-] (g2)--(g1);
		    \draw[decorate,decoration={coil,segment length=4pt}] (f1)--(g1) node[midway,above] {$+$};
			\draw[decorate,decoration={coil,segment length=4pt}] (g2)--(f2)node[midway,above] {$-$};
	\end{tikzpicture}
\;
}

\def\simpAFundOne[#1]{
\;
	\begin{tikzpicture}[baseline=0]
		    \node at (0,0) (g1) [flavor, black] {$1^#1$};
		    \node at (0,-1.5) (f2) [flavor,black] {$1$};
		    \draw[->-] (f2)--(g1);
	\end{tikzpicture}
\;
\mathbb{I}^{[1^#1]}
\;
}

\def\simpBifLOneMirror[#1]{
\;
	\begin{tikzpicture}[baseline=0]
		\node at (-1.5,0) (f1) [flavor,black] {$1^#1$};
		    \node at (0,0) (g) [gauge, black] {$#1$};
                \draw[red] (0,-.6) node {$_{(-\frac{1}{2}, #1-\frac{1}{2})}$};
		    \node at (1.5,0) (f2) [flavor,black] {$1^#1$};
		 	 \node at (-1.5,1) (f3) [flavor,black] {$1$};
		    \draw[decorate,decoration={coil,segment length=4pt}] (f1)--(g) node[midway,above] {$+$};
			\draw[decorate,decoration={coil,segment length=4pt}] (g)--(f2)node[midway,above] {$-$};
		\draw[->-] (f3)--(g);
	\end{tikzpicture}
\;
}

\def\simpBifROneMirror[#1]{
\;
	\begin{tikzpicture}[baseline=0]
		\node at (-1.5,0) (f1) [flavor,black] {$1^#1$};
		    \node at (0,0) (g) [gauge, black] {$#1$};
         \draw[red] (-.5,-.6) node {$_{(-\frac{1}{2}, #1-\frac{1}{2})}$};
		    \node at (1.5,0) (f2) [flavor,black] {$1^#1$};
		 	 \node at (1.5,-1) (f3) [flavor,black] {$1$};
		    \draw[decorate,decoration={coil,segment length=4pt}] (f1)--(g) node[midway,above] {$+$};
			\draw[decorate,decoration={coil,segment length=4pt}] (g)--(f2)node[midway,above] {$-$};
		\draw[->-] (g)--(f3);
	\end{tikzpicture}
\;
}

\tikzset{cross/.style={cross out, draw=black, minimum size=5*(#1-\pgflinewidth), inner sep=0pt, outer sep=0pt},
cross/.default={2pt}}

\tikzset{snake it/.style={decorate, decoration=snake}}

\tikzset{mid arrow/.style={postaction={decorate,decoration={
        markings,
        mark = at position .55 with {\arrow[#1]{Straight Barb[width=5pt]}}
      }}}}
      
\tikzset{mid arrowsm/.style={postaction={decorate,decoration={
        markings,
        mark = at position .55 with {\arrow[#1]{Straight Barb[width=3pt]}}
      }}}}
      
\tikzset{middx arrowsm/.style={postaction={decorate,decoration={
        markings,
        mark = at position .7 with {\arrow[#1]{Straight Barb[width=3pt]}}
      }}}}
      
\tikzset{midsx arrowsm/.style={postaction={decorate,decoration={
        markings,
        mark = at position .4 with {\arrow[#1]{Straight Barb[width=3pt]}}
      }}}}

\title{\boldmath A Chiral-Planar dualization algorithm for $3d$ $\mathcal{N}=2$ Chern-Simons-matter theories}

\author[a]{Sergio Benvenuti}
\author[b,c]{Riccardo Comi}
\author[b,c]{Sara Pasquetti}
\author[a,d]{Gabriel Pedde Ungureanu}
\author[a,d]{Simone Rota}
\author[a,d]{Anant Shri}
\affiliation[a]{INFN, Sezione di Trieste, Via Valerio 2, I-34127 Trieste, Italy}
\affiliation[b]{Dipartimento di Fisica, Università di Milano-Bicocca, Piazza della Scienza 3, I-20126 Milano, Italy}
\affiliation[c]{INFN, Sezione di Milano-Bicocca, Piazza della Scienza 3, I-20126 Milano, Italy}
\affiliation[d]{SISSA, Via Bonomea 265, I-34136 Trieste, Italy}

\emailAdd{benve79@gmail.com} \emailAdd{r.comi2@campus.unimib.it} \emailAdd{sara.pasquetti@gmail.com} \emailAdd{gpeddeun@sissa.it} \emailAdd{srota@sissa.it}  \emailAdd{ashri@sissa.it}

\abstract{
We show that a broad class of three-dimensional $\mathcal{N}=2$ chiral Chern-Simons gauge theories admit an abelian and planar dual description. These chiral-planar dualities emerge by performing real mass deformations on known $\mathcal{N}=4$ mirror pairs, using the $\mathcal{N}=2^*$ setup to flow to chiral theories on the electric side. While identifying the correct dual vacuum is subtle due to the rich structure of the Coulomb branch, we develop a mirror dualization algorithm that streamlines this process and systematically provides the abelian-planar duals of chiral quivers.}

\begin{document}
\maketitle
\flushbottom

\section{Introduction}
\label{sec: introduction}




In this work, we show that a broad class of 3d $\mathcal{N}=2$ Chern-Simons (CS) theories admit a purely abelian and planar dual description.

These abelian-planar duals are generated by performing real mass deformations that break $\mathcal{N}=4 \to \mathcal{N}=2$, starting from known 3d $\mathcal{N}=4$ mirror pairs \cite{Intriligator_1996}.

More precisely, we work in the $\mathcal{N}=2^*$ setup, with an abelian R-symmetry
$U(1)_R = U(1)_{C+H}$ and a global symmetry $U(1)_\tau = U(1)_{H-C}$,
where $U(1)_C \subset SU(2)_C$ and $U(1)_H \subset SU(2)_H$ are the Cartan subgroups of the non-abelian $\mathcal{N}=4$ R-symmetry $SU(2)_C \times SU(2)_H$ \cite{Tong:2000ky}. We then perform a real mass deformation \cite{Aharony:1997bx} 
for the axial $U(1)_\tau$, thus breaking the R-symmetry to $U(1)_R$, hence supersymmetry from $\mathcal{N}=4 \to \mathcal{N}=2$.

Naively, this deformation renders all fields massive. However, by combining it with suitable vacuum expectation values (VEVs) for the real scalars in vector multiplets—i.e., by moving along the Coulomb branch—we can reach vacua where some fields remain massless, yielding a nontrivial infrared theory.

For instance, starting from 3d $\mathcal{N}=4$ $U(N)$ SQCD with $F$ hypermultiplets, one can combine the $U(1)_\tau$ real mass deformation with additional deformations to flow to a chiral theory: $\mathcal{N}=2$ 
 $U(N)$ SQCD with $F$ fundamental chirals and CS levels $\left(-\frac{F}{2}+N,-\frac{F}{2}\right)$ for the $SU(N)$ and $U(1)$ components of the gauge groups.

\begin{figure}[h!] \centering \includegraphics[width=.9\textwidth]{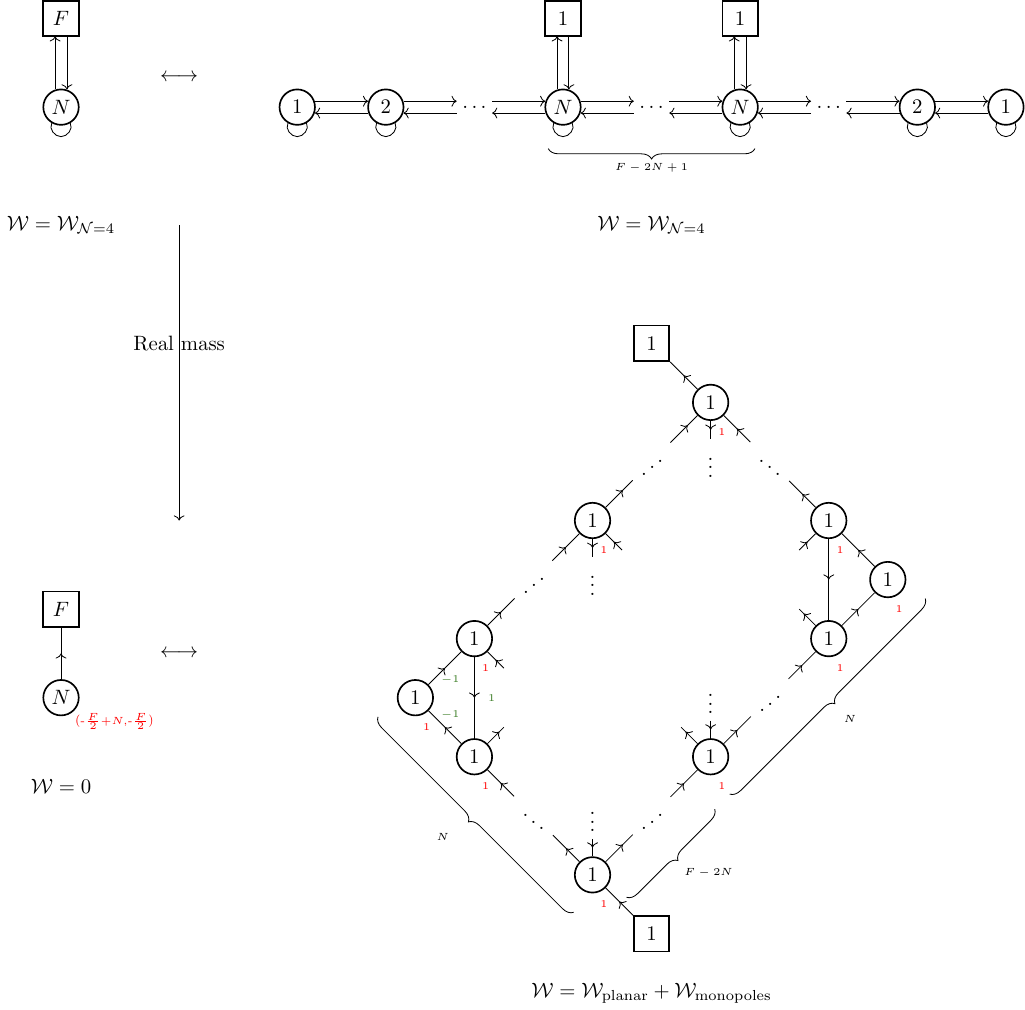} \caption{Schematic depiction of the $\mathcal{N}=2$ mirror-like duality (bottom row) observed between $U(N)$ SQCD with $[F,0]$ fundamental fields and a planar, abelian quiver gauge theory. The quivers are depicted in the $\mathcal{N}=2$ notation. The labels in red denote CS levels while those in green denote the level of mixed CS interaction. The superpotential in the planar quiver consist of mesonic and monopole terms that we explain in detail below Figure \ref{planar_mirror_general}.} \label{fig:PlanarSQCDWeb} \end{figure}

We can then apply the same real mass deformation on the mirror dual side \cite{Hanany_1997},
shown in the first line of Figure \ref{fig:PlanarSQCDWeb}.
For SQCD, the mirror dual is a quiver with $F-1$ gauge nodes, meaning there are $F-1$ vector multiplets that can acquire VEVs in various ways. This opens up a multitude of possible directions on the Coulomb branch to explore, each potentially leading to a nontrivial low-energy theory. As a result, identifying the vacuum corresponding to the dual of the chiral SQCD becomes highly nontrivial.

In general, starting from a 3d $\mathcal{N}=4$ mirror pair of quivers and turning on a $U(1)_\tau$ real mass deformation to produce a chiral quiver on the electric side leads to the subtle and intricate problem of identifying the correct dual theory—comparable to finding a needle in a haystack.

A powerful guiding principle emerges from the study of the $S^3_b$ partition function \cite{Kapustin:2009kz, Hama_2011}.
As discussed in \cite{Aharony:2013dha,Benini:2011mf}, turning on real masses in the UV theory translates into taking certain limits of the partition function where the mass parameters are sent to infinity.

In this limit, the partition function of the original theory flows to that of the interacting IR theory, multiplied by a highly oscillatory phase.
Crucially, this phase is sensitive to the VEVs of the real scalars in vector multiplets—that is, it captures how we moved along the Coulomb branch.
Therefore, if we start from a known partition function identity between a 3d $\mathcal{N}=4$ mirror pair and perform the appropriate real mass limits on both sides, we can match the resulting IR theories by identifying the terms with identical oscillatory phases.

This technique helps us pinpoint the correct dual of the chiral $U(N)$ SQCD.
Somewhat surprisingly, we find that the mirror dual is a planar, abelian quiver,
shown in the bottom right corner of Figure \ref{fig:PlanarSQCDWeb}.
An earlier account of part of our findings appeared in \cite{Benvenuti:2024seb}, where we focused on the case of $\mathrm{SU}(N)$ SQCD with fundamental flavors.

In general, we find that if we deform an $\mathcal{N}=4$ mirror pair of quivers to produce a chiral quiver on the electric side, in the dual vacuum, every $U(k)$ gauge node in the $\mathcal{N}=4$ mirror is Higgsed down to $U(1)^k$.

The chiral-planar mirror pair inherits various features from the parent $\mathcal{N}=4$ mirror pair. In particular, mesonic symmetries are mapped to topological symmetries, and we observe nontrivial IR enhancement of topological symmetries. This motivates the terminology: chiral-planar \textit{mirror} pairs.

While this deformation procedure is effective, it is also tedious and lacks clear intuition.
It is far from obvious, a priori, that one should search for a planar abelian dual.
Indeed, the idea of deriving $\mathcal{N}=2$ dualities via real mass deformations of $\mathcal{N}=4$ mirror pairs was already anticipated in \cite{Tong:2000ky}, where the $\mathcal{N}=2^*$ framework was introduced. At the time, such deformations appeared to lead to chiral-chiral dualities. However, we now find that, with refined tools such as exact partition function techniques and a more systematic analysis of Coulomb branch vacua, these flows can instead give rise to chiral-planar dualities.

What we ultimately seek is a more systematic and streamlined strategy for identifying mirror planar duals of chiral quivers.

Recently, it has been shown that 3d $\mathcal{N}=4$ (and recently some $\mathcal{N}=2$) mirror dualities can be derived via the mirror dualization algorithm \cite{Hwang:2021ulb, Comi:2022aqo}.
In this paper, we extend this approach to the chiral-planar case.

Thanks to the algorithmic construction, we are able to generate the abelian planar duals of chiral quiver theories across a broad range of 
ranks of the gauge groups and matter content, with specific CS levels.
For example, we provide the abelian planar dual of $U(N)$ SQCD with high enough, but otherwise arbitrary number of fundamentals and antifundamentals, generic $N$ and specific CS levels. 
This is a significant generalization of the dualities proposed in \cite{Benvenuti:2024seb} by the same authors, where the abelian planar dual of $SU(N)$ SQCD with only fundamental matter was first proposed.

Combining this algorithmic approach with the direct analysis of real mass deformations of $\mathcal{N}=4$ mirror pairs, we can find planar duals for an even larger family of theories.
For example, we can find a planar dual for  $\mathcal{N}=2$ $U(N)/SU(N)$ CS-SQCD with $[n_f,n_a]$ fundamentals/antifundamentals and more general CS level, for more general linear chiral quivers, for circular quivers, and for  $\mathcal{N}=2$ $USp(2N)$ CS-SQCD.\\

A natural question is whether every 3d $\mathcal{N}=2$ quiver theory admits a dual description in terms of a planar, abelian quiver.
The chiral-planar dualities we construct here emerge from a specific class of theories—those obtainable via real mass deformations from 3d $\mathcal{N}=4$ mirror pairs, possibly in combination with additional operations such as Aharony-like dualities and Witten's $SL(2,\mathbb{Z})$ action \cite{witten2003sl2zactionthreedimensionalconformal}. 
However, it remains an open question to determine whether this strategy can be extended, or generalized, to encompass arbitrary $\mathcal{N}=2$ gauge theories. \\

The mirror dualization algorithm was originally developed for $\mathcal{N}=4$ mirror dualities with unitary gauge groups.
3d $\mathcal{N}=4$ theories can be engineered in Hanany-Witten brane setups \cite{Hanany_1997}, and mirror symmetry arises as a consequence of type IIB S-duality, which swaps NS and D5 branes.

The mirror algorithm is a field theory counterpart of the local S-dualization of 5-branes.
However, the algorithm has been extended to other situations where a brane realization is either not known (as in the case of the 4d mirror pairs discussed in \cite{Hwang:2020wpd}) or only partially known, as for the 3d $\mathcal{N}=2$ mirror pairs of  generalized quivers discussed in \cite{Benvenuti:2023qtv}. The latter are conjectured to be associated with brane setups preserving four supercharges, involving NS, NS', D5, and D5' branes, for which no clear prescription exists to extract the correct low-energy theory.

In this paper, we will focus on the field theory construction of an algorithm to dualize chiral quivers into planar ones.
The structure of the mirror dualization algorithm—rooted in the logic of S-duality and the manipulation of local building blocks—hints at an underlying brane realization for these chiral-planar dualities. 
This possibility is further supported by the resemblance between the planar quivers and the dimer models discussed in \cite{Hanany:2005ve, Franco:2005rj}.
This resemblance may in fact provide the natural setting for a string-theoretic realization via brane constructions.
\\

In \cite{Comi:2022aqo}, the mirror dualization algorithm was extended to generate all the $SL(2,\mathbb{Z})$ duals of $\mathcal{N}=4$ unitary linear quiver theories.
It would be interesting to study the effect of the $\mathcal{N}=4 \to \mathcal{N}=2$  breaking axial real mass deformations also across these duality frames.
\\

Furthermore, by compactifying 3d $\mathcal{N}=2$ gauge theories on a circle, one flows to 2d $\mathcal{N}=(2,2)$ GLSMs. In this context, it is natural to ask whether every 2d non-abelian GLSM admits an abelian-planar dual, in the sense described in this work.\\

Finally, a natural direction for generalization is the further breaking of supersymmetry. Our construction crucially relies on $\mathcal{N}=2$ supersymmetry to control the RG flows and establish dualities, but it is natural to ask whether non-supersymmetric analogs of chiral-planar dualities exist. We address this question in our upcoming work \cite{Benvenuti:2025qnq}.
\\

\subsection*{Organization of the paper}

In Section \ref{subsec: SQCD_example}, we focus on the chiral-planar duality for
SQCD $U(N)_{-\frac{F}{2}+N,-\frac{F}{2}}$ with $F$ fundamental chirals, as shown in Figure \ref{fig:PlanarSQCDWeb}.
We discuss the global symmetries and the operator map, and describe in detail how to derive this duality via real mass deformation of the $\mathcal{N}=4$ SQCD duality, both at the level of the partition function and through the semiclassical equations of motion (EOM).
We also show how, starting from the duality for $U(N)_{-\frac{F}{2}+N,-\frac{F}{2}}$ SQCD, one can obtain planar duals for
$SU(N)_{-\frac{F}{2}+N}$ and for more general Chern-Simons levels of the form $U(N)_{-\frac{F}{2}+N,-\frac{F}{2}+\Delta_{\ell} N}$.
As a consistency check, we show that the planar dual of $SU(2)_0$ with $F=4$ sequentially s-confines to a WZ model.

In Section \ref{sec: FTUN_mirror}, we discuss an interesting web of chiral-planar dualities originating from the $\mathcal{N}=4$ $FT[SU(N)]$ theory \cite{Gaiotto_2007}, which is self-dual under mirror symmetry.
Starting from this self-duality and performing a real mass deformation, we obtain a new $\mathcal{N}=2$ duality between a chiral quiver and a planar quiver.
Moreover, $FT[SU(N)]$ admits an additional duality frame known as the flip-flip dual frame \cite{Aprile:2018oau}.
Starting from the flip-flip duality and performing real mass deformations, we derive two new $\mathcal{N}=2$ dualities: a chiral-chiral duality relating two chiral quivers, and a planar-planar duality relating two planar quivers.

In Section \ref{sec: S_walls,QFT_blocks_Duality_Moves}, after briefly reviewing the mirror dualization algorithm for $\mathcal{N}=4$ theories,
we extend it to the chiral-planar case by defining the chiral-planar $\mathcal{S}$-wall and introducing the basic duality moves.

In Section \ref{sec: Examples}, we use the chiral-planar algorithm to find the planar dual of
SQCD $U(N)_{\left(-\frac{F_1+F_2}{2}, -\frac{F_1+F_2}{2}-N\right)}$ with $[F_1+N, F_2+N]$ chirals.
We also identify planar abelian duals for various other chiral quivers, highlighting different patterns of topological symmetry enhancement.

In Section \ref{furtherexamples}, we present further examples not covered by the current version of the algorithm.
While we leave the extension of the algorithm to these cases for future work, we show that these examples can nonetheless be derived directly from $\mathcal{N}=4$ mirror pairs using real mass deformations.

In Appendix \ref{app: monopole}, we review how to compute the R-charge, global and gauge charges of monopole operators.
Finally, in Appendix \ref{App:B}, we demonstrate how to derive the chiral-chiral and planar-planar dualities via local dualization.

\newpage


\section{A Planar Abelian Dual for CS-SQCD$_3$ with Fundamental Matter} \label{subsec: SQCD_example}

In this section, we explicitly derive the $\mathcal{N}=2$ mirror of $U(N)_{(-\frac{F}{2}+N, -\frac{F}{2})}$ SQCD with $F$ fundamental chiral multiplets by starting from the $\mathcal{N}=4$ mirror duality for the $U(N)$ SQCD with $F$ flavors.

Our main tool is the $\mathbf{S}^3_b$ partition function 
\cite{Kapustin:2009kz, Hama_2011} which allows us to follow the effect of the real mass deformations efficiently. We verify these results by checking that the resulting vacuum satisfies the F- and D-term equations \cite{Intriligator_2013}.

\subsection{The $\mathcal{N}=4$ Mirror Pair}\label{subsec: N=4Mirror}

We start from the mirror duality for the $U(N)$ $\mathcal{N}=4$ SQCD:
\begin{figure}[h]
    \centering
    \includegraphics[width=1\linewidth]{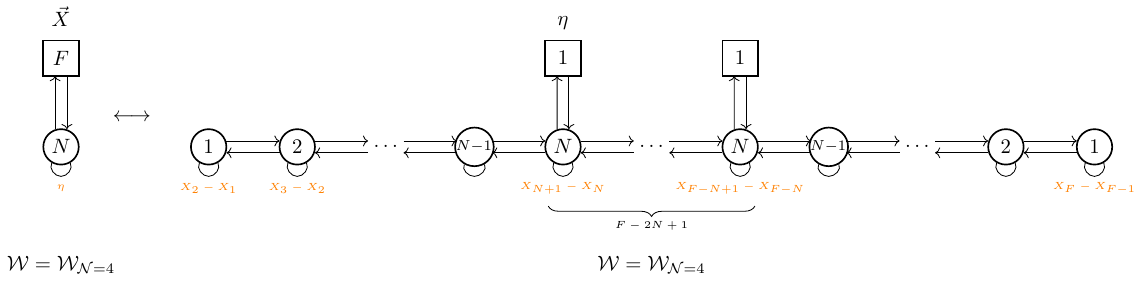}
    \caption{Mirror duality for $\mathcal{N}=4$ SQCD with gauge group $U(N)$ and $F$ flavors. Notice that the theories are depicted in the $\mathcal{N}=2$ notation.
    Round nodes denote gauge groups and square nodes denote flavor groups, while arrows denote chiral multiplets charged under the nodes that they connect.
    Black labels denote fugacities associated to flavor global symmetries while orange labels
     denote the FI of the corresponding gauge group.}
    \label{u_n_mirror}
\end{figure}

To study possible deformation breaking supersymmetry from $\mathcal{N}=4$ to $\mathcal{N}=2$ we work in the so-called $\mathcal{N}=2^*$ set-up \cite{Tong:2000ky}
taking $U(1)_R = U(1)_{C+H}$ as the R-symmetry group, where $U(1)_C\in SU(2)_C$ and $U(1)_H\in SU(2)_H$ of the non-abelian $\mathcal{N}=4$ R-symmetry $SU(2)_H\times SU(2)_C$. The commutant of $U(1)_R$, which is $U(1)_\tau=U(1)_{C-H}$, is a global symmetry from the point of view of the $\mathcal{N}=2$ theory.

However, we have the freedom to redefine $U(1)_R$ up to abelian flavor symmetries, meaning that we can shift the R-charge by $U(1)_\tau$ transformations. Using this freedom we pick a convention in which
we assign trial R-charge $1$ and $U(1)_\tau$ charge $-1/2$ to the fundamental and anti-fundamental chiral fields while adjoint chiral fields have R-charge $0$ and $U(1)_\tau$ charge $1$. 
The charges of the fields under the symmetries of the theories are reported in the tables below where we adopt an $\mathcal{N}=2$
notation denoting by $\Phi$, $Q$, $\tilde Q$ the adjoint and the fundamental/antifundamental chirals of the SQCD and  on the mirror side denoting
respectively by $\Phi_J$, ${b}_{J,J+1}$,  $\widetilde{b}_{J+1,J}$, 
the adjoints and the bifundamental chirals
and the left and right flavors  $q,\tilde q$ and $p,\tilde p$.

\begin{equation}
\begin{array}{c|c|c|c|c}
	 & SU(F) & U(1)_{\tau} & U(1)_R & U(1)_\eta \\
	\hline
	\Phi & 1 & 1 & 0 & 0 \\
	Q & \overline{\square} & -\frac{1}{2} & 1 & 0 \\
	\tilde{Q} & {\square} & - \frac{1}{2} & 1 & 0  
\end{array}
\qquad 
\begin{array}{c|c|c|c|c}
	 & SU(F) & U(1)_\tau &U(1)_R& U(1)_{\eta} \\
	\hline
	\Phi_J & 1 & -1 & 2 & 0 \\
	b_{J,J+1},\widetilde{b}_{J,J+1} & 1 & \frac{1}{2} &0& 0 \\
	q, \widetilde{q} & 1 & \frac{1}{2} & 0 & -1/+1 \\
	{p},\widetilde{p} & 1& \frac{1}{2}& 0 & 0
\end{array}
\end{equation}
Notice that  the R-charges and the  $U(1)_\tau$ charges in the mirror theory are obtained from the definitions given before, taking into account that mirror symmetry swaps $SU(2)_C \leftrightarrow SU(2)_H$.

\subsubsection{Global symmetries and Operator Map}
The global symmetry of the $U(N)$ SQCD with $F$ flavors, depicted on the l.h.s.~of \eqref{u_n_mirror}, is $SU(F) \times U(1)_\eta \times U(1)_\tau$\footnote{The actual faithful global symmetry of the theories is $PSU(F) \times U(1)_\eta \times U(1)_\tau$. We will not discuss the features related to the discrete factors of the global symmetry group. Moreover, the $U(1)_\eta$ symmetry enhances to $SU(2)$ in the IR if $F=2N$. Indeed in the mirror theory for $F=2N$ there is a manifest $SU(2)$ flavor symmetry instead of $U(1)$.
This will not have an effect in the computations that we will perform and we refer to the topological symmetry of the SQCD as being generically $U(1)_\eta$.}, where $U(1)_\eta$ is the topological symmetry. On the other hand, the manifest global symmetry of the mirror theory, depicted on the r.h.s.~of \eqref{u_n_mirror}, is $\prod_{j=1}^{F-1} U(1)_{X_{j+1}-X_j} \times U(1)_\eta \times U(1)_\tau$, where the first term is the collection of topological symmetries, while $U(1)_\eta$ is a flavor symmetry. In the IR this global symmetry enhances, in particular all the topological symmetries combine to give an enhanced $SU(F)$ global symmetry group.
We therefore see that the IR global symmetries of the two theories are in agreement and topological symmetries are mapped into flavor symmetries and vice-versa, as expected in a mirror duality.

It is also instructive to see how the operators of the electric theory are mapped to those of the mirror theory.
\begin{itemize}
    \item The fundamental monopoles $\mathfrak{M}^{\pm}$  of the SQCD are mapped into the long mesons on the mirror quiver side.
    \item The electric mesons, in  the traceless adjoint of $SU(F)$, are mapped to a collection of monopole operators in the mirror theory. The latter is  a  traceless  adjoint  matrix  of $SU(F)$ constructed with the $F-1$ traces of the adjoints and with monopoles ($\mathfrak{M}^{\ldots,0,\pm1,0,\ldots}$) with flux $\pm 1$  under one of the $F-1$ gauge nodes, $(F-2)$ monopoles  ($\mathfrak{M}^{\ldots,\pm 1,\pm1,0,\ldots}$) carrying GNO flux $\pm 1$ under two adjacent gauge nodes, $(F-3)$ monopoles ($\mathfrak{M}^{\ldots,\pm1,\pm1,\pm1,\ldots}$) carrying GNO flux $+ 1$  under three adjacent gauge nodes, and so on with the final monopole operators carrying GNO flux $\pm1$ under all $F-1$ gauge nodes.
\end{itemize}

\subsection{Real Mass Deformation to the $\mathcal{N}=2$ Chiral-Planar Mirror Pair}\label{sec:chiral-planar-sccd}

We now move on the SUSY-breaking deformation.
We perform a real mass deformation for the $U(1)_\tau$ symmetry, meaning that we turn on a non-zero VEV for the real scalar component of the background vector multiplet associated to the $U(1)_\tau$ symmetry. The VEV induces a mass term for every chiral superfield proportional to its charge under the symmetry. Recalling that $U(1)_\tau$ is a flavor subgroup of the $\mathcal{N}=4$ R-symmetry, the deformation performed has the effect of breaking the R-symmetry from $SU(2)_C \times SU(2)_H \to U(1)_R$, indeed this means that supersymmetry is broken from $\mathcal{N}=4$ to $\mathcal{N}=2$. \\
As we shall discuss in depth, in the most generic vacua this deformation triggers a flow to a TFT. However, we will try to suitably follow this deformation by moving on special points of the Coulomb branch, so to reach an interacting vacua. Both the SQCD and its mirror can have many possible interacting vacua that can be reached and the task of understanding how these are mapped correctly among them is the problem we attempt to solve in the following section. We will start by giving a qualitative description of the deformation of both the SQCD and its mirror in the following subsection, discussing the features of the resulting $\mathcal{N}=2$ mirror duality. In the following subsections we will show how this duality can be derived from two perspectives: the $\mathbf{S}^3_b$ partition function and the 1-loop corrected EOM. \\

\begin{figure}
    \centering
    \includegraphics[width=.85\textwidth]{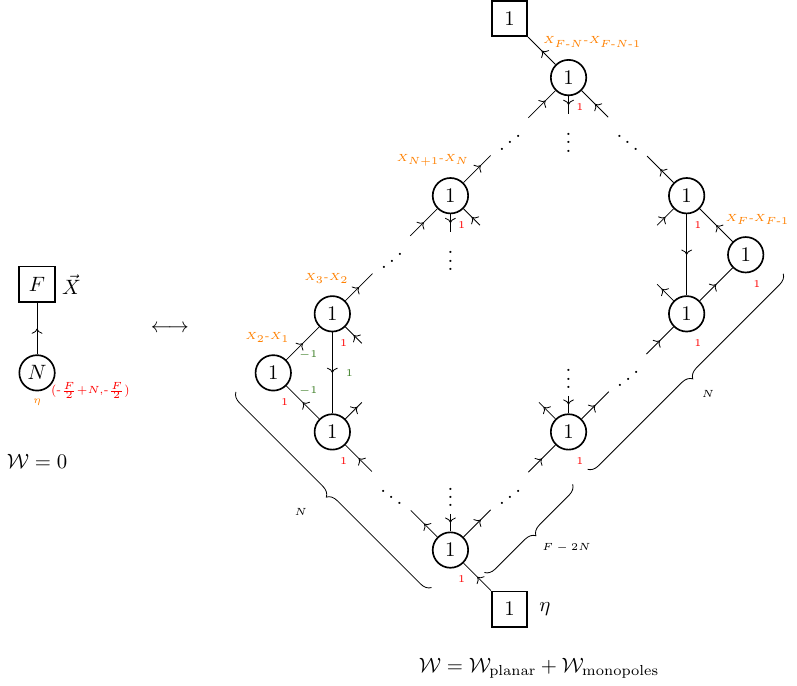}
    \caption{
    On the l.h.s.~it is reported the $\mathcal{N}=2$ $U(N)_{-\frac{F}{2}+N,-\frac{F}{2}}$ SQCD with $F$ fundamental fields obtained from the real mass deformation of the $\mathcal{N}=4$ SQCD. 
    On the r.h.s.~it is reported the planar mirror-like dual of the SQCD. Note that every gauge node has a CS level (red labels) $1$ and there are mixed CS terms (green labels) for every pair of nodes connected by a line, with level $-1/+1$ if the line is diagonal/vertical. 
    FI terms are indicated in orange.
    On the SQCD side we assign trial R-charge $1$ to the fundamental chirals and on the planar side the diagonal chiral fields have trial R-charge $0$ and the vertical ones $2$. The diagonal chirals pointing from SE to NW connect $N$ nodes, while the diagonal chirals  pointing from SW to NE connect $F-N$ nodes. }
    \label{planar_mirror_general} 
\end{figure}

Let us first discuss the SQCD side, on the l.h.s.~of Figure \ref{u_n_mirror}. As already commented, a real mass for the $U(1)_\tau$ symmetry naively renders every chiral field massive, and no massless fields are left in the deep IR. We then choose to move on the Coulomb branch in such a way to reach an interacting vacuum where all the fundamental chiral fields $Q$ remain massless while the adjoint $\Phi$ and the $F$ anti-fundamental chiral fields $\widetilde{Q}$ acquire a mass, negative and positive respectively, and are integrated out\footnote{One can also consider different interacting vacua where a certain number of fundamental and antifundamental fields remain massless. While this problem is tractable, we will tackle the case of SQCD theories with more general matter content in a different, algorithmic way in Section \ref{sec: Examples}.}. 
The procedure of integrating out chiral fields produces CS interactions due to fermionic modes becoming massive.
The resulting theory is the $\mathcal{N}=2$ $U(N)_{-\frac{F}{2}+N,-\frac{F}{2}}$ SQCD depicted on the l.h.s.~of Figure \ref{planar_mirror_general}.
Here, $U(N)_{k,k+l N}$\footnote{We follow the standard notation for Chern-Simons levels: $U(N)_{(k_1,k_2)} = \frac{SU(N)_{k_1} \times U(1)_{N k_2} }{\mathbb{Z}_{N}}$. To avoid anomalies $k_1 \in \mathbb{Z}$ and the $U(1)$ level must be of the form $k_2 = k_1 + l N$, with $l \in \mathbb{Z}$.} denotes a $U(N)$ gauge group with CS terms:
\begin{equation}
- i\frac{k}{4\pi} \int \text{tr} \left( A \wedge dA \right)
- i\frac{l}{4\pi} \int \text{tr} \left( A\right) \wedge \text{tr} \left( dA \right) + \text{SUSY completion}
\end{equation}
where $A$ is the $U(N)$ gauge field.
Each massive antifundamental multiplet, with negative mass, contributes as $(-\frac{1}{2},-\frac{1}{2})$ to the CS level, and the massive adjoint chiral, with positive mass, contributes as $(N,0)$.
Notice in particular that we have $l=-1$.
\\ 

We now discuss the result of the deformation in the mirror theory, depicted on the r.h.s.~of Figure \ref{u_n_mirror}. As already pointed out, many possible interacting vacua can be reached with the $U(1)_\tau$ real mass, each lying at different points of the Coulomb branch. We claim that 
the vacuum dual to the chiral SQCD theory described above is a special Coulomb branch point where each gauge node $U(k)$ of the original $\mathcal{N}=4$ mirror is Higgsed to its maximal torus $U(1)^k$. Also, some components of the chiral fields in the original theory remain massless. The complete description of the theory is depicted as a quiver on the r.h.s.~of Figure \ref{planar_mirror_general}. \\
A few comments regarding the planar dual on the r.h.s.~of Figure \ref{planar_mirror_general} are in order:
\begin{itemize}
    \item There is a cubic superpotential term for every triangular loop. There is one term with $-1$ ($+1$) coefficient for each clock-wise (anti-clock-wise) closed triangle. The cubic terms in the superpotential are a remnant of the cubic $\mathcal{N}=4$ superpotential that are preserved by the deformation. We denote these superpotential terms in short as $\mathcal{W}_{\text{planar}}$. The diagonal/vertical chiral multiplets have R-charge $0/+2$, compatible with the superpotential.

    \item $\mathcal{W}_{\text{monopole}}$ are the monopole terms in the superpotential generated by the Polyakov mechanism \cite{Polyakov1977} due to the Higgsing of a $U(k)$ gauge symmetry to $U(1)^k$. The exact collection of terms can be obtained with the following schematic rule. For each vertical arrow, there is a linear superpotential for the monopole with GNO flux $-1$ and $+1$ under the nodes connected by the arrow, from top to bottom.
    As a consequence of these monopole superpotential terms, the topological symmetry is broken to $U(1)^{F-1}$ which is expected to enhance in the infrared to $SU(F)$,
    thus matching the global symmetry on the electric side.
    The FI terms of the $U(1)$ gauge nodes are compatible with the monopole superpotential terms. With the choice of trial R-symmetry
    described above, the $\alpha^{th}$ gauge node in the $I^{th}$ column of the planar quiver ($\alpha=1,\dots, |G^{(I)}|$, and $|G^{(I)}|$ is the rank of the corresponding gauge node in the $\mathcal{N}=4$ quiver, or equivalently, the number of $U(1)$ gauge nodes in the $I^{th}$ column of the $\mathcal{N}=2$ quiver) has FI terms equal to $X_{I+1}-X_I+\frac{iQ}{4}(\delta_{\alpha,1}-\delta_{\alpha,|G^{(I)}|})$. In Figure  \ref{planar_mirror_general} we only reported the $X_I$-deopendent part of the FIs.
    
    \item CS interactions are present. Self and mixed-interactions can be encoded in a matrix $k_{ab}$ such that:
    \begin{equation}
        - i \frac{k_{ab}}{4\pi} \int A_a \wedge d A_b + \text{SUSY completion}
    \end{equation}
    where $A_a$ is the vector field of the $a$-th $U(1)$ gauge field. Indeed, by construction $k_{ab}$ is symmetric. For the theory in Figure \ref{planar_mirror_general} the rule is that each gauge node has (self) CS interaction with level $1$, denoted by the red labels in Figure \ref{planar_mirror_general}. Nodes connected by diagonal/vertical lines have a mixed-CS interaction with level $-1/+1$, that is $k_{ab} = \mp \tfrac{1}{2}$. Mixed CS levels are denoted by the green labels along the corresponging line in Figure \ref{planar_mirror_general}.
\end{itemize}

\subsubsection{Superconformal Index, global symmetries and operator map}\label{sec:SQCDmaps}
We now comment on the features of the duality described before, highlighting the tests that this duality passes.

The first test that we performed consists in the matching of the numerical expansion of the Superconformal Index. This guarantees that a number of properties are correctly reproduced by the dual theory such as: the presence of conserved currents associated to the correct global symmetry, the properties of the operators in the chiral ring and more.
We performed this test for  $N=2, \, F=4,5,6$ and $N=3 \,, F=6$, the resulting indices are reported in Table \ref{tab:SQCD_index}.
Notice that in order to perform the matching it is important to have the mirror theory parameterized as in Figure \ref{planar_mirror_general}. In particular the mixing between the R-symmetry and the topological symmetries is crucial for the enhancement of the global symmetry to be seen from the Superconformal Index. \\

\begin{table}
\centering
\begin{tabular}{||c | c | c | c||} 
 \hline
 N& F & r & Index \\
 \hline\hline
 2 & 4 & 1/5 & $1+ \eta \,x^{4/5} +\eta^2 \,x^{8/5}-16x^2+\eta^3 \, x^{12/5} +\eta^4 \, x^{16/5} +(\eta^5+88 )x^4 +O(x^{21/5}) $ \\ 
 \hline
 2 & 5 & 1/5 & $1+\eta  x+\left(\eta ^2-25\right) x^2+\eta ^3 x^3+\left(\eta ^4+250\right) x^4+\left(\eta ^5-100 \eta \right) x^5+O\left(x^6\right)$ \\
 \hline
 2 & 6 & 1/5 & $1+\eta  x^{6/5}-36 x^2+\eta ^2 x^{12/5}+\eta ^3 x^{18/5}+558 x^4+\eta ^4 x^{24/5}+O\left(x^{26/5}\right)$ \\
 \hline
 3 & 6 & 1/7 & 
 $\begin{array}{c}1+\eta  x^{6/7}+\eta ^2 x^{12/7}-36 x^2+\eta ^3 x^{18/7}+\eta ^4 x^{24/7}+558 x^4\\ +\eta ^5 x^{30/7}-225 \eta  x^{34/7}+O\left(x^{36/7}\right) \end{array}$
 \\
 \hline
\end{tabular}
\caption{Index of $U(N)_{(-\frac{F}{2}+N, -\frac{F}{2})}$ SQCD with $F$ fundamental chiral multiplets. The chiral multiplets are assigned trial R-charge $1-r$. We checked that the indices match with the corresponding mirror abelian quiver in Figure \ref{planar_mirror_general}. }
\label{tab:SQCD_index}
\end{table}

We now proceed by describing the matching of the global symmetries and of the chiral ring generators. 

The global symmetry of the SQCD theory is, up to discrete factors, $SU(F)\times U(1)_\eta$. On the mirror side we can match the rank of the global symmetry as follows. 
In the absence of the monopole superpotential there is a $U(1)$ topological symmetry for each $U(1)$ gauge node. The monopole superpotential breaks the topological symmetries associated to the nodes in a column of the quiver to a diagonal $U(1)$. Therefore there is a $U(1)^{F-1}$ unbroken topological symmetry associated to the $F-1$ columns of the planar quiver. Furthermore there is a single $U(1)_\eta$ flavor symmetry which is preserved by the planar superpotential. In total the rank of the global symmetry if $F$, matching the SQCD side.
The mirror-like duality implies that the $U(1)^F$ symmetry of the quiver theory is enhanced to $SU(F)\times U(1)$.

The chiral ring of the planar theory is generated by the meson built out of $F$ diagonal bifundamentals starting from the $U(1)_{\eta}$ flavor node and ending at the other flavor node, this operator has trial R-charge 0 and  and $U(1)_\eta$ charge $1$. 
There are many possible paths which are all identified by the F-terms due to the planar superpotential $\mathcal{W}_{\text{planar}}$.

We expect this meson to be mapped to some monopole operator generating the SQCD chiral ring, as observed from the Index. The bare monopole operators $\mathfrak{M}^{\pm 0\dots 0}$ are gauge variant but can be dressed to obtain gauge-invariant operators (as can be verified by following the discussion on monopole charges in Appendix \ref{app: monopole}). We argue that the relevant operators are dressed with the gaugino as follows. In a monopole background, i.e.~when we turn on GNO fluxes, the gauge group is effectively broken and the breaking pattern depends on the magnetic flux. In the case of fundamental monopoles the breaking pattern is $U(N) \to U(1) \times U(N-1)$. Consequently, all gauge variant operators decompose into representations of the residual group. 
The branching rules for irreducible representations of $U(N)\supset U(N-1)\times U(1)$ are given below:
\begin{align}
    \Box_N & = (\Box_{N-1},1)\oplus (\mathbf{1}, 1-N) \nonumber \\
    \Box_N\otimes \overline{\Box}_N & = (\Box_{N-1} \otimes \overline{\Box}_{N-1}, 0)\oplus (\Box_{N-1}, N)\oplus (\overline{\Box}_{N-1},-N)\oplus(\mathbf{1},0)
\end{align}
where ($\overline{\Box}_K$) $\Box_K$ denotes the (anti-) fundamental representation of $U(K)$ and $(\cdot, \cdot)$ denotes the representation under the residual group $U(N-1)\times U(1)$.
We see, in particular, that the gaugino of $U(N)$ decomposes into the gaugino of $U(N-1)$, a singlet, a fundamental operator and an anti-fundamental operator. 
In general, for a $U(N)$ gauge theory with $F$ fundamentals and generic CS levels $(k,k+lN)$ the bare monopoles $\mon^{\pm,0,\ldots,0}$ transform in 
a representaton of $U(N-1)$ with $(N-1)$-ality of $\mp l(N-1)$ and with 
charge $(-\tfrac{F}{2} \mp k \mp l)$ of $U(1)$ \cite{Benini_2011}. Dressing with $N-1$ powers of the fundamental component of the gaugino compensates these charges to:
\begin{equation}
    Q_{U(1)} (\mon^{\pm,0,\ldots,0}) = 
   ( N-1) - \left(\frac{F}{2} \pm k \pm l \right) \,.
\end{equation}
In the case studied in this section we have  $k=-\tfrac{F}{2}+N$ and $l=-1$ and the monopole $\mon^{+,0,\ldots,0}$ can be dressed $N-1$ times with the gaugino components in the fundamental representation of $U(N-1)$ to obtain a gauge invariant operator\footnote{If instead $k=\tfrac{F}{2}-N$ and $l=1$ the $\mon^{-,0,\ldots,0}$ monopole dressed $N-1$ times with the gaugino  is gauge invariant.}.
We expect that this operator is the generator of the chiral ring of $U(N)$ SQCD with levels $\left(-\frac{F}{2} +N, -\frac{F}{2} \right)$ and $F$ fundamentals, and is mapped on the planar side to the meson discussed above.
To support this claim, we compute the R-charge of this dressed monopole. The fundamental monopole $\mathfrak{M}^{+,0,\dots,0}$ has R-charge:
\begin{align}
     R(\mathfrak{M}^{+,0,\dots,0}) = (1-R_Q) \frac{F}{2}-(N-1) = 
             -(N-1) \,,
\end{align}
where $R_Q=1$ is the trial R-charge of the fundamental chirals.
The factor $-(N-1)$ is the contribution to the R-charge coming from the zero modes of the gaugini.
Dressing $N-1$ times with the gaugino contributes an additional factor of $N-1$ to the R-charge:
\begin{equation}
R(\mathfrak{M}^{+, 0, \dots, 0}\lambda^{N-1} )=
0 \,,
\end{equation}
thus matching the gauge invariant meson of the planar theory.

It is interesting to notice that by carefully analyzing the SQCD theory, it is possible to find that there is also a monopole with high topological charge which is naturally gauge invariant, without requiring any dressing. This monopole is $\mon^{+,+,\ldots,+}$, that has topological charge $N$ and trial R-charge $0$.
On the other hand in the theories discussed here, namely for $k=-F/2 +N$ and $l=-1$, we find that the  dressed fundamental monopole $\mathfrak{M}^{+, 0, \dots, 0}\lambda^{N-1}$ is in the chiral ring as well.
While this is non-trivial to establish from the SQCD point of view, the presence of this operator is manifest in the mirror dual where it is  mapped to the single mesonic operator with $U(1)_\eta$ charge $1$. Then the monopole $\mon^{+,+,\ldots,+}$ is the $N$-th power of the fundamental monopole in the chiral ring.

\subsubsection{The view from $\mathbf{S}^3_b$ partition function}
We now show how to derive the  chiral-planar mirror duality for the SQCD, depicted in Figure \ref{planar_mirror_general}, via an analysis of the $\mathbf{S}^3_b$ partition function \cite{Kapustin:2009kz, Hama_2011} following the strategies of \cite{Aharony:2013dha,Benini:2011mf}.

We start from the  $\mathbf{S}^3_b$ partition functions of the two theories in Figure \ref{u_n_mirror}. The partition function of the SQCD is:
\begin{equation}\label{u_n_el}
    \begin{aligned}
        Z(\vec{X},\eta,\tau) = &
        \int \prod_{\alpha = 1}^N \; d u_{\alpha} \; 
        \Delta_{(N)} (\vec{u},\tau)
        \; e^{2 \pi i \eta \sum_{\alpha = 1}^N u_\alpha} \; 
        \prod_{j=1}^{F} \; s_b\bigg(
        \frac{\tau}{2}\pm (u_{\alpha}-X_j)\bigg) \,;
    \end{aligned}
\end{equation}
while that of its mirror is:
\begin{equation}\label{mirror_un_Z}
    \begin{aligned}
        \hat{Z}(\eta,\vec{X},\tau) = & \; e^{2\pi i \eta \sum_{j=1}^{N} X_j} \; \int \; \prod_{I=1}^{F-1} \; \bigg( \prod_{\alpha=1}^{|G^{(I)}|} \; dz_{\alpha}^{(I)} e^{2\pi i z_{\alpha}^{(I)}(X_{I+1}-X_I)} 
              \Delta_{(|G^{(I)}|)} (\vec{z}^{(I)},iQ - \tau)
              \bigg) 
	\\
        & 
        \prod_{I = 1}^{F-2} \bigg[ \prod_{\alpha=1}^{|G^{(I)}|} \prod_{\beta=1}^{|G^{(I+1)}|} s_b \bigg( 
         \frac{iQ}{2}-\frac{\tau}{2} \pm (z_\alpha^{(I)} - z_\beta^{(I+1)}) \bigg) \bigg] \\
        & 
        \prod_{\alpha=1}^{|G^{(N)}|} \; s_b\bigg( 
         \frac{iQ}{2}-\frac{\tau}{2} \pm (z_\alpha^{(N)} - \eta) \bigg) \; \prod_{\alpha=1}^{|G^{(F-N)}|}  s_b\bigg(
          \frac{iQ}{2}-\frac{\tau}{2} \pm z_\alpha^{(F-N)} \bigg) \,;
    \end{aligned}
\end{equation}
where in the partition function of the mirror theory $\vec{z}^{(I)}$ denotes the fugacity of the $I^{th}$ gauge node whose rank is $|G^{(I)}|$. We also use the short notation for a $\mathcal{N}=4$ vector multiplet:
\begin{equation}
\Delta_{(N)}(\vec{z},\tau) = 
\frac{1}{N!} \;
\frac
    {\prod_{\alpha,\beta = 1}^N s_b (\frac{iQ}{2}-\tau + z_\alpha - z_\beta)}
    {\prod_{\beta < \alpha}^N s_b ( \frac{iQ}{2} \pm (z_\alpha - z_\beta))}
    \label{eq:n4measure}
\end{equation}
To write the $\mathbf{S}^3_b$ we followed the conventions of \cite{Jafferis:2010un, Kapustin:2009kz, Hama_2011}, wherein a chiral field of $R$-charge $r$ and charge $+1$ under an abelian flavor symmetry $U(1)_{x}$ contributes to the partition function as $s_b(\frac{iQ}{2} (1-r)-x)$, and $Q=b+b^{-1}$ is the squashing parameter. 

Indeed, since the SQCD and its mirror are IR-dual the two partition functions must be equal:
\begin{align}
    Z(\vec{X},\eta,\tau) = \hat{Z}({\eta},\vec{X},\tau) \,.
\end{align}
The two partition functions match as functions of the real mass parameters $\vec{X},\tau,\eta$. Here $\vec{X}$ is the set of real mass associated to the $SU(F)$ flavor symmetry of the SQCD and they thus satisfy $\sum_{j=1}^F X_j = 0$. From the mirror point of view the $X_j$ masses parameterize the $F-1$ topological symmetries in a convenient way so that they reflect how the Cartan subgroup of $SU(F)$ maps across the duality. Also, $\eta$ is the real mass associated to the $U(1)$ topological/flavor symmetry of the SQCD/mirror\footnote{Indeed, for $F=2N$ the parameter $\eta$ becomes the real mass of the $SU(2)$ enhanced topological symmetry. This subtlety does not spoil any of the analysis done in this section}. Finally, $\tau$ is the real mass of the $U(1)$ flavor subgroup of the $\mathcal{N}=4$ R-symmetry. \\

At the level of the partition function, the real mass deformation can be implemented as follows \cite{Aharony:2013dha,Benini:2011mf}: we perform the  $\tau \to +\infty$ limit combined with a suitable shift in gauge fugacities. Let us start the analysis considering the SQCD. The gauge shift performed is:
\begin{equation}
     \vec{u} \rightarrow \vec{u} + \frac{\tau}{2}
\end{equation}
accompanied by a shift of the FI parameter:
\begin{equation}
    \eta \rightarrow \eta - \frac{F}{2} \tau \,.
\end{equation}
Performing the limit $\tau \to +\infty$ takes us to the non-trivial Coulomb branch vacuum where the fundamental chiral fields remain massless. 
The limit is implemented using the asymptotic behavior of the double sine function:
\begin{equation}  
\lim_{\xi\to\pm\infty}\;s_b(x+\xi) \sim e^{\pm \frac{\pi i}{2}(x+\xi)^2} \,.
\label{sbl}
\end{equation}

In general, if $x$ is a global parameter then the limit produces highly oscillating phases as well as finite phases corresponding to background terms for global symmetries such as BF couplings. If instead $x$ is a gauge parameter, the limit produces CS interactions and divergent FI parameters together with a highly oscillating phase. The redefinition of $\eta$ is chosen to cancel exactly every divergent FI parameter produced by the limit. 
In the end, we obtain a highly oscillating phase $e^{i \Phi(\tau,\eta)}$ independent of the gauge fugacities which therefore factorizes outside the integral:
\begin{equation}
    Z(\vec{X},\eta,\tau) \to e^{\frac{i\pi}{2} \Phi( \tau,\eta)} Z(\vec{X},\eta) \,,
\end{equation}
with
\begin{equation}
    \Phi(\tau,\eta)=
     -N^2 \tau^2 
   + 2N \tau \left(
     \eta + \frac{iQ}{2}N
   \right)  \,,
\end{equation}
and\footnote{
Our sign convention is as follows - a $U(N)$ gauge node with FI parameter $\eta$ and CS level $(k_1,k_2)$ contributes to the $\mathbf{S}^3_b$ partition function as, with $\vec{u}$ denoting the gauge fugacity:
\begin{equation}
    e^{2\pi i \eta \sum_j u_j}\;e^{-\pi i k_1 \sum_j u_j^2}\; e^{-\frac{\pi i (k_2-k_1)}{N}(\sum_j u_j)^2}.
\end{equation}
A mixed CS (BF) level $k_{ij}$ contributes to the $\mathbf{S}^3_b$ partition function as: $e^{-\pi i k_{ij}u_iu_j}$. If $k_1=k_2\equiv \kappa$, we specify the CS level as $(\kappa)$.}:

\begin{equation}\label{u_n_chiral}
    \begin{aligned}
         Z(\vec{X},{\eta}) = & 
         e^{-\frac{N\pi i}{2} \sum_{j=1}^F X_j^2} \; 
         \int \prod_{\alpha=1}^N \; du_{\alpha} \; 
         	\frac{1}{\prod_{\alpha<\beta}s_b\bigg(\frac{iQ}{2}\pm(u_{\alpha}-u_{\beta})\bigg)} \\ 
         & 
         e^{2\pi i \eta \sum_{\alpha} u_{\alpha} 
         + \pi i (\sum_{\alpha = 1}^N u_{\alpha})^2 
         + \pi i (\frac{F}{2}-N)\sum_{\alpha}u_{\alpha}^2}
         \prod_{j=1}^{F} \;
        		s_b \bigg( X_j-u_{\alpha}  \bigg) \,.
    \end{aligned}
\end{equation}

The phase will be crucial in determining the mirror dual theory. Indeed, as explained in \cite{Aharony:2013dha,Benini:2011mf} the vacua related by duality under the deformation must have the same oscillating phase. 
In general, different shifts in the gauge fugacities (corresponding to moving to different points on the Coulomb branch) give different oscillating phases so there can be multiple vacua on the electric and magnetic sides and the match of the phases is a necessary criterion to map vacua related by duality.

We now move to the analysis of the partition function of the mirror theory.

We study the limit partition function of the mirror theory $\check{Z}(\eta,\vec{X},\tau)$ for large $\tau$ and with $\eta \to -\tfrac{F}{2}\tau + \eta$. There is a large number of possible $z^{(I)}_\alpha$ shifts that can be performed, each corresponding to different points on the Coulomb branch and each producing (possibly) a different highly oscillating phase times the partition function of an interacting theory describing the vacuum. 
We look for a vacuum that
reproduces the same oscillating prefactor as the SQCD. Schematically this means that:
\begin{equation}
    Z(\vec{X},\eta,\tau) = \check{Z}(\eta,\vec{X}, \tau)
    \quad\xrightarrow[]{\tau \to + \infty}\quad
    e^{\frac{i\pi}{2} \Phi(\tau)} Z(\vec{X},\eta) 
    = e^{\frac{i\pi}{2} \Phi(\tau)} \check{Z}(\eta,\vec{X})
\end{equation}
implying the equality of the partition function of the two $\mathcal{N}=2$ theories living in the vacua.
We found that the correct shift for the $z^{(I)}_\alpha$ is such that:
\begin{equation}\label{eq:mirzshift}
\begin{aligned}
    & z^{(I)}_\alpha - z^{(I)}_{\alpha+1} \to z^{(I)}_\alpha - z^{(I)}_{\alpha+1} -\tau \\
    & z^{(I)}_\alpha - z^{(I+1)}_\alpha \to z^{(I)}_\alpha - z^{(I+1)}_\alpha +\frac{\tau}{2},\qquad I< F-N  \\
    & z^{(I)}_\alpha - z^{(I+1)}_\alpha \to z^{(I)}_\alpha - z^{(I+1)}_\alpha -\frac{\tau}{2},\qquad I\geq F-N  \\
    & z^{(N)}_1 \to z^{(N)}_1 -\frac{F-1}{2} \tau\,.
\end{aligned}
\end{equation}

As a result, each gauge node is fully Higgsed to a Cartan subgroup and the chiral adjoint and bifundamentals decompose accordingly. Most of these fields are massive and are integrated out, resulting in the  $\mathcal{N}=2$ quiver gauge theory shown on the r.h.s.~of Figure \ref{planar_mirror_general}. \\

Notice that when we performed the $\tau$ limit, the shift of the gauge fugacities in \eqref{eq:mirzshift} is not invariant under a Weyl transformation $S_k$ associated to any gauge group $U(k)$. Therefore if we take a different shift by reshuffling the $k$ gauge parameters associated to a gauge symmetry we obtain the same result after the limit. Each of these $k!$ possibilities represent a different point of the Coulomb branch where the correct magnetic theory lives. One should then sum over all these vacua, all reproducing the same highly oscillating prefactor, effectively producing a $k!$ factor that cancel that at the denominator in the definition of the $\Delta_{(N)}$ measure in \eqref{eq:n4measure}.

By taking the real mass limit on the partition function on the mirror side we obtain the partition function of the quiver:

\begin{equation}\label{mirpartfun}
    \begin{aligned}
    \hat{Z}(\eta,\vec{X})=  
    & e^{2 \pi i \eta (- (2N-1)\frac{iQ}{4} + \sum_{i=1}^N X_i)} 
    e^{-\frac{i \pi}{2}\eta^2}  
    \int \prod_{I=1}^{F-1} \prod_{\alpha=1}^{|G^{(I)}|} dz_{\alpha}^{(I)} 
    e^{2\pi i z_{\alpha}^{(I)}(X_{I+1}-X_I
    )}
    e^{-i \pi (z_{\alpha}^{(I)})^2}\\
    & 
    \prod_{I=1}^{F-1} \prod_{\alpha}
    s_b\left(-\frac{iQ}{2}+z_{\alpha+1}^{(I)}-z_{\alpha}^{(I)} \right) 
    e^{-i\pi z_{\alpha}^{(I)} z_{\alpha+1}^{(I)}} \\
    & 
    \prod_{I=1}^{N-1}\prod_{\alpha}
    s_b\left(\frac{iQ}{2} +z_{\alpha}^{(I)} - z_{\alpha}^{(I+1)} \right)
    s_b\left(\frac{iQ}{2} -z_{\alpha}^{(I)} + z_{\alpha+1}^{(I+1)} \right) 
    e^{i\pi z_{\alpha}^{(I)} z_{\alpha+1}^{(I+1)}}\\
    & 
    \prod_{I=N}^{F-1}\prod_{\alpha}
    s_b\left(\frac{iQ}{2} -z_{\alpha}^{(I)} +z_{\alpha}^{(I+1)} \right)
    s_b\left(\frac{iQ}{2} +z_{\alpha}^{(I)} - z_{\alpha+1}^{(I+1)} \right) 
    e^{i\pi z_{\alpha}^{(I)} z_{\alpha+1}^{(I+1)}}\\
    & 
    s_b\left(\frac{iQ}{2}-\eta+z_{1}^{(N)}\right)
    s_b\left(\frac{iQ}{2}-z_{N}^{(F-N)}\right)
    e^{i\pi z_{1}^{(N)}\eta}
    \end{aligned}
\end{equation}
Where the products run over all the $\alpha$ such that the fugacities $z$ in the corresponding double sine functions are within the ranges described above.
Notice that in the first line are encoded the background terms, that are not included in the Figure \ref{planar_mirror_general}.

\subsubsection{The view from the semiclassical EOM}
We now show that the flow from $\mathcal{N}=4$ to $\mathcal{N}=2$ mirror dualities described in Subsection \ref{sec:chiral-planar-sccd} is consistent with the equation of motion of the $\mathcal{N}=4$ theories. In particular, we show that the vacua corresponding to the shifts in gauge fugacities satisfies the F-terms and D-terms equations. Furthermore, we show that fluctuations around these vacua reproduce the EOM of the resulting $\mathcal{N}=2$ theories.
This proves to be an important check of our proposal.

We start our analysis by first considering the $\mathcal{N}=4$ $U(N)$ SQCD with $F$ flavors. Recall that, as an $\mathcal{N}=2$ supersymmetric theory, it has $F$ (anti-)fundamental fields $Q$ ($\tilde{Q}$) and an adjoint chiral multiplet $\Phi$. Additionally, we also turn on a real FI parameter $\eta$. Upon diagonalizing the real scalar $\sigma$ in the vector multiplet, the equations for the F- and D-terms are as follows (note that each field represents a matrix)\footnote{N.B.: Latin letters $i,j=1,\ldots,F$ run over flavor indices, and Greek letters $\alpha,\beta=1,\ldots,N$ run over gauge indices. The sum over indices is never implied.} \cite{Intriligator_2013}:
\begin{equation}\label{eq:SQCD_EOM}
\begin{aligned}
    & \sum_{i=1}^F Q^i_\alpha \tilde{Q}_i^\beta = 0 \,,
    \qquad \sum_{\alpha = 1}^N Q^i_\alpha \Phi^\alpha_\beta = 0 \,, 
    \qquad \sum_{\alpha = 1}^N \Phi^\beta_\alpha \tilde{Q}^\alpha_i = 0 \,, \\ 
    & (\sigma_\alpha - X_i - \tfrac{\tau}{2}) Q^i_\alpha = 0 \,, 
    \qquad (X_i - \sigma_\alpha - \tfrac{\tau}{2}) \tilde{Q}_i^\alpha = 0 \,,
    \qquad [\sigma,\Phi]_\alpha^\beta + \tau \Phi_\alpha^\beta = 0 \,, \\  
    & [\Phi,\Phi^{\dagger}]_\alpha^\beta + \sum_{i=1}^{F}(Q^{i,\dagger}_\alpha Q^i_\beta  - \tilde{Q}^\alpha_i \tilde{Q}^{\beta,\dagger}_i ) = 
    \frac{\delta_\alpha^\beta}{2\pi} \bigg[ 
    - \eta +\frac{1}{2} \sum_{i=1}^F \big(  \big| \sigma_\alpha - X_i - \frac{\tau}{2} \big| - \big| X_i - \sigma_\alpha - \frac{\tau}{2} \big| \big) + \\
    & + \frac{1}{2}\sum_{\gamma=1}^N ( |\sigma_\alpha - \sigma_\gamma + \tau | - |\sigma_\gamma - \sigma_\alpha + \tau| ) 
    \bigg] \,.
\end{aligned}
\end{equation}
Where the values of $Q_\alpha^i, \tilde{Q}^\alpha_i, \Phi^\alpha_\beta$ are the Vacuum Expectation Values (VEVs) of the (scalar component of) chiral fields of the theory. $\sigma_\alpha$ is the VEV for the real scalar in the $\mathcal{N}=2$ vector multiplet, noting that in principle it is a matrix in the adjoint representation of the $U(N)$ gauge symmetry but we can always perform a gauge transformation such that it is diagonal and we employ the single index notation for the diagonal components $\sigma_\alpha^\beta = \delta_\alpha^\beta \sigma_\alpha$. 
Lastly, the real parameters $X_i, \tau, \eta$ are the real masses associated to the $SU(F)$/$U(1)_\tau$/$U(1)_\eta$ global symmetries, i.e.~VEVs of the real scalars in background $\mathcal{N}=2$ vector multiplets of the global symmmetries. Since $X_i$ parameterizes a $SU(F)$ symmetry they satisfy $\sum_{i=1}^F X_i = 0$. \\
The set of equations in \eqref{eq:SQCD_EOM} represent the following: In the first line the F-term constraint imposed by the $\mathcal{N}=4$ superpotential $\mathcal{W} = \sum_{i=1}^F \sum_{\alpha,\beta=1}^N Q^i_\alpha \Phi^\alpha_\beta \tilde{Q}^\beta_i $; in the second line the mass term for the chiral fields; in the third and last line the 1-loop corrected D-term equation. \\

We now consider a mass deformation for the axial symmetry $U(1)_\tau$, which corresponds to taking a positive large value of $\tau$ in \eqref{eq:SQCD_EOM}.
While it would be interesting to study the most general solution to the vacuum equations under this deformation, we do not attempt to solve this problem here.
Instead, we look for a vacuum where the fundamental fields remain massless.
In order to do so we consider a large $\tau$ real mass, and set the real masses $X_i$ to zero.

The effective mass of the field $Q^i_\alpha$ is given by $(\sigma_\alpha - \tfrac{\tau}{2})$, therefore we set $\sigma_\alpha =  \tfrac{\tau}{2}$. The first line of equations in \eqref{eq:SQCD_EOM} then requires that $\tilde{Q}^\alpha_i = \Phi^\alpha_\beta = 0$. Also, although the condition $\sigma_\alpha = \tfrac{\tau}{2}$ is enough to solve the first equation in the first line, we also require $Q_\alpha^i$ to be zero, in order for these fields to not acquire any VEV and to therefore be light degrees of freedom of the theory in the vacuum.  The D-term equation therefore reads:
\begin{equation}        
    0 = \frac{1}{2\pi} \bigg[ 
    - \eta - \frac{F}{2} \tau 
    \bigg] \; \longrightarrow \;
    \eta = - \frac{F}{2} \tau
\end{equation}
Therefore, we are performing a real mass deformation for both the $U(1)_\tau$ and $U(1)_\eta$ symmetries, keeping fixed the combination $\tfrac{F}{2} U(1)_\tau + U(1)_\eta$.
This corresponds to the deformation considered in the previous subsection. \\

Now to conclude, we would like to study the theory that lives in the vacuum solution that we found. To do so, we study small fluctuations around the solution:
\begin{equation}
\begin{aligned}
    & Q^i_\alpha = \check{Q}^i_\alpha \\
    & X_i = \check{X}_i \\
    & \sigma_\alpha = \tfrac{\tau}{2} + \check{\sigma}_\alpha \\
    & \eta = - \frac{F}{2} \tau + \check{\eta} 
\end{aligned}
\end{equation}
Notice that the equations in the second line of \eqref{eq:SQCD_EOM} impose that the $\tilde{Q}$ and $\Phi$ fields do not fluctuate, as expected since they do not describe light degrees of freedom in the chosen vacuum. \\
Expanding the equations in the light modes we get:
\begin{equation}
\begin{aligned}
    & (\check{\sigma}_\alpha - \check{X}_i)\check{Q}^i_\alpha = 0 \\
    & \sum_{i=1}^{F}Q_{\alpha}^{i, \dagger}Q^{\beta}_i =  
    \frac{\delta_{\alpha\beta}}{2\pi}\bigg[
    - \underbrace{\check{\eta}}_{=\,\eta_{eff}} 
    +\underbrace{ \bigg( -\frac{F}{2}+N \bigg) \check\sigma_{\alpha} - \sum_{\gamma=1}^N\check\sigma_{\gamma}}_{\text{CS interactions}} + \frac{1}{2}\sum_{i=1}^{F}|\check\sigma_{\alpha}-\check{X}_i|\bigg]
\end{aligned}
\end{equation}
Which can be interpreted as the EOM of a $U(N)$ gauge theory with $F$ massless fundamental chirals and with CS level $(-\tfrac{F}{2} +N, - \tfrac{F}{2})$ by comparing with the generic formula:
\begin{equation}
\begin{aligned}
    \sum_{i=1}^{F} Q_{\alpha}^{i, \dagger} Q^{\beta}_i =  \frac{\delta_{\alpha\beta}}{2\pi}\bigg[ -\eta_\text{eff} + k_\text{eff} \sigma_\alpha + l_\text{eff} \sum_{\gamma=1}^N \sigma_{\gamma} + \frac{1}{2}\sum_{i=1}^{F}|\sigma_{\alpha}-\check{X}_i| \bigg]
\end{aligned}
\end{equation}
where the CS level is $(k,k+lN)$. \\

We now turn to the mirror of the $\mathcal{N}=4$ $U(N)$ SQCD theory in Figure \ref{u_n_mirror}. Studying the vacuum of this theory using the equations of motion is a highly non-trivial task and is not attempted here. The difficulty of this task lies clearly in the high number of equations that must be analysed. Another issue is that for the same problem, i.e.~same values of the real mass parameter in input, there can be multiple solutions and therefore multiple possible vacua. 
\\

Instead, we employ mirror duality to know how the real mass parameters are mapped across the duality so that we can fix correctly the problem we attempt to study. 
The system of equations for the mirror theory are\footnote{The index between the round brackets labels the number of the gauge node. The latin indices are used for gauge symmetries.}:
\begin{equation}\label{eq:mir_massterms}
\begin{aligned}
    & 
    (\sigma_i^{(a)} - \sigma_j^{(a+1)} + \tfrac{\tau}{2} ) b^{(a,a+1)}_{i,j} = 0 \,; \qquad
     (- \sigma_i^{(a)} + \sigma_j^{(a+1)} + \tfrac{\tau}{2} ) \tilde{b}^{(a,a+1)}_{i,j} = 0 \\
    & 
    (\sigma^{(N)}_i - \eta + \tfrac{\tau}{2} ) q_i = 0 \,; \qquad 
    (- \sigma^{(N)}_i + \eta + \tfrac{\tau}{2} ) \tilde{q}_i = 0 \\
    & 
    (\sigma^{(F-N)}_i + \tfrac{\tau}{2}) p_i = 0 \,; \qquad 
    (-\sigma^{(F-N)}_i + \tfrac{\tau}{2}) \tilde{p}_i = 0 \\
    & 
    [\sigma^{(a)},\Phi^{(a)}]_i^j - \tau {\Phi^{(a)}}_i^j = 0
\end{aligned}
\end{equation}
All the $\sigma$'s are taken to be diagonal after an appropriate gauge fixing.
Also, we have the set of F-terms equations, that we do not attempt to report here since they will not be important for the following discussion. And lastly we have the D-terms equations that are:
\begin{equation}
\begin{aligned}
    & [\Phi^{(a)},\Phi^{(a),\dagger}]_{i,j} + 
    \ldots (\text{sum over $b^{(a,a+1)}$'s and eventual $q,p$})
    = \frac{\delta_{i,j}}{2\pi} F^{(a)}_i \,, \\
    & F^{(a)}_i = X_{a+1} - X_a + \sum_{k=1}^{a+1} \big( \big| \sigma^{(a)}_i - \sigma^{(a+1)}_k + \frac{\tau}{2} \big| - \big| -\sigma^{(a)}_i + \sigma^{(a+1)}_k + \frac{\tau}{2} \big| \big) + \\
    & 
    + \sum_{k=1}^{a-1} \big(- \big| \sigma^{(a-1)}_k - \sigma^{(a)}_i + \frac{\tau}{2} \big| + \big| -\sigma^{(a-1)}_k + \sigma^{(a)}_i + \frac{\tau}{2} \big| \big) + \\
    & 
    + \sum_{k=1}^a (|\sigma^{(a)}_i - \sigma^{(a)}_k - \tau| - |-\sigma^{(a)}_i + \sigma^{(a)}_k - \tau|) \,.
\end{aligned}
\end{equation}
The solution that we are looking for have non-zero positive value of $\tau$, the value of $\eta$ is fixed to $- \frac{F}{2} \tau$ and $X_a=0$. Notice that the value of the external parameters is fixed by the analysis of the SQCD.
We observe that there is a solution for the unknown vairiables $\sigma^{(a)}_i$ obtained by taking:
\begin{equation}
\begin{aligned}
    &  \sigma^{(a)}_\alpha - \sigma^{(a)}_{\alpha+1} = -\tau \\
    &  \sigma^{(a)}_\alpha -  \sigma^{(a+1)}_\alpha = +\frac{\tau}{2},\qquad a< F-N  \\
    &  \sigma^{(a)}_\alpha -  \sigma^{(a+1)}_\alpha = -\frac{\tau}{2},\qquad a\geq F-N  \\
    & \sigma^{(N)}_1 = -\frac{F-1}{2} \tau
\end{aligned}
\end{equation}
This corresponds to the vacuum considered in the previous section \eqref{eq:mirzshift}, where the VEVs for $\sigma$'s indicate the pattern of Higgsing. The light DOFs can be found by computing the masses of the various fields from the expression \eqref{eq:mir_massterms}, which gives the same answer as the analysis of the $\mathbf{S}^3_b$ partition function. One can also try to perform a perturbative expansion of the equation of motion, as we did for the SQCD, to find that this analysis reproduces correctly all the properties of the chiral-planar mirror duality of Figure \ref{planar_mirror_general}, such as the self and mixed-CS interactions. The planar superpotential are the F-term relations surviving in the vacuum. On the other hand, the monopole superpotential can not be inferred easily from the analysis of the EOM.

To summarize, we have shown that the vacua corresponding to the real mass flows considered in the previous Section satisfy the EOM of the $\mathcal{N}=4$ theory. This, combined with the matching of the $\mathbf{S}_b^3$ partition function and the Superconformal Index discussed above provide a strong check of the chiral-planar mirror duality for $\mathcal{N}=2$ SQCD. \\

As a technical aside we would like to comment on the similarities and differences between the analysis of the $\mathbf{S}_b^3$ partition function and the EOM. 
It can be checked that, under a real mass deformation $\tau$, the requirement that the $\mathbf{S}_b^3$ partition function factorizes into a divergent phase and a finite integral is equivalent to solving the D-term equations up to orders $\mathcal{O}(\tau^0)$. This is due to the fact that large shifts in the gauge fugacities appearing in the $\mathbf{S}_b^3$ partition function correspond to moving on the Coulomb branch far away from the origin.
On the other hand, the $\mathbf{S}_b^3$ partition function is not sensitive to motion on the Higgs branch, therefore solving the other EOM provide an additional check of the deformation considered above.
When there are multiple vacua the analysis of the EOM does not provide a general prescription to understand the mapping of vacua across the duality. The divergent phase coming from the $\mathbf{S}_b^3$ partition function is sensitive to the matter that becomes massive at a given vacuum, and therefore can be leveraged to constrain the mapping of vacua across the duality.

As an example, both the $\mathcal{N}=4$ SQCD and its mirror have a \textit{chiral-like} and a \textit{planar-like} vacuum under the axial mass deformation. 
In \cite{Tong:2000ky}, where the $\mathcal{N}=2^*$ framework was introduced, the axial mass deformation appeared to lead to \textit{chiral-chiral} dualities.
In the previous Section we showed that the \textit{chiral-like} vacuum of one theory can be mapped to the \textit{planar-like} vacuum of the mirror theory because the divergent prefactors in the $\mathbf{S}_b^3$ partition function cancel between the two. Such a cancellation does not happen for pairs of \textit{chiral-like} vacua or \textit{planar-like} vacua, showing that such a mapping of vacua is not consistent.
Therefore, our results represent an example where refined precision tools such as the computation of the $\mathbf{S}_b^3$ partition function can be leveraged to understand qualitative properties of dualities between QFTs.

\subsection{A Planar Dual for $SU(N)_k$ and $U(N)_{(k,k+\ell N)}$ SQCD}\label{ga-ung-section}
In this paper, we primarily consider dualities involving $U(N)_{(k,k-N)}$ gauge theories (that is, theories where the abelian CS level $\ell$ is $-1$). One can obtain an $SU(N)_{k}$ CS theory by introducing a background BF term between the topological symmetry $U(1)_Y$ and a new $U(1)_B$ and then gauging the topological symmetry.
This operation is referred to as \textit{ungauging} of the diagonal $U(1)\subset U(N)$. 
In the notation used throughout this paper, this ungauging operation consists 
in 
\begin{equation}	\label{eq:ungauging_U1}
\int dY e^{2\pi i Y B N}
\begin{tikzpicture}[baseline=20pt]
\node[gauge,black] (g) at (0,1.5) {$N$};
\node[flavor,black] (f) at (0,0) {$F$};
\path[->-,draw] (g) -- (f);
\draw[red] (g)++(0.2,-0.1) node[anchor=north west] {\tiny$(k,k+N\ell )$};
\draw[red] (f)++(0.3,-0.2) node[anchor=north west] {\tiny$\xi$};
\draw[FIcolor] (g)++(0,0.3) node[anchor=south] {\tiny$Y+\Delta Y$};
\draw[black] (f)++(0.3,0.2) node[anchor=west] {$\vec{X}$};

\end{tikzpicture}
=
\begin{tikzpicture}[baseline=20pt]
\node[gauge,black,double] (g) at (0,1.5) {$N$};
\node[flavor,black] (f) at (0,0) {$F$};
\path[->-,draw] (g) -- (f);
\draw[red] (g)++(0.3,-0.2) node[anchor=north west] {\tiny$k$};
\draw[red] (f)++(0.2,-0.1) node[anchor=north west] {\tiny$\left(\xi, \xi + F\left(\frac{Nk + N^2 \ell - \xi F}{F^2}\right)\right)$};
\draw[FIcolor] (f)++(-0.2,-0.4) node[anchor=north] {\tiny$-\Delta Y\frac{N}{F}$};
\draw[black] (f)++(0.3,0.2) node[anchor=west] {$\vec{\widetilde{X}}$};

\end{tikzpicture}
,\qquad
{\color{black}\widetilde{X}_a = X_a + B}
\end{equation}
wherein the double circle denotes a $SU(N)$ gauge symmetry. In the $U(N)$ SQCD the flavor symmetry is $SU(F)$ and thus $\sum_{j=1}^F X_j = 0$ and $\xi$ is a possible background CS for the flavor symmetry.
In the $SU(N)$ theory $\sum_{j=1}^F \tilde{X}_j = F B$, where $U(1)_B$ is a baryonic symmetry.

Conversely, starting from SQCD with gauge group $SU(N)$ one can gauge the baryonic  symmetry  $U(1)_B$ to obtain a unitary gauge group:
\begin{equation}	\label{eq:gauging_U1}
\int d(BN) e^{-2\pi i \eta B N}
\begin{tikzpicture}[baseline=20pt]
\node[gauge,black,double] (g) at (0,1.5) {$N$};
\node[flavor,black] (f) at (0,0) {$F$};
\path[->-,draw] (g) -- (f);
\draw[red] (g)++(0.3,-0.2) node[anchor=north west] {\tiny$k$};
\draw[red] (f)++(0.3,-0.2) node[anchor=north west] {\tiny$\left(\xi_1, \xi_1 + F \xi_2\right)$};
\draw[FIcolor] (f)++(-0.2,-0.4) node[anchor=north] {\tiny$-\Delta 
\eta\frac{N}{F}$};
\draw[black] (f)++(0.3,0.2) node[anchor=west] {$\vec{\widetilde{X}}$};

\end{tikzpicture}
=
\begin{tikzpicture}[baseline=20pt]
\node[gauge,black] (g) at (0,1.5) {$N$};
\node[flavor,black] (f) at (0,0) {$F$};
\path[->-,draw] (g) -- (f);
\draw[red] (g)++(0.2,-0.1) node[anchor=north west] {\tiny$\left(k,k+N\left(\frac{F \xi_1 + F^2 \xi_2 - Nk}{N^2}\right)\right)$};
\draw[red] (f)++(0.3,-0.2) node[anchor=north west] {\tiny$\xi_1$};
\draw[FIcolor] (g)++(0,0.3) node[anchor=south] {\tiny$\eta+\Delta \eta$};
\draw[black] (f)++(0.3,0.2) node[anchor=west] {$\vec{X}$};
\end{tikzpicture}
,\qquad
{\color{black}\widetilde{X}_a = X_a + B}
\end{equation}

Where $\xi_{1}$ and $\xi_{2}$ are possible background CS levels for the $U(F)$ global symmetry of the SQCD on the r.h.s. and $-\Delta\eta \tfrac{N}{F}$ is a background FI term.
Consistently, one can check that by performing both transformations \eqref{eq:ungauging_U1} and \eqref{eq:gauging_U1}, a theory is mapped back to itself.

\begin{figure}[ht]
    \centering
    \includegraphics[width=.85\textwidth]{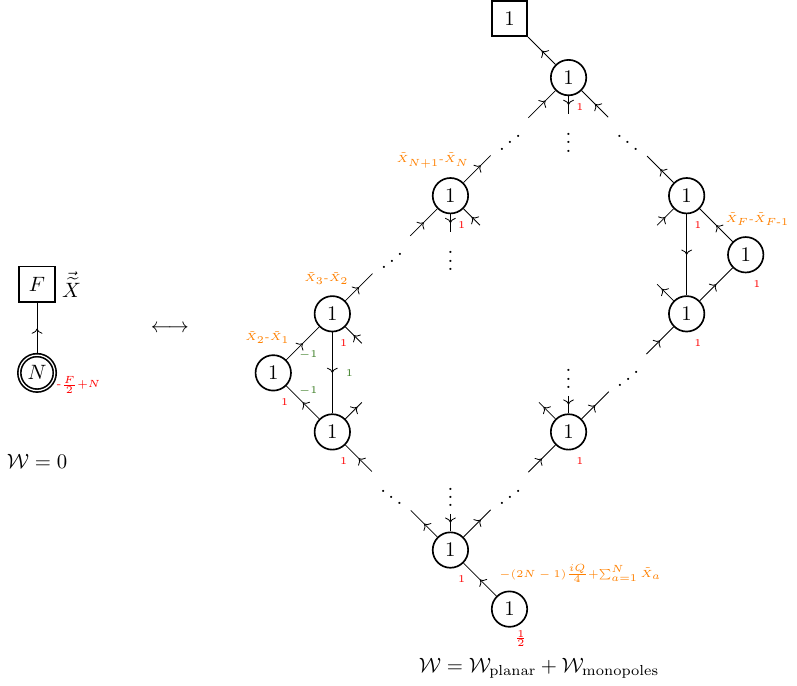}
    \caption{The $\mathcal{N}=2$ planar mirror dual of $SU(N)_{-\frac
    {F}{2}+N}$ SQCD with $F$ fundamentl chiral multiplets.}
    \label{sun_sqcd}
\end{figure}

We may apply the ungauging operation \eqref{eq:ungauging_U1} to the chiral-planar  mirror duality of SQCD in Figure \ref{planar_mirror_general}.
On the electric side we obtain the $SU(N)_{-F/2+N}$ SQCD.
On the mirror side, the topological symmetry corresponds to a flavor symmetry acting on the bottom flavor node. Therefore, the operation described above corresponds to gauging the flavor node itself. In order to correctly perform this gauging it is important to keep track of possible background terms in the duality that are not provided in Figure \ref{planar_mirror_general}, but can be read from the partition function given in \eqref{mirpartfun}.
The resulting chiral-planar duality is given in Figure \ref{sun_sqcd}
and was first presented in \cite{Benvenuti:2024seb}\footnote{Notice that the new gauge node is a $U(1)_{1/2}$ with one chiral and thus it can be confined. This results in a chiral that is not charged under any gauge symmetry and it is coupled to the theory as a flipper of a gauge invariant monopole. This does not trigger a sequential confinement except for the case of $N=2$, see Subsubsection \ref{sec:SU2confinement}.}. 
\newpage

\begin{figure}[ht]
    \centering
    \includegraphics[width=.85\textwidth]{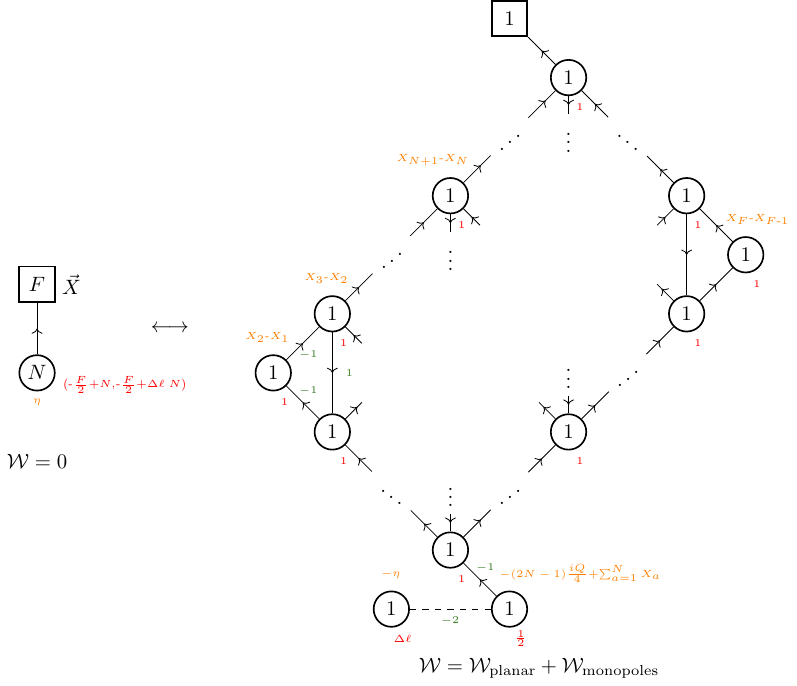}
    \caption{The $\mathcal{N}=2$ planar mirror dual of $U(N)_{(-\frac
    {F}{2}+N, -\frac{F}{2} + N\Delta \ell)}$ SQCD with $F$ fundamental chiral multiplets. }
    \label{Generic_l}
\end{figure}

By gauging the baryonic symmetry in Figure \ref{sun_sqcd}, using the strategy in \eqref{eq:gauging_U1}, we revert to the duality for $U(N)_{(k,k+N)}$ SQCD. At this stage, we may also introduce a CS term at level $\Delta \ell$ for the baryonic symmetry prior to gauging, resulting in a $U(N)_{(k,k+N(1+\Delta \ell))}$ gauge theory. The gauging/ungauging operations together with background CS terms correspond to Witten's $SL(2,\mathbb{Z})$ action \cite{witten2003sl2zactionthreedimensionalconformal} applied to the topological symmetry of $U(N)$. 
On the mirror side, this procedure introduces a new $U(1)_{\Delta_{\ell}}$ gauge node coupled to the rest of the quiver by a BF term, the result is the duality in Figure \ref{Generic_l}. 
\newpage

\begin{figure}[h]
    \centering
    \includegraphics[]{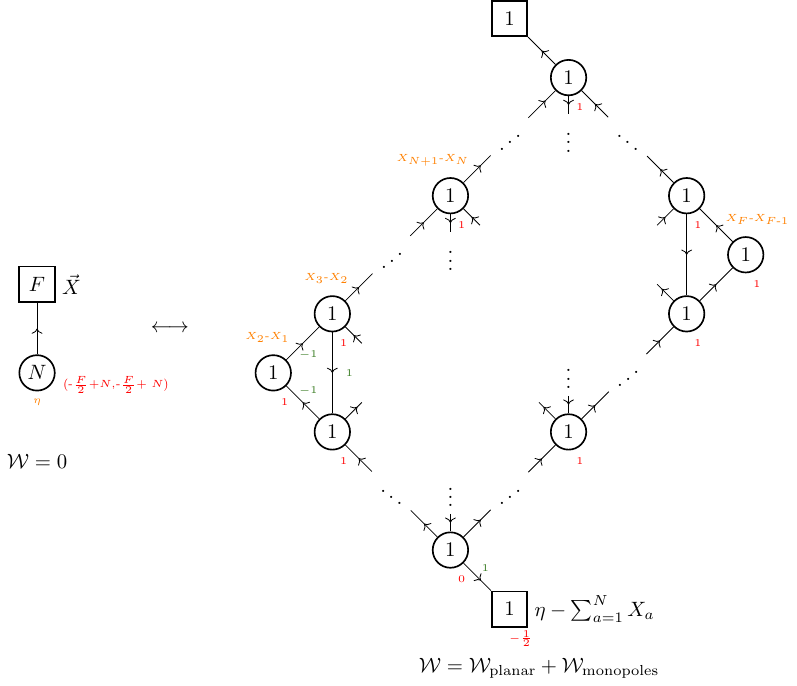}
    \caption{The $\mathcal{N}=2$ planar mirror dual of $U(N)_{(-F/2+N,-F/2+N)}$ SQCD with $F$ fundamental chiral multiplets. In the quiver we omitted the mixing of the topological symmetries with the R-symmetry, in particular the bottom-most gauge node has FI parameter $X_{N+1}-X_{N} - \frac{iQ}{8}$. The bottom chiral fundamental has trial R-charge $-N+1$. }
    \label{Planar_l}
\end{figure}

For $\Delta \ell = 0, \pm 1$ the quiver can be further simplified by explicitly integrating over the additional gauge node.
For $\Delta\ell=0$ the integration results in a delta function which further fixes the gauge field of the node on its right, which becomes a flavor node and we recover the original mirror for $U(N)_{-F/2+N,-F/2}$.
For $\Delta\ell=\pm 1$
the additional gauge node is a $U(1)_{\pm 1}$ sector, which is almost trivial and its path integral can be performed exactly as \cite{Kapustin:1999ha, witten2003sl2zactionthreedimensionalconformal}:
\begin{equation}\label{eq:tftintegration}
    \includegraphics[]{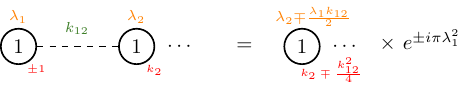}
\end{equation}
In particular, for $\Delta \ell=1$ we obtain a planar abelian dual for $U(N)_{(k,k)}$. Integrating out the additional node as above shifts the CS level of the node on its right by $-1$. Then the bottom-most node is a $U(1)_{-1/2}$ gauge node with one fundamental, which can be locally dualized to a chiral field, resulting in the duality shown in Figure \ref{Planar_l}.

\subsubsection{Consistency-check: S-Confinement of $SU(2)_0$ with 2 flavors }\label{sec:SU2confinement}

In the previous section we proposed the
planar mirror dual  for $SU(N)_{(-\frac
    {F}{2}+N)}$ SQCD with $[F, 0]$ chiral multiplets.
For $N=2$ and $F=4$ this theory
is known to S-confine to a Wess-Zumino model of $6+1$ chiral fields, interacting through a cubic superpotential \cite{Aharony:1997bx}.
As a further consistency check of our proposed dualities, we will show that indeed our planar dual theory confines to the expected Wess-Zumino model by applying local Aharony-like dualities \cite{Aharony:1997gp,Giveon:2008zn,Benini:2011mf}. 
More precisely we make use of the following s-confining dualities: $U(1)_{\tfrac{1}{2}}$ with one fundamental s-confines into a free chiral and $U(1)_{0}$ with one flavor is dual to the XYZ model.
Furthermore we use the following dualities: $U(1)_{1}$ with one flavor is dual to $U(1)_{-1}$ with one flavor, modulo flippers and $U(1)_{-\tfrac{1}{2}}$ with $[1,2]$ flavors is dual to $U(1)_{\tfrac{1}{2}}$ with $[2,1]$ flavors, modulo flippers.

\begin{center}
\tikzstyle{flavor}=[rectangle,draw=red!50,thick,inner sep = 0pt, minimum size = 6mm]
\tikzstyle{manifest}=[rectangle,draw=blue!50,thick,inner sep = 0pt, minimum size = 6mm]
\tikzstyle{gauge}=[circle,draw=black!50,thick,inner sep = 0pt, minimum size = 6mm]
\tikzstyle{gauge2}=[circle,draw=black!50,thick,inner sep = 0pt, minimum size = 4mm]
\tikzset{->-/.style={decoration={
  markings,
  mark=at position .5 with {\arrow{>}}},postaction={decorate}}}
  \begin{equation}
    \begin{tikzpicture}
        \node at (0,0) [gauge,black] (g1) {$2$};
        \node at (0,0) [gauge2,black]  {};
        \node at (0,1.5) [flavor,black]  (f1) {$4$};
        \draw[->-] (g1) -- (f1);
        \draw (f1)+(0.5,0) node [black] {\tiny$\vec{X}$};
        \draw (0,-1) node {$\mathcal{W}=0$};
        
        \draw (0.3,-0.4) [red] node {$_0$};
        \draw (2,0) node {$\longleftrightarrow$};

        \begin{scope}[shift={(-1,0)}]
        \node at (5,2) [flavor, black] (f) {$1$};
        \nodeCS(5,0)(g11,$1$,$1$);
        \nodeCS(6,1)(g21,$1$,$1$);
        \nodeCS(6,-1)(g22,$1$,$1$);
        \nodeCS(7,0)(g31,$1$,$1$);
        \nodeCS(7,-2)(gSU,$1$,$\frac{1}{2}$);

        \arrowBFlr(g11,g21)($-1$,left);
        \arrowBFlr(gSU,g22)($-1$,left);
        \arrowBFlr(g21,g22)($1$,left);
        \arrowBFlr(g31,g21)($-1$,left);
        \arrowBFlr(g22,g11)($-1$,left);
        \arrowBFlr(g22,g31)($-1$,right);
        \arrowBFlr(g21,f)(  ,right);

        
        \draw (6,-3.5) node {$\mathcal{W}=\mathcal{W}_{\text{planar}}+\;  \mathfrak{M}^{\left(\:\hspace{-2pt}
        \resizebox{30pt}{!}{%
        \begin{tabular}{ccc} &-&0\\0&+&0
        \end{tabular}}\right)
        } $};
        \end{scope}

        \begin{scope}[shift={(-.5,0)}]
        \draw (9,0) node {$\longleftrightarrow$};
        \draw (11.5,.5) node {7 chirals $Y, B_{i,j}$};
        \draw (11.5,-.5) node {$\mathcal{W}= Y \; \text{Pfaf}(B)$};
        \end{scope}
    \end{tikzpicture}
\end{equation}
\end{center}

Below we give the  local dualization sequence  where at each step we highlight in blue the gauge node that we dualize.

In the first step we use the confining duality for $U(1)_{1/2}$ with one chiral, which confines to a chiral field.
Notice that because of background CS levels produced in the dualization  the CS level of the lower node is now $\tfrac{1}{2}$. We also produce a singlet flipping a monopole.

\begin{center}
\tikzstyle{flavor}=[rectangle,draw=red!50,thick,inner sep = 0pt, minimum size = 6mm]
\tikzstyle{manifest}=[rectangle,draw=blue!50,thick,inner sep = 0pt, minimum size = 6mm]
\tikzstyle{gauge}=[circle,draw=black!50,thick,inner sep = 0pt, minimum size = 6mm]
\tikzstyle{gauge2}=[circle,draw=black!50,thick,inner sep = 0pt, minimum size = 4mm]
\tikzset{->-/.style={decoration={
  markings,
  mark=at position .5 with {\arrow{>}}},postaction={decorate}}}
  \begin{equation}
    \begin{tikzpicture}[baseline=(current bounding box).center]
        \node at (0,2) [flavor, black] (f) {$1$};
        \nodeCS(0,0)(g11,$1$,$1$);
        \nodeCS(1,1)(g21,$1$,$1$);
        \nodeCS(1,-1)(g22,$1$,$1$);
        \nodeCS(2,0)(g31,$1$,$1$);
        \begin{scope}[black/.style={blue}] 
        \nodeCS(2,-2)(gSU,$1$,$\frac{1}{2}$);
        \end{scope}

        \arrowBFlr(g11,g21)($-1$,left);
        \arrowBFlr(gSU,g22)($-1$,left);
        \arrowBFlr(g21,g22)($1$,left);
        \arrowBFlr(g31,g21)($-1$,left);
        \arrowBFlr(g22,g11)($-1$,left);
        \arrowBFlr(g22,g31)($-1$,right);
        \arrowBFlr(g21,f)(  ,right);

        \draw (1,-3.5) node {$\mathcal{W}=\mathcal{W}_{\text{planar}}+\;  \mathfrak{M}^{\left(\:\hspace{-2pt}
        \resizebox{30pt}{!}{%
        \begin{tabular}{ccc} &-&0\\0&+&0
        \end{tabular}}\right)
        } $};
    \end{tikzpicture}
    \quad \longrightarrow \quad 
    \begin{tikzpicture}[baseline=(current bounding box).center]
        \node at (0,2) [flavor, black] (f) {$1$};
        \begin{scope}[black/.style={blue}] 
        \nodeCS(0,0)(g11,$1$,$1$);
        \end{scope}
        \nodeCS(1,1)(g21,$1$,$1$);
        \nodeCS(1,-1)(g22,$1$,$\frac{1}{2}$);
        \nodeCS(2,0)(g31,$1$,$1$);

        \arrowBFlr(g11,g21)($-1$,left);
        \arrowBFlr(g21,g22)($1$,left);
        \arrowBFlr(g31,g21)($-1$,left);
        \arrowBFlr(g22,g11)($-1$,left);
        \arrowBFlr(g22,g31)($-1$,right);
        \arrowBFlr(g21,f)(  ,right);

        \draw (2,-3.5) node {$\mathcal{W}=\mathcal{W}_{\text{planar}}+\;  B_{12}\mathfrak{M}^{\left(\:\hspace{-2pt}
        \resizebox{30pt}{!}{%
        \begin{tabular}{ccc} &-&0\\0&+&
        \end{tabular}}\right)
        } $};
        \draw (3,0) node[anchor=west] {$+ \; B_{12}$};
    \end{tikzpicture}
    \longrightarrow
\end{equation}
\end{center}

In the second step we use Giveon-Kutasov duality  of $U(1)_{1}$ with a flavor
for the left node which removes the vertical chiral and modifies the superpotential.

\begin{center}
\tikzstyle{flavor}=[rectangle,draw=red!50,thick,inner sep = 0pt, minimum size = 6mm]
\tikzstyle{manifest}=[rectangle,draw=blue!50,thick,inner sep = 0pt, minimum size = 6mm]
\tikzstyle{gauge}=[circle,draw=black!50,thick,inner sep = 0pt, minimum size = 6mm]
\tikzstyle{gauge2}=[circle,draw=black!50,thick,inner sep = 0pt, minimum size = 4mm]
\tikzset{->-/.style={decoration={
  markings,
  mark=at position .5 with {\arrow{>}}},postaction={decorate}}}
  \begin{equation}
    \begin{tikzpicture}[baseline=(current bounding box).center]
        \node at (0,2) [flavor, black] (f) {$1$};
        \nodeCS(0,0)(g11,$1$,$-1$);
        \nodeCS(1,1)(g21,$1$,$\frac{1}{2}$);
        \begin{scope}[black/.style={blue}] 
        \nodeCS(1,-1)(g22,$1$,$0$);
        \end{scope}
        \nodeCS(2,0)(g31,$1$,$1$);

        \arrowBFlr(g21,g11)($1$,right);
        \arrowBFlr(g31,g21)($-1$,right);
        \arrowBFlr(g11,g22)($1$,left);
        \arrowBFlr(g22,g31)($-1$,right);
        \arrowBFlr(g21,f)(  ,right);

        \draw (2,-2.5) node {$\begin{split}\mathcal{W}=& \; \mathcal{W}_{\text{quartic} \; +}
        \\
        +& \; B_{12}\mathfrak{M}^{\left(\:\hspace{-2pt}
        \resizebox{30pt}{!}{%
        \begin{tabular}{ccc} &-&0\\0&0&
        \end{tabular}}\right) }\mathfrak{M}^{\left(\:\hspace{-2pt}
        \resizebox{30pt}{!}{%
        \begin{tabular}{ccc} &0&0\\0&+&
        \end{tabular}}\right) }
         \end{split}$};
        \draw (3,0) node[anchor=west] {$+ \; B_{12}$};
    \end{tikzpicture}
    \quad \longrightarrow \quad
    \begin{tikzpicture}[baseline=(current bounding box).center]
        \node at (0,2) [flavor, black] (f) {$1$};
        \begin{scope}[black/.style={blue}] 
        \nodeCS(0,0)(g11,$1$,$-\frac{1}{2}$);
        \end{scope}
        \nodeCS(1,1)(g21,$1$,$\frac{1}{2}$);
        \nodeCS(2,0)(g31,$1$,$\frac{1}{2}$);

        \arrowBFlr(g21,g11)($1$,right);
        \arrowBFlr(g31,g21)($-1$,right);
        \arrowBFlr(g21,f)(  ,right);

        \path[draw] (g11) edge[->-,bend left=10] node[midway,above] {\tiny$M$} (g31);
        \path[draw] (g31) edge[->-,bend left=10,red] node[midway,below,red] {\tiny$T_-$} (g11);

        \path (g31) edge[draw=none] node[midway,BFcolor] {\tiny$0$} (g11);

        \draw (2,-2.5) node {$\begin{split}\mathcal{W}= &
        \; B_{12}T_{-}\mathfrak{M}^{\left(\:\hspace{-2pt}
        \resizebox{30pt}{!}{%
        \begin{tabular}{ccc} &-&\\ 0&&0
        \end{tabular}}\right) } \; + 
        \\
        +& \; \mathcal{W}_{\text{cubic}} \; + \; MT_{-}B_{23}
        \end{split}$};
        \draw (3,0) node[anchor=west] {$+B_{12}, B_{23}$};
    \end{tikzpicture}
    \quad 
    \longrightarrow
\end{equation}
\end{center}

In the third step we use the confining duality for $U(1)_0$ with a flavor to confine the bottom node producing a singlet.

\begin{center}
\tikzstyle{flavor}=[rectangle,draw=red!50,thick,inner sep = 0pt, minimum size = 6mm]
\tikzstyle{manifest}=[rectangle,draw=blue!50,thick,inner sep = 0pt, minimum size = 6mm]
\tikzstyle{gauge}=[circle,draw=black!50,thick,inner sep = 0pt, minimum size = 6mm]
\tikzstyle{gauge2}=[circle,draw=black!50,thick,inner sep = 0pt, minimum size = 4mm]
\tikzset{->-/.style={decoration={
  markings,
  mark=at position .5 with {\arrow{>}}},postaction={decorate}}}
\begin{equation}
    \begin{tikzpicture}[baseline=(current bounding box).center]
        \node at (0,2) [flavor, black] (f) {$1$};
        
        \nodeCS(0,0)(g11,$1$,$\frac{1}{2}$);
        \nodeCS(1,1)(g21,$1$,$1$);
        \begin{scope}[black/.style={blue}] 
        \nodeCS(2,0)(g31,$1$,$0$);
        \end{scope}

        \arrowBFlr(g11,g21)($-1$,left);
        \arrowBFlr(g21,f)(  ,right);

        \path[draw] (g11) edge[->-,bend left=10] node[midway,above] {\tiny$M$} (g31);
        \path[draw] (g31) edge[->-,bend left=10,red] node[midway,below,red] {\tiny$\tilde{T}_-$} (g11);

        \path (g31) edge[draw=none] node[midway,BFcolor] {\tiny$0$} (g11);

        \draw (2,-2.5) node {$\begin{split}\mathcal{W}=& \;
        \mathfrak{M}^{\left(\:\hspace{-2pt}
        \resizebox{30pt}{!}{%
        \begin{tabular}{ccc} &0&\\0&&-
        \end{tabular}}\right)
} \bigg(B_{12}\mathfrak{M}^{\left(\:\hspace{-2pt}
        \resizebox{30pt}{!}{%
        \begin{tabular}{ccc} &+&\\+&&+
        \end{tabular}}\right)
}
        \\
        -& \; B_{13}\mathfrak{M}^{\left(\:\hspace{-2pt}
        \resizebox{30pt}{!}{%
        \begin{tabular}{ccc} &0&\\+&&+
        \end{tabular}}\right)
} \; + \; B_{23}\mathfrak{M}^{\left(\:\hspace{-2pt}
        \resizebox{30pt}{!}{%
        \begin{tabular}{ccc} &0&\\0&&+
        \end{tabular}}\right)
}\bigg)
        \end{split}$};
        \draw (3,0) node[anchor=west] {$+B_{12},B_{23},B_{13}$};
    \end{tikzpicture}
    \quad \longrightarrow \quad 
    \begin{tikzpicture}[baseline=(current bounding box).center]
        \node at (0,2) [flavor, black] (f) {$1$};
        \begin{scope}[black/.style={blue}] 
        \nodeCS(0,0)(g11,$1$,$\frac{1}{2}$);
        \end{scope}
        \nodeCS(1,1)(g21,$1$,$1$);

        \arrowBFlr(g11,g21)($-1$,right);
        \arrowBFlr(g21,f)(  ,right);

        \draw (2,-2.5) node {$\begin{split}\mathcal{W}=& \; 
        Y \bigg(
        - \;B_{13} \mathfrak{M}^{\left(\:\hspace{-2pt}
        \resizebox{20pt}{!}{%
        \begin{tabular}{cc} &0\\+&
        \end{tabular}}\right)
        }
        \\&
        + \; B_{12} \mathfrak{M}^{\left(\:\hspace{-2pt}
        \resizebox{20pt}{!}{%
        \begin{tabular}{cc} &+\\+&
        \end{tabular}}\right)
        } \;
        + \; B_{23}B_{14}
        \bigg)
        \end{split}$};
        \draw (2,0) node[anchor=west] {$+B_{12},B_{23},B_{13},B_{1,4},Y$};
    \end{tikzpicture}
    \longrightarrow
\end{equation}
\end{center}

In the fourth step we use a flipped version of  the duality \eqref{quiv:U1_[1,2]} on the left node. This removes the bifundamental chiral connnecting the right and the bottom nodes.

\begin{center}
\tikzstyle{flavor}=[rectangle,draw=red!50,thick,inner sep = 0pt, minimum size = 6mm]
\tikzstyle{manifest}=[rectangle,draw=blue!50,thick,inner sep = 0pt, minimum size = 6mm]
\tikzstyle{gauge}=[circle,draw=black!50,thick,inner sep = 0pt, minimum size = 6mm]
\tikzstyle{gauge2}=[circle,draw=black!50,thick,inner sep = 0pt, minimum size = 4mm]
\tikzset{->-/.style={decoration={
  markings,
  mark=at position .5 with {\arrow{>}}},postaction={decorate}}}
  \begin{equation}
    \begin{tikzpicture}[baseline=(current bounding box).center]
        \node at (0,2) [flavor, black] (f) {$1$};
         \begin{scope}[black/.style={blue}] 
        \nodeCS(1,1)(g21,$1$,$\frac{1}{2}$);
        \end{scope}

        \arrowBFlr(g21,f)(  ,right);
        \draw (3,0) node {$\mathcal{W}=Y(-B_{13}B_{24}+B_{12}\mathfrak{M}^{+} +B_{23}B_{14})$};
        \draw (4.5,1.5) node {$+ \quad B_{12},B_{23},B_{13},B_{14},B_{24},Y$};
    \end{tikzpicture}
    \quad \longrightarrow \quad
    \begin{array}{l}
    \prod_{i<j}B_{ij}\quad+\quad Y
    \\
    \mathcal{W}=Y \; \text{Pfaf}(B_{ij})
    \end{array}
\end{equation}
\end{center}

In the fifth and sixth steps we use the confining duality for $U(1)_{1/2}$ with one chiral, to confine the quiver to a collection of singlets interacting with a superpotential.

The fields $B_{ij}$ in the last step are mapped to the 6 baryons of electric theory, while $Y$ is the (single) gauge-invariant monopole of the electric theory.

\section{A Chiral-Planar duality web}\label{sec: FTUN_mirror}

In this section we discuss an interesting web of chiral-planar dualities which originates from the $\mathcal{N}=4$ $FT[U(N)]$ theory\footnote{This theory only differs from the $T[SU(N)]$ theory introduced in \cite{Gaiotto_2007} by the presence of a singlet field in the adjoint of the $SU(N)$ symmetry flipping the Higgs branch moment map.}. 

In Subsection \ref{subsec: S-Wall_G} starting from the mirror self-duality of the $FT[U(N)]$ theory \cite{Aprile:2018oau} and performing a real mass deformation, we obtain an  $\mathcal{N}=2$ duality relating a linear, chiral quiver gauge theory to a planar abelian theory.

The $FT[U(N)]$ theory is also know to have another self-dual frame (modulo flips), the so called flip-flip dual frame \cite{Aprile:2018oau}.
In Subsection \ref{ffccpp} we start from this flip-flip duality and performing real mass deformations, we obtain two $\mathcal{N}=2$ dualities: a chiral-chiral duality, relating two chiral quivers and  a planar-planar duality, relating two planar quivers. 

Notice that starting from the flip-flip duality we don't expect to find a chiral-planar duality, since this is a peculiarity of mirror dualities  exchanging flavor and topological symmetries.
As shown in \cite{Aprile:2018oau} the flip-flip duality is not a mirror duality but is more akin to an Aharony duality, and indeed it is possible to prove it by iterative local applications of Aharony duality. In appendix \ref{App:B}
we discuss how the chiral-chrial
and planar-planar dualities can be similarly derived by iterative local applications of Given-Kutasov-like dualities. \\

We begin with a quick review of the $FT[U(N)]$ theory. The $FT[U(N)]$ theory is an SCFT with global symmetry ${\color{black}{SU(N)_{\vec{X}}}}\times {\color{black}{SU(N)_{\vec{Y}}}}\times U(1)_{\tau}$\footnote{Note that we work in an off-shell parametrization wherein the manifest symmetry is $U(N)$ and not $SU(N)$, as in \cite{Benvenuti:2023qtv}.}. We denote this theory with the compact notation:
\begin{equation} \label{quiv:ftun}
\begin{tikzpicture}
 \node at (-4,0) (f1) [flavor, black] {$N$};
    \node at (-2,0) (f2) [flavor,black] {$N$};
    \draw[dashed] (f1) -- (f2);
    \draw (f1)+(1,+0.2) node {$-$};
    \draw[black] (f1)+(0,-.6) node {$\vec{X}$};
    \draw[black] (f2)+(0,-.6) node {$\vec{Y}$};
\end{tikzpicture} 
\end{equation}
Where $\vec{X}$, $\vec{Y}$ are fugacities for the global symmetries and the label ``$-$" is included for later convenience, see Section \ref{sec: S_walls,QFT_blocks_Duality_Moves}. 
The $FT[U(N)]$ theory has two operators in its spectrum $A_{\vec{X}}$ and $A_{\vec{Y}}$, which transform in the adjoint representation of $SU(N)_{\vec{X}}$ and $SU(N)_{\vec{Y}}$.
The $FT[U(N)]$ SCFT admits a UV completion given by the following UV quiver theory: 
\begin{equation} \label{quiv:ftun}
    \includegraphics[]{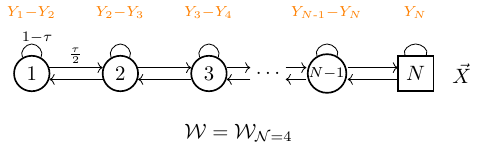} 
\end{equation}
The UV global symmetry is $SU(N)_{\vec{X}} \times \prod_{j=1}^{N-1}U(1)_{Y_j-Y_{j+1}} \times U(1)_{\tau}$, which enhances to ${\color{black}{SU(N)_{\vec{X}}}}\times {\color{black}{SU(N)_{\vec{Y}}}}\times U(1)_{\tau}$ in the IR. 
In this UV completion the operator $A_{\vec{X}}$ coincides with the traceless singlet  matrix $A_{\vec{X}} := a_N$.
The other operator $A_{\vec{Y}}$
is instead constructed  by collecting the $N(N-1)$ gauge invariant monopoles with magnetic fluxes $+1$ ($-1$) under a sequence of gauge nodes
and the $N-1$ traces $\text{Tr}(a_{i})$ 
.
For example, for $N=4$ we have:
\begin{equation}
A_{\vec{Y}}=
    \begin{bmatrix}
         \text{Tr} (a_1)  & \mathfrak{M}^{+,0,0}  & \mathfrak{M}^{+,+,0} & \mathfrak{M}^{+,+,+}  \\
         \mathfrak{M}^{-,0,0} &  \text{Tr} (a_2)  & \mathfrak{M}^{0,+,0} & \mathfrak{M}^{0,+,+}\\
        \mathfrak{M}^{-,-,0} & \mathfrak{M}^{0-,,0} & \text{Tr} (a_3)  & \mathfrak{M}^{0,0,+}\\
        \mathfrak{M}^{-,-,-}  & \mathfrak{M}^{0,-,-}  & \mathfrak{M}^{0,0,-}  & - \sum_{i=1}^3 \text{Tr} (a_i) \\
    \end{bmatrix} \,.
\end{equation}

\begin{figure}[h!]
    \centering
    \includegraphics[width=1\textwidth]{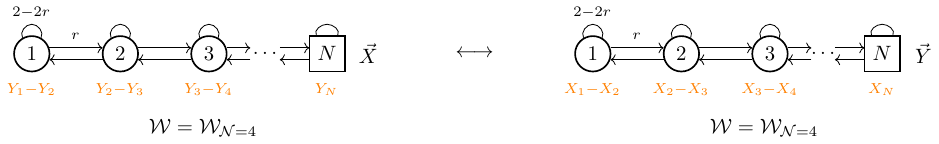}
    \caption{
       The $FT[U(N)]$ quiver gauge theory is self-mirror under the exchange of the manifest and topological symmetries. We also indicate the assignment of trial R-charges for the bifundamental and adjoint chiral fields. Notice that the bifundamental fields have trial R-charge $r$ and adjoint chiral fields have trial R-charge $2-2r$ on both sides of the duality. 
       As a consequence, the two moment maps $A_{\vec{X}}, A_{\vec{Y}}$ have both trial R-charge $2-r$.}
    \label{fig:FT_self_mirror}
\end{figure}

The $FT[U(N)]$ admits a second UV completion,
as shown in  Figure \ref{fig:FT_self_mirror}. In each completion only one of the two global $SU(N)$ symmetries is realized manifestly as the flavor symmetry rotating the last bifundamental. The second $SU(N)$ global symmetry instead enhances from the $N-1$ topological symmetries due to all the nodes being balanced. The two UV completions are related by mirror duality, in this case a self-duality.

The $\mathbf{S}^3_b$ partition function of 
the first UV completion
theory is defined recursively as follows:
\begin{equation}\label{tsu}
\begin{aligned}
    Z_{FT[U(N)],I}(\vec{X},\vec{Y},\tau) &:= e^{2\pi i Y_N\sum_j^NX_j}\;
    \prod_{j,k=1}^N\;
    s_b\bigg(\tau-\frac{iQ}{2}(1-2r)
    + X_j - X_k \bigg)\\& \int\;\frac{ \prod_{\alpha=1}^{N-1} du_{\alpha}\;e^{-2\pi i Y_N u_{\alpha}}}{(N-1)! \; \prod_{\alpha<\beta}\;s_b\big(\frac{iQ}{2}\pm (u_{\alpha}-u_{\beta})\big)}\;
    s_b\bigg(\frac{iQ}{2}(1-r)-\frac{\tau}{2}
    \pm (u_{\alpha}-X_{j})\bigg)\\& Z_{FT[U(N-1)],I}(\vec{u},\{Y_i\}^{N-1}_{i=1},\tau) \,.
\end{aligned}
\end{equation}
with 
\begin{equation}
    Z_{FT[U(1)],I}(X,Y,\tau) := e^{2\pi i X Y}\;s_b\bigg(\tau-\frac{iQ}{2}(1-2r)\bigg) \,.
\end{equation}
The partition function of the second UV completion can be obtained by simply swapping $\vec{X} \leftrightarrow \vec{Y}$.
\begin{equation}
   Z_{FT[U(N)],II} (\vec{X}, \vec{Y}, \tau) =   Z_{FT[U(N)],I} (\vec{Y}, \vec{X}, \tau) \,,
\end{equation}
Notice in the first UV completion the first slot $FT[U(N)](\cdot \; ,\cdot\, ; \tau)$ is for the manifest $U(N)$, the second slot for  emergent $U(N)$. On the other hand in the second UV completion the 
second slot is for the manifest and the first one for the emergent $U(N)$.

Since both the UV completions describe the same SCFT, their partition functions are equal and we can define:
\begin{equation}
   Z_{FT[U(N)],I} (\vec{X}, \vec{Y}, \tau) =   Z_{FT[U(N)],II} (\vec{Y}, \vec{X}, \tau) :=  Z_{FT[U(N)]} (\vec{X}, \vec{Y}, \tau) \,,
\end{equation}

Furthermore, the partition function satisfies:
\begin{equation}
   Z_{FT[U(N)]}(\vec{X},\vec{Y},\tau)=
 Z_{FT[U(N)]}(-\vec{X},-\vec{Y},\tau)
\end{equation}
as can be seen through a redefinition of the gauge fields. 

For the rest of the paper, we consider the $FT[U(N)]$ theory with a $U(N)_{\vec{X}}\times U(N)_{\vec{Y}}$ global symmetry instead (in practice, this corresponds to relaxing the constraints $\sum_{i}X_i=\sum_{i}Y_i=0$ on the fugacities in the $\mathbf{S}^3_b$ partition function).

\subsection{A Chiral-Planar mirror dual} \label{subsec: S-Wall_G}

\begin{figure}
    \centering
    \includegraphics[width=.8\textwidth]{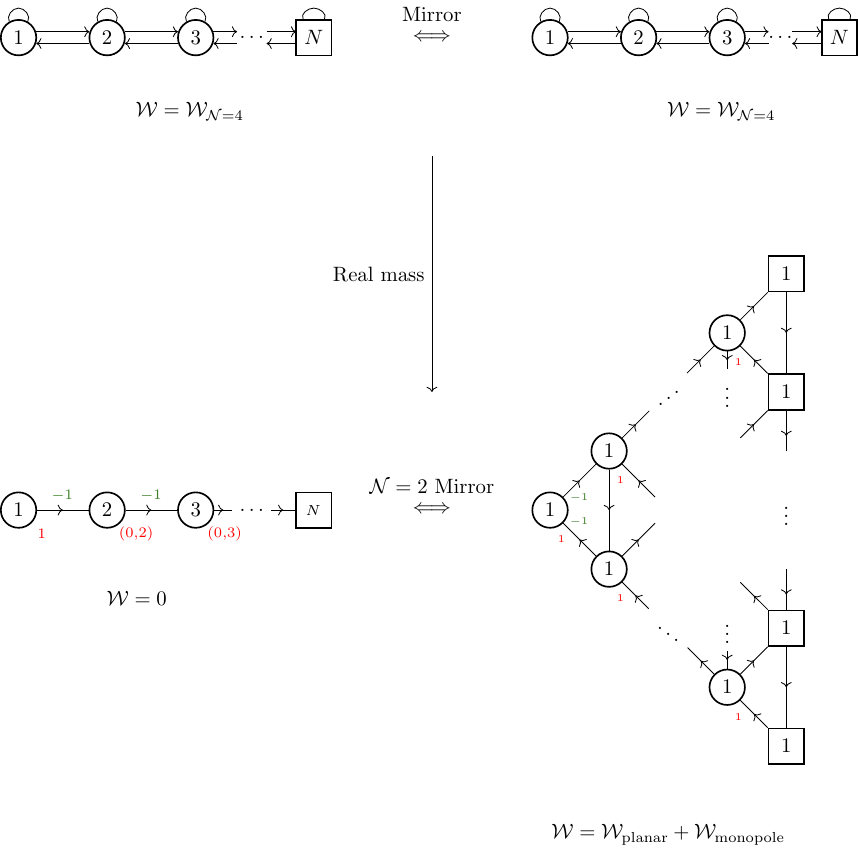}
    \caption{On the top row the mirror self-duality relates the two UV completions of the $FT[U(N)]$ theory.
    The real mass deformation yield a new $\mathcal{N}=2$  mirror duality relating the two UV completions of the $G[U(N)]$ theory depicted in the second row.}
    \label{fig:FTUWeb}
\end{figure}

Starting from the self-mirror pair relating the two UV completion of the $FT[U(N)]$ SCFT in Figure \ref{fig:FT_self_mirror}, we will now consider a real mass deformation 
to produce a new duality relating a chiral quiver to a planar quiver which is depicted in Figure \ref{fig:FTUWeb}.

As in the previous section, if
on the electric side we implement the real mass deformation landing on a chiral non-abelian quiver theory, on the mirror dual side, the matching vacuum we flow to, requires to move to a point on the Coulomb where all the gauge symmetries are broken to their maximal torus subgroup. 
The chiral and the planar quivers on the bottom of Figure \ref{fig:FTUWeb} are IR dual and can be regarded as two UV completions of a new SCFT which we name $G[U(N)]$ with $U(1)^{N-1}\times SU(N)$ global symmetry, which we denote in compact form as:
\tikzstyle{flavor}=[rectangle,draw=red!50,thick,inner sep = 0pt, minimum size = 6mm]
\tikzstyle{manifest}=[rectangle,draw=blue!50,thick,inner sep = 0pt, minimum size = 6mm]
\tikzstyle{gauge}=[circle,draw=black!50,thick,inner sep = 0pt, minimum size = 6mm]
\tikzset{-<-/.style={decoration={
  markings,
  mark=at position .5 with {\arrow{<}}},postaction={decorate}}}
\tikzset{->-/.style={decoration={
  markings,
  mark=at position .5 with {\arrow{<}}},postaction={decorate}}}
\begin{equation}
G[\vec{X},\vec{Y}] =\;
    \begin{tikzpicture}[baseline=(current bounding box.center)]
        \node at (-4,0) (f10) [flavor, black] {$N$};
        \node at (-2,0) (f11) [flavor,black] {$1^N$};
        \draw[] (f10)++(-.3,-.6) node[anchor=west]{$\vec{X}$};
        \draw[] (f11)++(-.3,-.6) node[anchor=west]{$\vec{Y}$};
        \draw[decorate,decoration={coil,segment length=4pt}] (f10)--(f11) node[midway,above,yshift=5pt] {$-$};
    \end{tikzpicture} 
\end{equation}

We stress that in our notation the first argument of the $G[\cdot \,;\cdot]$ theories corresponds to fugacities for an $SU(N)$ global symmetry, while the second argument corresponds to a $U(1)^{N-1}$ global symmetry.

We now describe in detail this deformation
beginning from the first UV completion of the $FT[U(N)]$, the electric theory,
where the $U(N)_X$ is manifest.
We consider the  real mass deformation defined at the level of the fugacities by
\begin{equation}    
\left\{
\begin{array}{l}
Y_J \to Y_J + \frac{\tau}{2} (2J -N -1), \qquad J=1,\dots,N
\\
\vec{X} \to \vec{X} + \frac{\tau}{2}
\end{array}
\right.,
\qquad\qquad 
\tau \to +\infty
\label{eq:chiral_mass_FTSUN1}
\end{equation}
In addition to this
we also perform  the following shifts in gauge fugacities:
\begin{equation}
\vec{u}^{(J)} \to \vec{u}^{(J)} - \frac{\tau}{2} (N-1-J).
\end{equation}
Here $\vec{u}^{(J)}$ are the fugacities of the $J$-th gauge group, starting from the left.
Integrating out the massive fields we obtain the following $\mathcal{N}=2$ quiver gauge theory:
\begin{equation}\label{g_n_chiral}
    \includegraphics[]{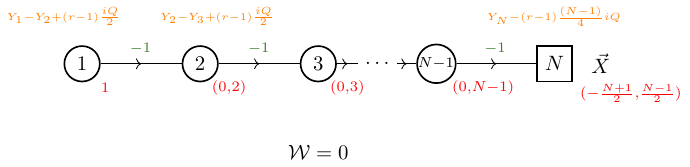}
\end{equation}
All the chirals in the theory have trial R-charge $r$ which should be possible, in principle, to fix by performing F-extremization to determine the superconformal R-charge. However, we will not perform F-extremization in this paper and keep $r$ to be generic. Notice also that the FI parameters depend on the value of $r$.
This is the first UV completion of the $G[U(N)]$ theory with UV symmetry ${\color{black}U(1)^{N-1}_Y}\times {\color{black}{SU(N)_{\vec{X}}}}$.

The $\mathbf{S}^3_b$ partition function in the $\tau \to +\infty$ limit is given by:
\begin{equation}
\lim_{\tau \to +\infty} Z_{FT[U(N)],I}(\vec{X},\vec{Y},\tau)
=
\lim_{\tau \to +\infty} e^{\frac{\pi i }{2} H^{\text{chiral}}(\vec{X},\vec{Y},\tau) }
Z_{G[U(N)],I}(\vec{X},\vec{Y})
\end{equation}
where $Z_{G[U(N)],I}(\vec{X},\vec{Y})$ is the partition function of the first UV completion of the $G[U(N)]$ theory in \eqref{g_n_chiral} and:
\begin{equation}
H^{\text{chiral}}(\vec{X},\vec{Y},\tau) = H_{\text{div}}^{\text{chiral}}(\vec{X},\vec{Y},\tau) + H_{\text{res}}^{\text{chiral}} 
\end{equation}

\begin{equation}
\begin{split}
H_{\text{div}}^{\text{chiral}}(\vec{X},\vec{Y},\tau) =& 
\frac{\tau^2}{4} \left( \frac{4}{3}N^3 + 2 N^2 + \frac{2}{3} N \right)
\\&
	+ \frac{\tau}{2} \left(	
    8 \sum_{J=1}^N J Y_J
    -4 \sum_{J=1}^N Y_J 
    -i Q (N+N^2) +i Q r \left(\frac{2}{3} N^3+ 2N^2 + \frac{4}{3} N\right)	\right),
\end{split}
\end{equation}

and $H_{\text{res}}^{\text{chiral}}$ is a finite term independent on the global parameters $\vec{X}$ and $\vec{Y}$, that only depends on the squashing parameter $Q$.\\

\begin{figure}[h!]
    \centering
    \includegraphics[]{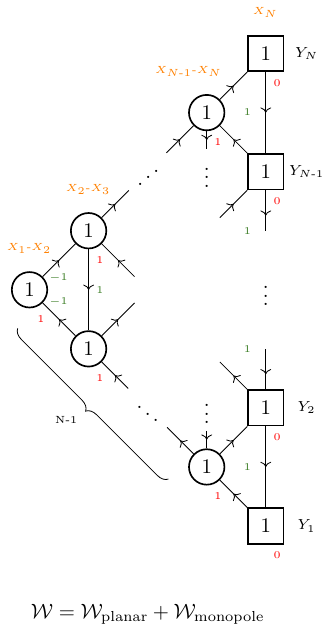}
    \caption{
    The $\mathcal{N}=2$ planar mirror dual of the $G[U(N)]$ theory. 
    CS levels are indicated in red and the level of mixed CS interactions is indicated in green.
    On top of each column we indicate in orange the FIs of all the gauge nodes of the column, up to shifts that are discussed in the text.
    }
    \label{planar_mirror_ftu}
\end{figure}

We now consider the second UV completion of the $FT[U(N)]$ theory, the mirorr theory.
As in the case of the planar mirror of the chiral SQCD discussed in the previous section, here in addition to the real mass \eqref{eq:chiral_mass_FTSUN1}
to flow to the vacumm dual to the chiral quiver  we also need to perform the following shifts in gauge fugacity: 
\begin{equation}
\omega^{(J)}_{\alpha} \to \omega^{(J)}_{\alpha}  + \tau \left( \alpha - \frac{J+1}{2} \right)
\end{equation}
and take the $\tau \to +\infty$ limit. 
Here $\omega_{\alpha}^{(J)}$ is the $\alpha$-th fugacity of the $J$-th gauge group, starting from the left. The resulting $\mathcal{N}=2$x  planar quiver theory in Figure \ref{planar_mirror_ftu} is IR dual to the chiral theory in Figure 
\eqref{g_n_chiral}, and provides the second UV completion of the $G[U(N)]$ SCFT.

A few comments regarding the superpotential of the quiver in Figure \ref{planar_mirror_ftu} are in order:
\begin{itemize}
    \item There is a cubic superpotential term for every closed triangle. This superpotential originates from the cubic $\mathcal{N}=4$ interactions that are not broken by the real mass deformation.
    We denote these superpotential terms as $\mathcal{W}_{\text{planar}}$. 
    The assignment of trial R-charges is $r$ for each diagonal chiral field and $2-r$ for each vertical one. Indeed this assigment of R-charges is compatible with the planar superpotential. Notice also that the FI parameters depend on the value of $r$, encoding a mixing between the topological and R-symemtries as described in Subsection \ref{sec:chiral-planar-sccd}.
    \item Due to adjoint Higgsing of the gauge nodes, there are monopole terms in the superpotential \cite{Polyakov1977}. There is a linear monopole superpotential for each monopole with charges $+1/-1$ under the upper/lower gauge nodes connected by a vertical line.
    We denote these linear monopole superpotential terms as $\mathcal{W}_{\text{monopole}}$. These are essential for the topological symmetry to correctly enhance to $SU(N)$, as we will comment later.
    
    \item In the planar theory there are self and mixed CS interactions. They are such that each gauge node has CS level $+1$ and each pair of gauge nodes connected by a diagonal/vertical line have a mixed CS interaction with level $-1/+1$. All these interactions arise from the integrated fermionic modes that became massive upon real mass deformation.

    \item The are shifts in the FIs associated to the mixing of the topological symmetry of the various nodes and the R-symmetry. The FIs of the gauge nodes in the $J$-th colum from the right are:
    \begin{equation}
        \text{FIs: }
        X_{J}-X_{J+1} - \frac{iQr}{4},\;X_{J}-X_{J+1},\;
        \dots,\; X_{J}-X_{J+1} + \frac{iQr}{4}
    \end{equation}
    from top to bottom, while the background FIs for the $j$-th flavor node associated to $Y_j$ are:
    \begin{equation}
        \text{FIs: }
        X_N + \frac{iQr}{2} (2j-N-1) + \delta_{j,1} \frac{iQ}{2} - \delta_{j,N}\frac{iQ}{2}
    \end{equation}
    where $j=1$ corresponds to the top flavor node in Figure \ref{planar_mirror_ftu} and $j=N$ corresponds to the bottom node.
    In Figure \ref{planar_mirror_ftu} we suppressed these shifts and only report the contribution to the FIs that depends on $X_i$.
    
\end{itemize}
The $\mathbf{S}^3_b$ partition function in the $\tau \to +\infty$ limit is given by:
\begin{equation}
\lim_{\tau \to \infty} Z_{FT[U(N)],II}(\vec{X},\vec{Y},\tau)
=
\lim_{\tau \to \infty} e^{\frac{\pi i }{2} H^{\text{planar}}(\vec{X},\vec{Y},\tau) }
Z_{{G}[U(N)],II}(\vec{X},\vec{Y})
\end{equation}
where $Z_{{G}[U(N)],II}(\vec{X},\vec{Y})$ is
the partition function of the second UV completion of the $G[U(N)]$ theory in \eqref{planar_mirror_ftu} and:
\begin{equation}
H^{\text{planar}}(\vec{X},\vec{Y},\tau) = H_{\text{div}}^{\text{planar}}(\vec{X},\vec{Y},\tau) + H_{\text{res}}^{\text{planar}}
\end{equation}
Here $H_{\text{res}}^{\text{planar}}$ is a constant phase independent of the global parameters $\vec{X}$ and $\vec{Y}$, that only depends on the squashing parameter $Q$.
$H_{\text{div}}^{\text{planar}}$ is a divergent phase that is equal to $H_{\text{div}}^{\text{chiral}}$ and therefore it cancels against the chiral dual theory, providing a strong check that the theories in \ref{planar_mirror_ftu} and \eqref{g_n_chiral} are dual. We found also that $H_{\text{res}}^{\text{planar}} = H_{\text{res}}^{\text{chir}}$, therefore taking the limit $\tau + \infty$ we find:
\begin{equation}
    Z_{G[U(N)],I}(\vec{X},\vec{Y}) = Z_{G[U(N)],II}(\vec{X},\vec{Y})
\end{equation}

\subsubsection{Global Symmetries and Operator Map}
We now comment on the duality map between global symmetries and the chiral rings of the two theories. 

Let us start from the global symmetries. As already stated at the beginning of the section, the $G[U(N)]$ SQCFT has a $U(1)^{N-1} \times SU(N)$ global symmetry. In the chiral UV completion \eqref{g_n_chiral}
, the two global symmetries are realized manifestly in the UV. $SU(N)$ is the flavor symmetry while $U(1)^{N-1}$ are the topological symmetries of the $N-1$ nodes. In the planar UV completion, in Figure \ref{planar_mirror_ftu}, instead the $U(1)^{N-1}$ is realized as the flavor symmetry that is unbroken by the planar superpotential and up to gauge transformations. The $SU(N)$ global symmetry is instead obtained from the enhancement of the $N-1$ topological symmetries. We recall that the presence of $\mathcal{W}_{\text{monopole}}$ has the effect of breaking the $U(1)^k$ topological symmetries in a colum with $k$ gauge nodes down to a diagonal $U(1)$, such that effectively each column of gauge nodes contributes with a single $U(1)$ topological symmetry.

Let us now move to the operator map.
The chiral ring of the planar $G[U(N)]$ theory in Figure \ref{planar_mirror_ftu} is generated by the $N-1$ vertical bifundamental fields connecting adjacent vertical flavor nodes. These can be taken as the only indpendent chiral ring generators up to F-term relations imposed by $\mathcal{W}_{\text{planar}}$. \\
We expect these operators to be mapped to monopole operators in the chiral $G[U(N)]$ theory \eqref{g_n_chiral}. This is the case, but the fundamental monopoles are gauge variant and must be appropriately dressed by appropriate powers of fundamental chiral multiplets. 
Focusing on a section of the chiral $G[U(N)]$ quiver:
\tikzset{->-/.style={decoration={
  markings,
  mark=at position .5 with {\arrow{>}}},postaction={decorate}}}
\begin{equation}
\begin{tikzpicture}[xscale=1.3]
\node [gauge, black] (g1) at (0,0) {$_{k-1}$};
\draw[red] (0,-.6) node {$_{(0,\; k-1)}$};
\draw[red] (2,-.6) node {$_{(0,\; k)}$};
\draw[red] (4,-.6) node {$_{(0,\; k+1)}$};
\node [gauge, black] (g2) at (2,0) {$k$};
\node [gauge, black] (g3) at (4,0) {$_{k+1}$};
\path[->-,draw] (g1) ++(-1,0) -- (g1);
\path[->-,draw] (g3)  -- ++(1,0);
\path[draw,->-] (g1)--(g2);\path[draw,->-] (g2)--(g3);
\arrowBFlr(g1,g2)($-1$,left)
\node at (1,-.5) {$Q$}; \node at (3,-.5) {$P$};
\arrowBFlr(g2,g3)($-1$,left)
\end{tikzpicture}
\end{equation}
we notice that the fundamental monopole
of the $U(k)$ node can be dressed with $Q^{k-1}$ to ensure gauge invariance of the resulting operators.
The $N-1$ dressed fundamental monopoles of the chiral $G[U(N)]$ theory generate the chiral ring of the theory. 
We can also calculate that they have the correct charges
to be  mapped  to the $N-1$ vertical bifundamental chirals connecting adjacent
vertical flavor nodes in the mirror, which can be verified also via the superconformal index
Assume that each bifundamental in the electric theory ($Q,\; P \dots$) has trial R-charge $r$. 
Consequently, the vertical bifundamental chirals connecting adjacent
vertical flavor nodes in the mirror planar theory, that the monopoles map to, have R-charge $2-2r$. 
By using  eq. \eqref{monfor} dressed monopole of the $k-$th gauge node in the chiral theory $tr(Q^{k-1}\mathfrak{M}^{0,\dots,0,-,0,\dots,0})$ has R-charge:
\begin{equation}
\label{dressmoneq}
    \begin{aligned}
      R(Q^{k-1}\mathfrak{M}^{0,\dots,0,-,0,\dots,0})= & -(r-1) + \frac{1}{2}(k-1)(1-r)+ \frac{1}{2}(k+1)(1-r)-(k-1)+(k-1)r =\\
      = & (1-r)+k(1-r)-(k-1)+(k-1)r =\\
      = & 2-2r
    \end{aligned}
\end{equation}
where the first factor comes from the mixing between the topological symmetry and the R-symmetry\footnote{The mixing can be extracted as the coefficient of $\frac{i Q}{2}$ in the FI  of the $k$-th node: $Y_k-Y_{k+1} +\frac{i Q}{2}(r-1)$. The contribution to the charge is then obtained by  multiplying it by the topological charge.}, while the second and third terms in \eqref{dressmoneq} are the bifundamental fermions contributions, the $-(k-1)$ is the contribution from the gaugini, and $(k-1)r$ is the contribution from the dressing with $Q^{k-1}$. 
So the $N-1$ fundamental dressed monopoles map to the $N-1$ vertical bifundamental chirals connecting adjacent
vertical flavor nodes in the mirror.
In the rest of the paper we denote the gauge-invariant dressed monopoles as $\mon_d$, and we omit the dressing fields.

\subsection{Flip-Flip chiral-chiral and planar-planar dualities}
\label{ffccpp}

The $FT[U(N)]$ theory enjoys another UV completion known as the Flip-Flip dual \cite{Aprile:2018oau}.
The Flip-Flip dual quiver $ffFT[U(N)]$ is obtained starting from the $FT[U(N)]$ theory with two
extra sets of singlets which flip the operators $A_{X}, A_Y$:

\begin{equation}
\mathcal{W}_{ffFT}= 
 \mathcal{W}_{FT}+
        O_X A_{\vec{X}}+O_Y A_{\vec{Y}} 
\end{equation}
Since in the $FT[U(N)]$ theory the meson matrix is already flipped by the singlet $a_N$, this further flip give mass to $a_N$.
As a result the spectrum of the Flip-Flip theory consist in a the adjoint $SU(N)_X$ matrix is realised as the meson matrix while adjoint matrix $SU(N)_Y$  coincides with the monopole flip matrix $O_Y$.

\begin{figure}[ht]
    \centering
    \includegraphics[]{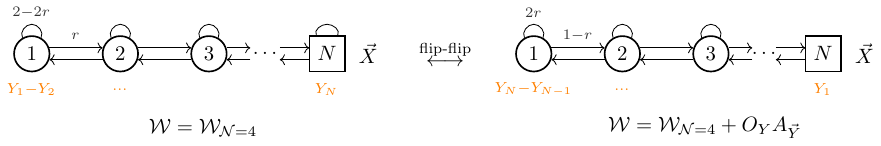}
    \caption{The flip-flip duality for the FT[U(N)] theory. On the electric side the bifundamentals have $U(1)_\tau$ charge $\frac{1}{2}$ and trial R-charge $r$, while on the magnetic side the bifundamentals have $U(1)_\tau$ charge $-\frac{1}{2}$ and trial R-charge $1-r$, as enphasized by the labels in black.}
    \label{n4ff}
\end{figure}

The Flip-Flip duality is represented by the following equality of $\mathbf{S}^3_b$ partition functions:
\begin{equation}
\begin{aligned}
\label{flipflip}
    Z_{ffFT[U(N)]}(\vec{X},\vec{Y},\tau) & = \prod_{j,k=1}^{N}s_b\left(\frac{iQ}{2}(2r-1)+\tau + X_j-X_k \right)s_b\left(\frac{iQ}{2}(2r-1)+\tau + Y_j-Y_k \right)\\& 
    \left. Z_{FT[U(N)]}(\vec{X},\vec{Y},-\tau)
    \right|_{r \to 1-r}
    \end{aligned}
\end{equation}
As shown in \cite{Hwang:2020wpd} it is possible to prove the flip-flip duality by iterative applications of the Aharony duality.
The idea of the proof is the following.
Starting from the left-most $U(1)$ node in the $FT[U(N)]$ theory (whose adjoint chiral is just a singlet) we apply Aharony duality \cite{Aharony:1997gp} which leaves the rank invariant but  gives mass to the  adjoint chiral field of the adjacent $U(2)$, hence we can apply again the Aharony duality on it.
This will remain a $U(2)$ node but the dualization will alaso give mass to the 
the  adjoint chiral field of the adjacent $U(3)$ node.
We continue applying iteratively the Aharony duality until we reach the last $U(N-1)$ node. Notice that since every $U(k)$ node sees $2k$ flavors, ranks do not change when we apply the Aharony duality.
We then perform another sequence of dualizations starting from the leftmost $U(1)$ node and stopping at the  second last node $U(N-2)$.
In the  third sequence of dualization we start again from the $U(1)$ node and proceed along the tail stopping at the $U(N-3)$ node. We iterate this procedure for a total of $N-1$ times,  applying Aharony duality $N(N-1)/2$ times.
The singlet fields flipping the mesons and the monopoles appearing in the Aharony duality reconstruct the singlet matrix $O_Y$
and give mass to the singlet matrix $O_X$.

We will now show that we  can derive similar  self-dualities (modulo flips) for the two  ${G}[U(N)]$ UV completions.
In analogy with the $\mathcal{N}=4$ case we denote these dualities as flip-flip dualities. These dualities map chiral theories to chiral theories and  planar theories to planar theories, in contrast to the mirror-like dualities described in Section \ref{sec: FTUN_mirror}. This is expected because the $\mathcal{N}=4$ flip-flip does not exchange flavor and topological symmetries, therefore in the $\mathcal{N}=2$ case the $SU(N)$ and $U(1)^{N-1}$ factors of the global symmetries of ${G}[U(N)]$ will not be exchanged.

\subsubsection{Chiral-Chiral flip-flip duality}

Starting from the flip-flip duality in Figure \ref{n4ff}, we take the real mass deformation  \ref{eq:chiral_mass_FTSUN1} 
that takes the $FT[U(N)]$ on the l.h.s.~to the first UV completion of $G[U(N)]$, the chiral UV completion. 

Contrary to the case discussed in the previous section, where we started from the two UV completions related by mirror duality, now the same deformation can be taken on the dual side in \ref{n4ff}.
Indeed in this case we find that in the dual vacuum is again a chiral quiver, so we obtain the following new chiral-chiral  duality relating the first chiral UV completion of the $G[U(N)]$ theory \eqref{g_n_chiral}:
\begin{equation} 
    \includegraphics[]{FiguresSec3/GUN_chiral.pdf}
\end{equation}
to its flip-flip chiral dual:
\begin{equation} 
    \includegraphics[]{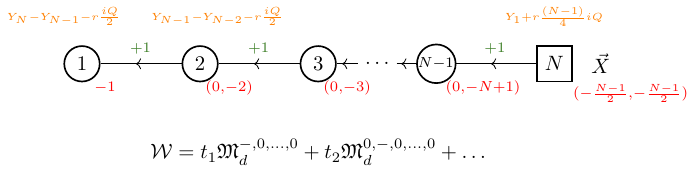}
\end{equation}

Notice that in the flip-flip dual theory
there is a superpotential 
containing flippers $t_i$ for all the dressed monopoles $\mon_d^{0,\dots,-,\dots,0}$ with negative GNO flux under a single gauge node.
In addition all the arrows and  CS coupling  signs are flipped 
and the FIs are reshuffled. The bifundamental fields in the flip-flip theory have trial R-charge $1-r$.
On the the flip-flip side there is also a background CS term at level $-\tfrac{1}{2}$ for each $U(1)$ factor in the $U(1)^{N-1}_Y$ symmetry group.

Under the duality map the dressed monopoles with negative
charge under the topological symmetries of the electric theory map to the set of singlets $t_i$ in the flip-flip dual.

Also in  this case, similarly as for the $FT[U(N)]$ theory, the duality can be demonstrated by iterative applications of local dualities as shown in Appendix \ref{accff}.
Now at each step we need a to locally apply the ciral Giveon-Kutasov duality for $U(N)_{1,1}$ with $[N-1,N+1]$ flavors \cite{Giveon:2008zn,Benini:2011mf,Aharony:2014uya}.

\subsubsection{Planar-Planar flip-flip duality}

Let's start again from the duality in Figure \ref{n4ff}.
Now we take the real mass deformation that takes the $FT[U(N)]$ on the l.h.s.~to the second UV completion of $G[U(N)]$, the planar UV completion.
So we perform the real mass deformation:
\begin{equation}
\label{eq:chiral_mass_FTSUN}
\left\{
\begin{array}{l}
X_J \to X_J + \frac{\tau}{2} (2J -N -1), \qquad J=1,\dots,N
\\
\vec{Y} \to \vec{Y} + \frac{\tau}{2} N
\end{array}
\right.,
\qquad\qquad 
\tau \to +\infty
\end{equation}
and we also perform the following shifts in gauge parameters:
\begin{equation}
\omega^{(J)}_{\alpha} \to \omega^{(J)}_{\alpha}  + \tau \left( \alpha - \frac{J+1}{2} \right)
\end{equation}

\begin{figure}[ht]
    \centering
    \includegraphics[]{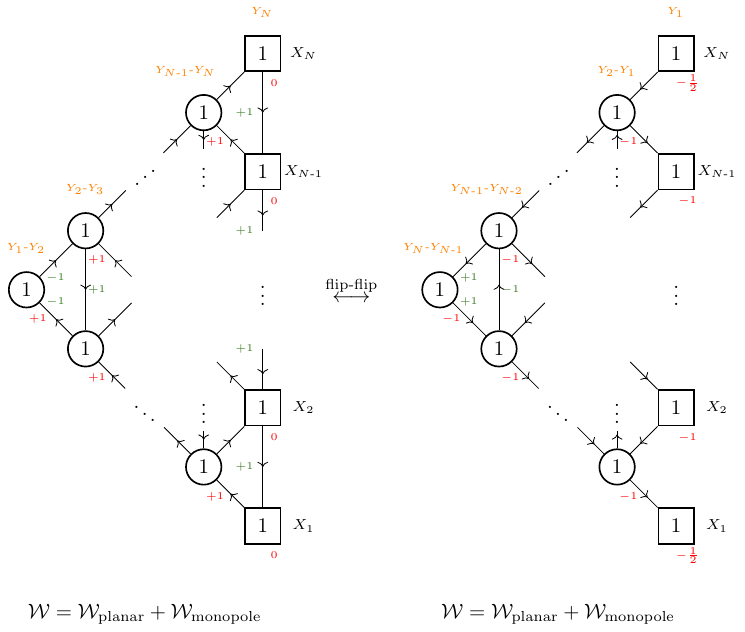}
    \caption{Self-duality \textit{modulo flips} displayed by the planar $G[U(N)]$ theory.
    The background CS terms for the flavor symmetry are different on the two sides of the duality,
    furthermore on the r.h.s.~there is a background CS term at level $-N$ for the emergent $SU(N)_Y$ global symmetry. Flowing from the $\mathcal{N}=4$ \textit{flip-flip} this is generated from integrating out the monopole flippers $O_Y$ in Figure \ref{n4ff}.
    On the r.h.s.~there are also shifts in the FI terms, discussed in the main text, which are not reported in the figure.
    }
    \label{quiv:flip-flip_ghat}
\end{figure}

Also in this case we can implement the same deformation on the dual side and we land on a pair of dual planar abelian quivers 
given in Figure \ref{quiv:flip-flip_ghat}.

Notice that in the flip-flip dual theory
all the arrows are flipped and 
the FIs are reshuffled.
Furthermore, CS and mixed CS levels have opposite sign with respect to the electric theory and the trial R-charge of the chirals is $1-r$. To avoid clutter here we set $r=0$, therefore the electric diagonal bifundamentals have trial R-charge $0$ and the magnetic diagonal bifundamentals have trial R-charge $1$.
On the the r.h.s.~there is also a background CS term at level $-N$ for the emergent $SU(N)_Y$ symmetry, as well as shifts for the FIs. On the flip-flip side the FIs for the gauge node of the $J$-th column from the right are:
\begin{equation}
\text{FI: } Y_J - Y_{J-1} + \frac{iQ}{4},\; Y_J - Y_{J-1},\; \dots,\; Y_J - Y_{J-1},\;Y_J - Y_{J-1} - \frac{iQ}{4}
\end{equation}
from top to bottom. In Figure \ref{quiv:flip-flip_ghat} we only reported the $Y$-dependent part of the FIs. 

The operator map is very simple: the 
vertical string of $N-1$ singlets in the last column of the electric theory maps to the  $N-1$ mesons
of the flip-flip theory theory 
constructed along the shortest path connecting two adjacent
$U(1)$ flavor nodes.

Also in this case the duality can be demonstrated by iterative applications of local Aharony-like dualities, as discussed in Appendix \ref{appff}.

\section{Towards an Algorithmic Approach: $\mathcal{S}-$walls, QFT Building Blocks, and Basic Duality Moves} \label{sec: S_walls,QFT_blocks_Duality_Moves}

Starting from the  $\mathcal{N}=4$ $U(N)$ SQCD with $F\geq 2N$ hypers we can also consider a more general real mass deformation, w.r.t.~the one discussed in the previous section, where we partially break the $SU(F)$ global symmetry to obtain an $\mathcal{N}=2$  $U(N)$ SQCD with $n_f$ fundamental and $n_a$ anti-fundamental chiral multiplets on the electric side: 
\begin{equation}    \label{eq:realmass_SQCD}
\begin{gathered}
\mathcal{N}=4 \quad U(N)
\\
F \text{ hypermultiplets}
\end{gathered}
\qquad\rightarrow\qquad
\begin{gathered}
\mathcal{N}=2 \quad U(N)_{-\frac{F}{2} + N, -\frac{F}{2}}
\\
[n_f,n_a] \text{ chiral fields,} \qquad F=n_f+n_a
\end{gathered}
\end{equation}
Here we consider flows that preserve the maximum amount of chiral matter multiplets
$F=n_f + n_a$, while giving mass to the adjoint chiral field. Therefore,
the number of (anti-) fundamentals $n_f$ and $n_a$ satisfy $F=n_f + n_a$ but are otherwise unconstrained and depend on the specific mass deformation that we consider. 
On the other hand, the choice of CS levels \textbf{is not arbitrary} and is fixed by the parent $\mathcal{N}=4$ theory: each massive (anti)fundamental multiple contributes as $(-\frac{1}{2},-\frac{1}{2})$ to the CS level and the massive adjoint chiral contributes as $(N,0)$. In the CS-level notation: $(k,k+lN)$, this corresponds to $k=-\tfrac{F}{2}+N$ and $l=-1$.

One can generalize to arbitrary values of $l$ by applying Witten's $SL(2,\mathbb{Z})$ action on both sides of the duality as discussed in Subsection \ref{ga-ung-section}.

We can also turn more general real masses and integrate out more chirals  to access a more general range of CS levels. 
Notice that this can also be achieved by first performing the real mass deformation that preserves the maximum number of chirals, discussed here, and then performing additional real masses for the resulting $\mathcal{N}=2$ theories.
We will address this generalization in future work \cite{RealMasses}.

One can then consider the same limit on the mirror side and try to identify the dual vacuum. In general, this is a non-trivial exercise as there are multiple potential vacua corresponding to possible ways of moving on the Coulomb branch of the $\mathcal{N}=4$ mirror (which, in turn, match how we shift gauge fugacities in the $\mathbf{S}^3_b$ partition function).

More generally, one can start on the electric side with a linear unitary $\mathcal{N}=4$ quiver with gauge group $\prod_{N_i} U(N_i)$. We consider a real mass deformation, paired with a VEV for the scalar in the $\mathcal{N}=2$ vector multiplets, such that the adjoint chiral fields and only one  of the $\mathcal{N}=2$ chiral multiplets in each matter hypermultiplet acquires a real mass. 
Such a deformation preserves the maximum total amount of chiral fields.
Integrating out the massive field we obtain a chiral linear quiver:
\begin{equation}
    \includegraphics[]{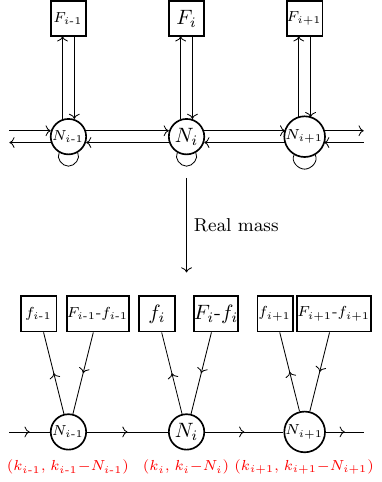}
\end{equation}
where $ f_i \leq F_i$ is the number of fundamental flavors for the $i$-th node, and $F_i-f_i$ is the number of antifundamental flavors.
The CS levels of the $i$-th gauge node $k_i$ are fixed by the parent theory:
\begin{equation}\label{eq:chiralCSrule}
    k_i = -\frac{N_{i-1} + N_{i+1} + F_{i}}{2} + N_{i} \,.
\end{equation}

One can also consider more general real mass flows such that in the resulting $\mathcal{N}=2$ quiver some of the bifundamental chiral fields point to the left.
As before, one can implement the same limit in the mirror dual quiver theory to reach the dual vacuum. This remains a non-trivial exercise.

Clearly, it is desirable to have a more systematic way to determine the planar mirror dual of a generic flavored $\mathcal{N}=2$ quiver without the need to resort to $\mathbf{S}^3_b$ partition functions. We do so by adapting the idea of the local dualization algorithm of \cite{Hwang:2021ulb,Comi:2022aqo}
which we quickly review now.

\subsection{Lightning Recap: The $\mathcal{N}=4$ Dualization Algorithm}\label{Subsec: N=4_alg}

The dualization algorithm for $\mathcal{N}=4$ linear quiver theories provides a purely field-theoretical proof of mirror duality assuming only basic Seiberg-like dualities.

It is well known that $\mathcal{N}=4$ linear quiver theories can be realized on Hanany-Witten brane set-ups and inherit mirror dualities from the S-duality of Type IIB set-ups \cite{Hanany_1997}. 
The idea of the algorithm originates from the observation \cite{Gaiotto_2007,Gulotta:2011si,Assel:2014awa} that on linear or circular brane setups, $\mathcal{S}-$duality can act locally on each 5-brane creating an $\mathcal{S}-$duality wall on its right and an $\mathcal{S}^{-1}$-duality  wall on its left:
$D5= \mathcal{S}\!\cdot\! NS5 \!\cdot\! \mathcal{S}^{-1}$ and $\overline{NS5}= \mathcal{S} \!\cdot\! D5 \!\cdot\! \mathcal{S}^{-1}.$
The dualization algorithm implements this local action of S-duality in field theory. \\
In this subsection we briefly review result presented in \cite{Bottini:2021vms,Hwang:2021ulb,Comi:2022aqo}, the reader can look at them for a more comprehensive discussion.

Following \cite{Gaiotto_2009}, we identify the S-duality wall with the  $FT[U(N)]$ theory:\footnote{The $FT[U(N)]$ theory differs from the $T[U(N)]$ theory introduced in \cite{Gaiotto_2009} for the presence of an extra set of singlets flipping the Higgs branch moment map.}
 \begin{equation}
 \begin{array}{l}
    \mathcal{S}_{\mathcal{N}=4}\text{-wall} :\; \mathcal{S}_{\mathcal{N}=4}[\vec X,\vec Y]:=FT[\vec{X},-\vec{Y}]=\;
    \begin{tikzpicture}[baseline=(current bounding box.center)]
        \node at (-4,0) (f10) [flavor, black] {$N$};
        \node at (-2,0) (f11) [flavor,black] {$N$};
        \draw (f10)+(1,-0.2) node {  };
        \draw[black] (f10)++(-.3,-.6) node[anchor=west]{$\vec{X}$};
        \draw[black] (f11)++(-.3,-.6) node[anchor=west]{$\vec{Y}$};
        \draw (-3,0.5) node {$+$};
        \draw[dashed] (f10) -- (f11);
    \end{tikzpicture} 
\\
    \mathcal{S}_{\mathcal{N}=4}^{-1}\text{-wall} :\; \mathcal{S}^-_{\mathcal{N}=4}[\vec X,\vec Y]:=FT[\vec{X},\vec{Y}] =\;
    \begin{tikzpicture}[baseline=(current bounding box.center)]
        \node at (-4,0) (f10) [flavor, black] {$N$};
        \node at (-2,0) (f11) [flavor,black] {$N$};
        \draw (f10)+(1,-0.2) node {};
        \draw[black] (f10)++(-.3,-.6) node[anchor=west]{$\vec{X}$};
        \draw[black] (f11)++(-.3,-.6) node[anchor=west]{$\vec{Y}$};
        \draw (-3,0.5) node {$-$};
        \draw[dashed] (f10) -- (f11);
    \end{tikzpicture} 
\end{array}
 \end{equation}
As discussed in Section \ref{subsec: S-Wall_G}, the $FT[U(N)]$ theory admits a Lagrangian UV completion Figure \ref{quiv:ftun} and is self-mirror under the exchange of emergent and manifest $U(N)$ symmetries. 

The $\mathcal{S}$-wall  satisfies the fusion to identity encoded in a partition function identity \cite{Bottini:2021vms}:
\begin{equation}
    \int d\vec{W} \Delta_{(N)}(\vec{W};\tau) Z_{\mathcal{S}_{\mathcal{N}=4}}^{(N)}(\vec{X},\vec{W};\tau)Z^{(N)}_{\mathcal{S}^{-1}_{\mathcal{N}=4}}(\vec{W},\vec{Y};\tau) = _{\vec{X}}\mathbb{I}_{\vec{Y}}(\tau) \,.
\end{equation}
interpretable as $\mathcal{S}\mathcal{S}^{-1} = 1$, where the identity operator on the r.h.s.~can be thought of as the identity element, with:
\begin{equation}
    _{\vec{X}}\mathbb{I}_{\vec{Y}}(\tau) = \frac{1}{N! \Delta_{(N)}(\vec{X},\tau)} \sum_{\sigma \in S_N} \prod_{j=1}^N \delta (X_j - Y_{\sigma(j)} ) \,,
\end{equation}
and the $\mathcal{N}=4$ measure
is defined in \eqref{eq:n4measure}.

This identity can pictorially be represented as follows:
\begin{equation} \label{n=4fusion_to_identity}
\begin{tikzpicture}
    \node at (-2,0) (m1) [manifest,black] {$N$};
    \node at (0,0) (g1) [gauge, black] {$N$};
    \node at (2,0) (f1) [flavor,black] {$N$};

    \draw[dashed] (m1) -- (g1);
    \draw[dashed] (f1) -- (g1);
    \draw[-] (g1) to[out=60,in=10] (0,0.5) to [out=180,in=120] (g1);
    \draw (-1,0.3) node {$+$};
    \draw (1,0.3) node {$-$};

    \draw[black] (-2,-.7) node {$\vec{X}$};
    \draw[black] (2,-.7) node {$\vec{Y}$};
    \draw (0,.8) node {$a$};
    
    \draw (4,0) node {$\longleftrightarrow$};

    \draw (6,0) node {$_{\color{black}\vec{X}}\mathbb{I}_{\color{black}\vec{Y}}(\tau)$};
\end{tikzpicture}\end{equation}
The adjoint chiral $a$ couples to the moment map operators $A_L$ and $A_R$ of the left and right  walls as $\mathcal{W}=a(A_L+A_R)$.

We then identify the QFT blocks associated to the different 5-brane configurations:
\begin{equation}
    \includegraphics[]{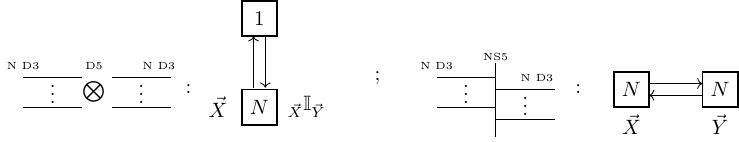}
\end{equation}
which are denoted the flavor block and the bifundamental block respectively.

\begin{figure}
    \centering
    \includegraphics[width=\textwidth]{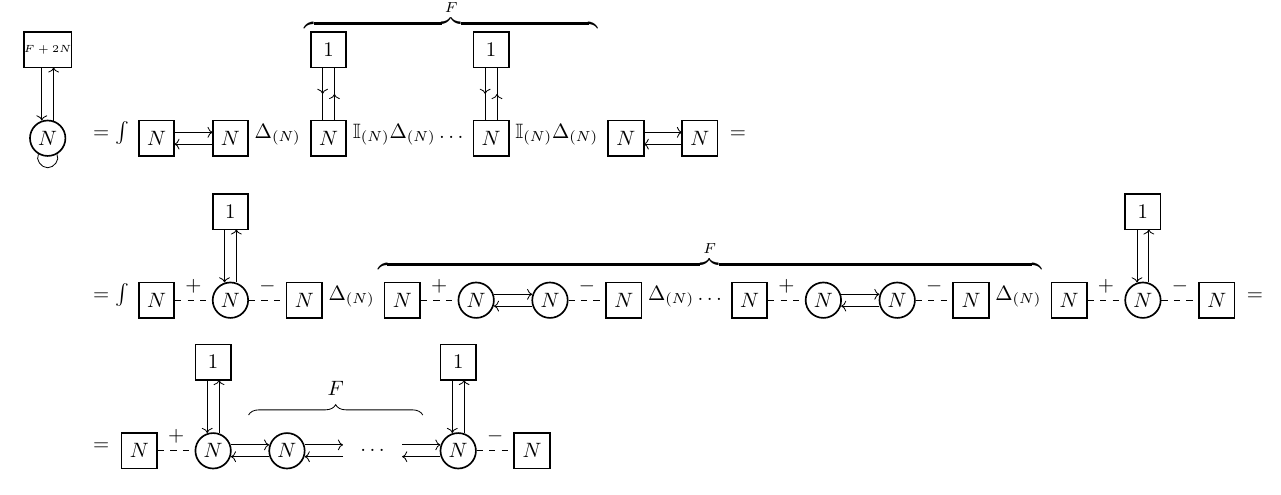}
    \caption{Applying the $\mathcal{N}=4$ dualization to the $\mathcal{N}=4$ $U(N)$ SQCD with $F+2N$ fundamental hypermultiplets proceeds as follows - in the first line the quiver is chopped into QFT blocks. In the second line each block is dualized using the basic moves. In the last line we used the fusion to Identity property of the $\mathcal{S}$-wall to obtain the known mirror dual theory. Notice that the two tails of gauge nodes with increasing ranks are condesed in the $\mathcal{S}$-wall notation. By carefully performing each step, it is also possible to keep track of all the parameters to find that we land precisely on the duality in Figure \ref{u_n_mirror}.
    We have also explicitly kept the integrals and integration measures to emphasize that gauging or ungauging nodes must be performed with the appropriate measure.}
    \label{fig:N=4Algorithm}
\end{figure}

We can now define the action of the S-duality wall on the basic QFT blocks:
\begin{enumerate}
\item The first basic move consists in
the $\mathcal{S}$-dualization of a bifundamental block into a fundamental flavor  block (the field-theoretic counterpart of  $\mathcal{S} \!\cdot\! D5 \!\cdot\! \mathcal{S}^{-1}=\overline{NS5})$:

\begin{equation}    \label{quiv:bif_block_N=4}
    \includegraphics[]{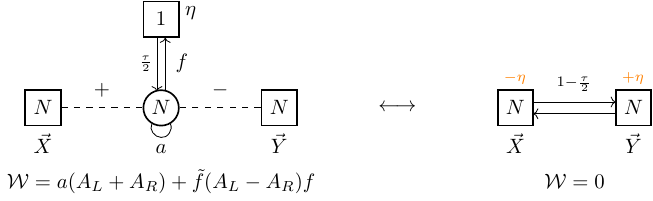}
\end{equation}
wherein we have explicitly reported the trial $R$-charge and the $U(1)_{\tau}$ charge of the various fields in the theory. On the r.h.s.~the orange fugacities denote the background FIs for the flavor nodes. This corresponds to the partition function identity:
\begin{equation}
   \begin{aligned}
     \int d\vec{W}\Delta_{(N)}(\vec{W},\tau) & \;
     Z_{\mathcal{S}_{\mathcal{N}=4}}^{(N)}(\vec{X},\vec{W},\tau) \; Z_{\mathcal{S}^{-1}_{\mathcal{N}=4}}^{(N)}(\vec{W},\vec{Y},\tau)\; \prod_{\alpha=1}^N\;s_b\bigg(\frac{iQ}{2}-\frac{\tau}{2}\pm(W_{\alpha}-\eta)\bigg) = \\
     & = e^{-2\pi i \eta\sum_{j=1}^N(X_j-Y_j)} \prod_{j,k=1}^N\;s_b\bigg(\frac{\tau}{2}\pm (X_j-Y_k)\bigg)\,.
   \end{aligned}
\end{equation}

\item The inverse  basic move consists of the $\mathcal{S}$-dualization of a fundamental block into a bifundamental  block (which is the field-theoretic counterpart  of $\mathcal{S} \!\cdot\! NS5 \!\cdot\! \mathcal{S}^{-1} = D5$)

\begin{equation}  \label{quiv:fund_block_N=4}
    \includegraphics[]{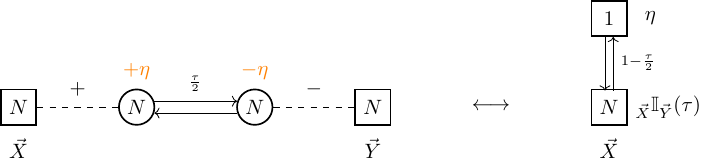}
\end{equation}
Notice that the gluing on the l.s.h.~does not include the addition of $U(N)$ adjoint multiplets coupling linearly to the moment map of the  $\mathcal{S}$-walls.
This inverse basic move is encoded in the following partition function identity:
\begin{equation}
    \begin{aligned}
      & \prod_{j=1}^{N} s_b\left(\frac{\tau}{2}\pm (X_j-\eta) \right)\, _{\vec{X}}\mathbb{I}_{\vec{Y}}(\tau)= \\
      & = \int d\vec{U}d\vec{V} \frac{e^{2\pi i \eta \sum_{a=1}^N(U_a-V_a)}}{\prod_{\overset{j,k=1}{j<k}}^N s_b \left(\frac{iQ}{2}\pm (U_j-U_k) \right)s_b \left(\frac{iQ}{2}\pm (V_j-V_k) \right)} \\
      &Z_{\mathcal{S}_{\mathcal{N}=4}}^{(N)}(\vec{X},\vec{U}; \tau)Z_{\mathcal{S}^{-1}_{\mathcal{N}=4}}^{(N)}(\vec{V},\vec{Y}; \tau) \prod_{a,b=1}^N s_b\left(\frac{iQ}{2}-\frac{\tau}{2} \pm (U_a-V_b) \right) \,.
    \end{aligned}
\end{equation}  

\end{enumerate}
We consider the case of $\mathcal{N}=4$ $U(N)$ SQCD with $2N+F$ flavors to show how the algorithm  works in Figure \ref{fig:N=4Algorithm}. \footnote{Here we use a short-cut, by treating $2N$ fundamental hypers as two sets of ``frozen" bifundamentals. We could have also proceeded as in \cite{Comi:2022aqo}
by chopping all the $F+2N$ flavors into fundamental blocks and adding a $(0,N)$ and an $(N,0)$ bifundamental block.}
When decomposing the theory into QFT blocks, it is important to keep in mind that each block effectively plays the role of a matrix element, with two sets of indices corresponding to the two $N$-dimensional vectors of parameters associated with the two $U(N)$ global symmetries. In particular, note that flavor blocks are accompanied by an identity wall, reflecting this matrix structure.

In this quick review, we have only discussed QFT blocks and Identity-walls with constant rank $N$. We can, for example, have quivers with non-constant ranks which are chopped into $U(N)\times U(M)$ \textit{asymmetric} bifundamental blocks.
The
dualization of such blocks requires the introduction of \textit{asymmetric} Identity-walls.
The algorithm in this case includes an addition step, a sequence of Hanany-Witten duality moves
to remove all the \textit{asymmetric} walls.
For the analysis of these more general cases we refer the reader to \cite{Comi:2022aqo}. \\

In \cite{Benvenuti:2023qtv} the local dualization algorithm approach has been extended to a class of generalized $\mathcal{N}=2$ quiver theories that can be realized in set-ups involving $NS, NS', D5, D5$ branes preserving four supercharges.
Unlike the $\mathcal{N}=4$ case where brane setups directly inform the structure of the low-energy field theory, here the situation is reversed: the algorithmic field theory approach yields an exact dual description, while the corresponding brane configurations are difficult to interpret directly. This inversion of logic highlights the power of the algorithmic method—not only does it provide precise duals, but it also offers a new lens through which to understand and possibly reconstruct the brane setups themselves. Motivated by this, our aim is now to develop a similar algorithm tailored to chiral-planar $\mathcal{N}=2$ theories.

The algorithm will allow us to construct mirror duals of $\mathcal{N}=2$ SQCD theories with both fundamentals and antifundamental chirals, as well as for chiral-planar quiver theories as we will show in Section \ref{sec: Examples}.
We provide a realization of the $\mathcal{N}=2$  $\mathcal{S}$-wall operator in Section \ref{subsec:N=2 S-wall} and construct the basic QFT building blocks and the corresponding duality moves in Section \ref{subsec: newBasic Duality Moves}.

\subsection{The chiral-planar $\mathcal{S}$-wall, fusion to Identity and gluing rules} 
\label{subsec:N=2 S-wall}

In this section we propose, in analogy $\mathcal{N}=4$ algorithm, to identify $G[U(N)]$ as the “chiral-planar $\mathcal{S}$-wall" and prove its fusion to Identity properties.

\begin{figure}[ht]
\centering
    \includegraphics[]{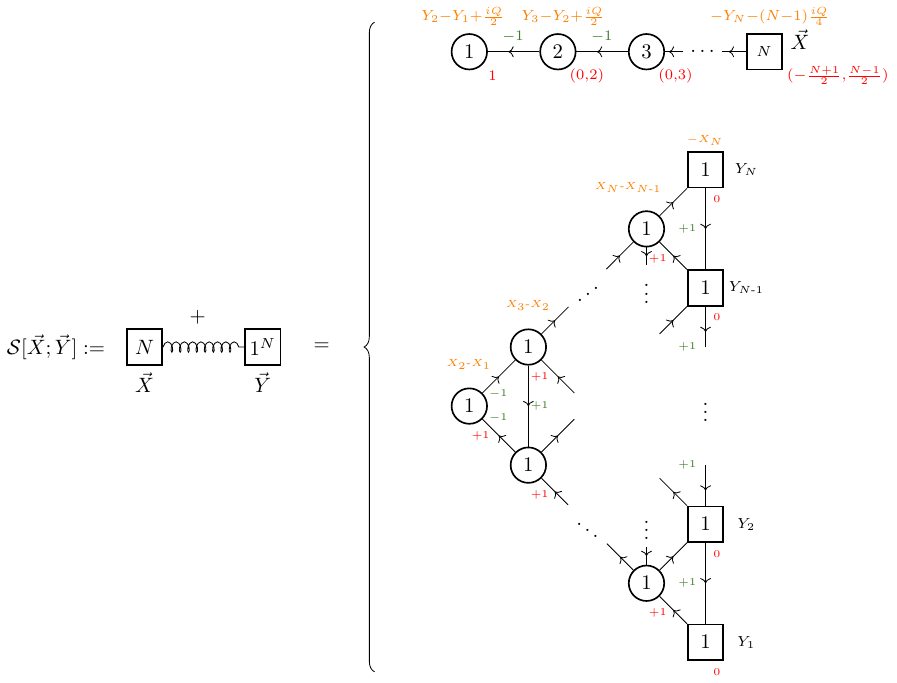}
    \caption{
        The planar and chiral UV completions of the $\mathcal{N}=2$ chiral-planar $\mathcal{S}$-wall coinciding with $G[-\vec{X};\vec{Y}]$ are shown here. Mixed CS interactions have been suppressed for brevity. Here, the trial R-charge $r$ is set to $0$. Therefore, in the chiral completion, on top, each bifundamental has trial R-charge 0. In the planar completion the vertical/diagonal chirals have trial R-charge $2/0$.}
    \label{fig:G-wall}
\end{figure}

\begin{figure}
\centering
    \includegraphics[]{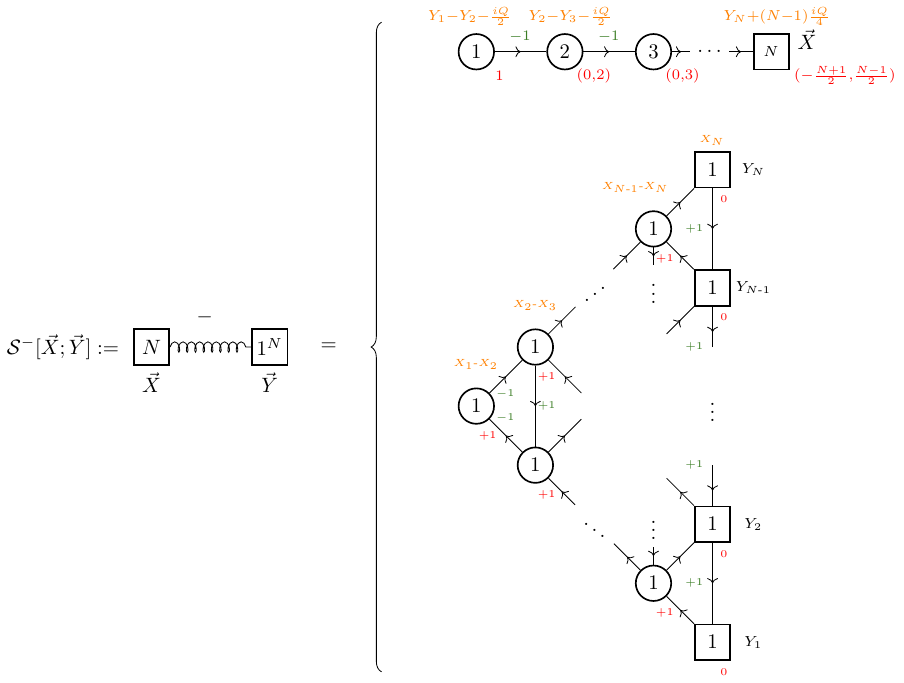}    
    \caption{
        The planar and chiral UV completions of the $\mathcal{N}=2$ $\mathcal{S}$-wall coinciding with $G[-\vec{X};\vec{Y}]$ are shown here. Mixed CS interactions have been suppressed for brevity. Here, the trial R-charge $r$ is set to $0$ and the trial R-charges of the chiral fields can be read as in the previous figure.}
    \label{fig:G-invwall}
\end{figure}

More precisely we define:
\begin{equation}
 \begin{array}{rl}
    \mathcal{S}\text{-wall} :&\;\mathcal{S}[\vec X,\vec Y]= G [-\vec{X},\vec{Y}]=\;
    \begin{tikzpicture}[baseline=(current bounding box.center)]
        \node at (-4,0) (f10) [flavor, black] {$N$};
        \node at (-2,0) (f11) [flavor,black] {$1^N$};
        \draw[] (f10)++(-.3,-.6) node[anchor=west]{$\vec{X}$};
        \draw[] (f11)++(-.3,-.6) node[anchor=west]{$\vec{Y}$};
        \draw[decorate,decoration={coil,segment length=4pt}] (f10)--(f11) node[midway,above,yshift=5pt] {$+$};
    \end{tikzpicture} 
\\
    \mathcal{S}^{-1}\text{-wall} :&\; \mathcal{S}^-[\vec X,\vec Y]=G[\vec{X},\vec{Y}] =\;
    \begin{tikzpicture}[baseline=(current bounding box.center)]
        \node at (-4,0) (f10) [flavor, black] {$N$};
        \node at (-2,0) (f11) [flavor,black] {$1^N$};
        \draw[] (f10)++(-.3,-.6) node[anchor=west]{$\vec{X}$};
        \draw[] (f11)++(-.3,-.6) node[anchor=west]{$\vec{Y}$};
        \draw[decorate,decoration={coil,segment length=4pt}] (f10)--(f11) node[midway,above,yshift=5pt] {$-$};
    \end{tikzpicture} 
\end{array}
 \end{equation}
 Both theories admit (at least) two UV completions whose difference lies in the direction of the arrows and the sign of FIs. For convenience the UV completions of both theories are reported in Figures \ref{fig:G-wall} and \ref{fig:G-invwall}.
\newpage

Notice that in the chiral UV completion 
of $\mathcal{S}$ the arrow direction is inverted w.r.t.~$\mathcal{S}^{-}$, where it emanates from the manifest $U(N)$, and the FI at each node are opposite: $Y_{i+1}-Y_i$.
In the planar UV completion of $\mathcal{S}$ the FI's associated to each string are $X_{i+1}-X_i$ rather than  $X_{i}-X_{i+1}$
as in $\mathcal{S}^{-}$.

In our notation the first argument of the $Z_{\mathcal{S}^\pm}[\cdot \, ;\cdot]$ theories corresponds to fugacities for an $U(N)$ global symmetry, while the second argument corresponds to a $U(1)^{N-1}$ global symmetry.

We will now prove the fusion to identity properties of the $\mathcal{N}=2$ $\mathcal{S}$-wall.

\clearpage

\subsubsection{Planar Fusion to Identity}

Our  starting point is the 
$\mathcal{N}=4$ fusion to identity (\ref{n=4fusion_to_identity})
where we consider the UV completion where we gauge the left and right  $\mathcal{S}_{\mathcal{N}=4}$-wall by gauging their manifest $U(N)$ symmetries.

\begin{figure}[ht]
    \centering
    \includegraphics[]{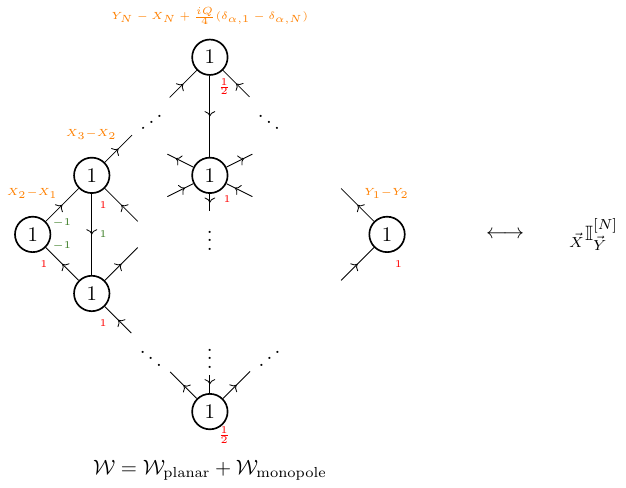}
    \caption{The planar fusion to identity observed upon gluing a $\mathcal{S}$ and $\mathcal{S}^-$ wall. Note that the $\delta_{\alpha,j}\frac{iQ}{4}$ indicates the $j^{th}$ node of the middle column has a shifted FI term, counting from below.}
    \label{eq:manplanfus}
\end{figure}

We  consider the large mass deformation: 
\begin{align}
\label{fusidshift}
    \begin{cases}     
     X_j \to X_j + N \frac{\tau}{2} \,, \\
     Y_j \to Y_j + N \frac{\tau}{2} \,,
     \end{cases} \qquad \; j=1,\dots,N
\end{align}
which preserves both the (emergent) $U(N)_{X,Y}$ global symmetries. We also perform the following shifts in gauge fugacities:
\begin{equation}
\label{fusidshift1}  
    \begin{cases} 
     u_{\alpha}^{(I)} \to u_{\alpha}^{(I)} + \frac{\tau}{2}(2\alpha-I-1) \,, \quad & \text{for}\; I=1,\dots,N\\
     u_{\alpha}^{(I)} \to u_{\alpha}^{(I)} + \frac{\tau}{2}(2\alpha+I-2N-1) \,, \quad & \text{for}\; I=N+1,\dots,2N-1
    \end{cases}
\end{equation}
with $\tau \to \infty$, where $u_{\alpha}^{(I)}$ are the gauge fugacities for the $I$-th gauge node in the quiver (starting from the left), while $\alpha=1,\dots, |G^{(I)}|$ with $|G^{(I)}|$ being the rank of the $I$-th gauge node. \\
This triggers a flow to the $\mathcal{N}=2$ planar fusion to identity depicted in eq.~\ref{eq:manplanfus}.
On the l.h.s.~we glue two $\mathcal{S}$-walls
in their planar abelian UV completion by gauging a diagonal $U(1)^N$ symmetry.
On the r.h.s.~we find an Identity-wall identifying  the $U(N)_X$ and $U(N)_Y$ manifest symmetries.

We encode the fusion to identity in the following partition function identity:
\begin{equation}\label{eq:chiral_gluing}
    \int d\vec{Z} \DeltaOne[\vec{Z}] Z_{\mathcal{S}}(\vec{X},\vec{Z}) Z_{\mathcal{S}^-}(\vec{Y},\vec{Z}) = \frac{1}{N!\DeltaN[\vec{X}] } \sum_{\sigma \in S_N} \delta \big( X_i-Y_{\sigma(i)} \big) \equiv \identityN[\vec{X},\vec{Y}]  \,.
\end{equation}
On the l.h.s.~we integrate over the abelian set of fugacities $\vec Z$ entering the second slot of $Z_{\mathcal{S}^\pm}[\cdot \,;\cdot]$.
The \textbf{Planar Measure} $\DeltaOne[\vec{Z}]$ in \eqref{eq:chiral_gluing} is defined as a limit of the $\mathcal{N}=4$ gluing measure  $\Delta_{(N)} (\vec{Z},\tau)$ \eqref{eq:n4measure}:
\begin{equation}
\lim_{\tau \to +\infty}
    \left.\Delta_{(N)} (\vec{Z},\tau) \right|_{Z_i \to Z_i + \frac{\tau}{2}(2i-N-1) }
    =
    \DeltaOne[\vec{Z}] \times (\text{divergent phase}) \,.
\end{equation}
Notice the shift on the gauge variables is consistent with eq. \eqref{fusidshift1} for $\vec Z=\vec{u}^{(I=N)}$.
We obtain:
\begin{equation}\label{planmeasure}
    \DeltaOne[\vec{Z}]=  e^{i \pi H^{[1^N]}(\vec{Z})} \prod_{i=1}^{N-1} s_b\bigg(\frac{iQ}{2}-Z_i+Z_{i+1}\bigg)
\end{equation}
with
\begin{equation}\label{planarcs}
        H^{[1^N]}(\vec{Z}):=  
        \frac{iQ}{2}(Z_1-Z_N)
        -\frac{1}{2}(Z_1^2+Z_N^2)-\sum_{i=2}^{N-1}Z_i^2+\sum_{j=1}^{N-1}Z_jZ_{j+1} \,.
\end{equation}
Note that the contribution from the vector multiplet of $U(N)$
 has disappeared, as expected from Higgsing the gauge group down to its maximal torus $U(1)^N$. The limit for the $U(N)$ adjoint gives rise to a string of vertical chiral multiplets connecting adjacent $U(1)^N$ nodes and flowing upward.
This string of vertical chiral fields couples
to the vertical chiral fields (flowing downward) connecting the $U(1)^N$ nodes of the left and right $\mathcal{S}$-walls.
As a result two strings of  vertical bifundamental (flowing in opposite directions)
give mass to each other and we are left with a single string of vertical bifundamentals in the middle flowing downward.
This is obviously the planar version of the 
$\mathcal{N}=4$ gluing $a(A_L+A_R)$ present in \eqref{n=4fusion_to_identity}.

The quadratic form $H^{[1^N]}(\vec{Z})$ encodes the CS levels and the BF couplings and FI shifts for the middle column of $U(1)^N$ gauge nodes.\\

\noindent On the r.h.s.~of the identity \eqref{eq:chiral_gluing}, we identify the 
$\mathcal{N}=2$ \textbf{Chiral Identity-wall}: 
\begin{equation}
\identityN[\vec{X},\vec{Y}]\equiv\frac{1}{N!\DeltaN[\vec{X}] } \sum_{\sigma \in S_N} \delta(X_i-Y_{\sigma(i)}) 
\end{equation}
identifying the
$U(N)_X$ and $U(N)_Y$ global symmetries.
\noindent The \textbf{Chiral Measure} $\DeltaN[\vec{X}]$ 
is defined by starting again from the definition of the $\mathcal{N}=4$ gluing measure $\Delta_{(N)}(\vec{X},\tau)$ in \eqref{eq:n4measure}, then shifting the flavor fugacities $\vec{X}$ as in \eqref{fusidshift} and taking the limit $\tau\to +\infty$.
We obtain (again after suppressing the diverging prefactor):
 \begin{equation}    \label{eq:DeltaN}
    \DeltaN[\vec{X}]= e^{i \pi H^{[N]}(\vec{X})}\prod_{\alpha = 1}^N  \prod_{\beta < \alpha} \frac{1}{s_b ( \frac{iQ}{2} \pm (X_\alpha - X_\beta))} \,;
\end{equation}
with
\begin{equation}
\begin{aligned}
    H^{[N]}(\vec{X}) & :=  
    -(N-1)(\sum_{i=1}^N X_i^2) +2 \sum_{i<j}^NX_iX_j = \\
    &
    =
    - N\sum_{i=1}^N X_i^2 + \left(\sum_{i=1}^N X_i\right)^2 \,,
\end{aligned}
\end{equation}
which corresponds to a CS coupling at level
 $(N,0)$.
 
Observe that in the integrand on the l.h.s.~of \eqref{eq:chiral_gluing} appears the Planar Measure $\DeltaOne[\vec{Z}]$, while at the denominator of the r.h.s.~of the same identity appears the Chiral Measure $\DeltaN[\vec{X}]$. This is consistent with our claim that mirror symmetry sends the $\mathcal{N}=2$ \textit{chiral} deformation of an $\mathcal{N}=4$ theory into the $\mathcal{N}=2$ \textit{planar} deformation of its $\mathcal{N}=4$ mirror dual and vice versa, and so the same happens for the basic moves.

Although we  suppressed  the divergent contributions to avoid clutter, we have checked
that the divergent pre-factors on l.h.s.~and r.h.s.~of the partition function identity
and cancel-out.

We schematically depict this planar fusion to identity as:
\begin{center}
\tikzstyle{flavor}=[rectangle,draw=red!50,thick,inner sep = 0pt, minimum size = 6mm]
\tikzstyle{manifest}=[rectangle,draw=blue!50,thick,inner sep = 0pt, minimum size = 6mm]
\tikzstyle{gauge}=[circle,draw=black!50,thick,inner sep = 0pt, minimum size = 6mm]
\tikzset{-<-/.style={decoration={
  markings,
  mark=at position .5 with {\arrow{<}}},postaction={decorate}}}
\tikzset{->-/.style={decoration={
  markings,
  mark=at position .5 with {\arrow{<}}},postaction={decorate}}}
  \begin{equation} \label{n=2_planar_fusion_to_identity}
\begin{tikzpicture}
    \node at (-2,0) (m1) [manifest,black] {$N$};
    \node at (0,0) (g1) [gauge, black] {$1^N$};
    \node at (2,0) (f1) [flavor,black] {$N$};

    \draw[decorate,decoration={coil,segment length=4pt}] (m1) -- (g1);
    \draw[decorate,decoration={coil,segment length=4pt}] (f1) -- (g1);
    \draw (-1,0.5) node {$+$};
    \draw (1,0.5) node {$-$};

    \draw[black] (-2,-.8) node {$\vec{X}$};
    \draw[black] (2,-.8) node {$\vec{Y}$};
    \draw[red] (g1)+(0.5,-0.95) node {\tiny$_{\def\arraystretch{0.8}
    \left(
    \begin{array}{c}
          \frac{1}{2}\\
          1 \\
          \vdots\\
          1 \\
          \frac{1}{2}
    \end{array}
    \right)}$};
    
    \draw (4,0) node {$\longleftrightarrow$};

    \draw (6,0) node {$\identityN[\color{black}\vec{X},\color{black}\vec{Y}]$};
\end{tikzpicture}\end{equation}
\end{center}
Where the vector 
in red indicates the CS levels for the middle column of $U(1)^N$ nodes in
\eqref{eq:manplanfus}.  Remember however that there also BF couplings and FI shitfs given by the planar measure \eqref{planarcs}.

\subsubsection{The Chiral Fusion to Identity}

Starting again from the
$\mathcal{N}=4$ fusion to identity (\ref{n=4fusion_to_identity})
where  we glue the left and right  $\mathcal{S}_{\mathcal{N}=4}$-wall by gauging their manifest $U(N)$ symmetries,
we perform a large mass deformations that breaks the $U(N)_{X,Y}$ global symmetries to their Cartan. This is encoded in the following shift in fugacities:
\begin{equation}
    \begin{cases}
     \vec{u}^{(I)} \to \vec{u}^{(I)} - \frac{\tau}{2}I \,, \quad &\text{for} \; I=1,\dots,2N-1 \\
     X_j \to X_j + \frac{\tau}{2}(2j-N) \,, \quad &\text{for} \; j=1,\dots,N \\
     Y_j \to Y_j + \frac{\tau}{2}(2j-N) \,, \quad &\text{for} \; j=1,\dots,N
    \end{cases}
\end{equation}
With this limit we flow to the $\mathcal{N}=2$ operator identity which admits the following Lagrangian UV completion:
\begin{equation} \label{eq:manchirfus}
    \includegraphics[]{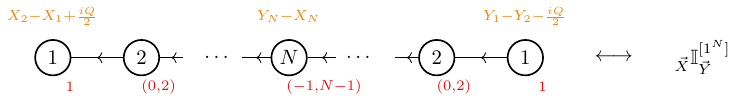}
\end{equation}

The associated  partition function identity reads: 

\begin{equation}\label{eq: planar gluing}
\int d\vec{Z}\DeltaN[\vec{Z}] Z_{\mathcal{S}}(\vec{Z},\vec{X}) Z_{\mathcal{S}^-}(\vec{Z},\vec{Y}) = \frac{1}{\DeltaOne[\vec{X}] } \prod_{i=1}^{N} \delta(X_i-Y_i) \equiv \identityOne[\vec{X},\vec{Y}],
\end{equation}
where we integrate over the first slot
of  $ Z_{\mathcal{S}}$ corresponding to the non-abelian symmetry.
On the l.h.s.~we have the chiral integration measure $\DeltaN[\vec{Z}]$ defined in \ref{eq:DeltaN} which carries the factor $H^{[N]}(\vec Z)$ contributing a CS coupling at level $(N,0)$.
The left and right $\mathcal{S}$-walls also carry a background (now becoming dynamical) CS level  $(-(N+1)/2, (N-1)/2)$ so the overal CS level of the middle $U(N)$ node is $(-1,N-1)$ in the the Lagrangian UV completion \eqref{eq:manchirfus}.
On the r.h.s.~of the identity  we identify the 
$\mathcal{N}=2$ \textbf{Planar Identity-wall}:
\begin{equation}
 \identityOne[\vec{X},\vec{Y}]\equiv
 \frac{1}{\DeltaOne[\vec{X}] } \prod_{i=1}^{N} \delta(X_i-Y_i) \,,
\end{equation}
identifying the two sets of $U(1)^N$ 
global symmetries of the left and right $\mathcal{S}$-walls.

Although we  suppressed  the divergent contributions to avoid clutter, we  checked
that the divergent pre-factors on l.h.s. and r.h.s. of the partition function identity
cancel out.

We schematically depict this chiral fusion to identity as:
\begin{center}
\tikzstyle{flavor}=[rectangle,draw=red!50,thick,inner sep = 0pt, minimum size = 6mm]
\tikzstyle{manifest}=[rectangle,draw=blue!50,thick,inner sep = 0pt, minimum size = 6mm]
\tikzstyle{gauge}=[circle,draw=black!50,thick,inner sep = 0pt, minimum size = 6mm]
\tikzset{-<-/.style={decoration={
  markings,
  mark=at position .5 with {\arrow{<}}},postaction={decorate}}}
\tikzset{->-/.style={decoration={
  markings,
  mark=at position .5 with {\arrow{<}}},postaction={decorate}}}
  \begin{equation} \label{n=2_chiral_fusion_to_identity}
\begin{tikzpicture}
    \node at (-2,0) (m1) [manifest,black] {$1^N$};
    \node at (0,0) (g1) [gauge, black] {$N$};
    \node at (2,0) (f1) [flavor,black] {$1^N$};

    \draw[decorate,decoration={coil,segment length=4pt}] (m1) -- (g1);
    \draw[decorate,decoration={coil,segment length=4pt}] (f1) -- (g1);
    \draw (-1,0.5) node {$+$};
    \draw (1,0.5) node {$-$};
    \draw[red] (g1)+(0.6,-0.5) node {$_{(-1,N-1)}$};

    \draw (-2,-.6) node {$\color{black}\vec{X}$};
    \draw (2,-.6) node {$\color{black}\vec{Y}$};
    \draw (4,0) node {$\longleftrightarrow$};

    \draw (6,0) node {$\identityOne[\color{black}\vec{X},\color{black}\vec{Y}]$};
\end{tikzpicture}\end{equation}
\end{center}

The two dualities schematically depicted in \eqref{n=2_planar_fusion_to_identity} and \eqref{n=2_chiral_fusion_to_identity} can be interpreted as the field theory equivalent of the multiplication between two operators $\mathcal{S}$ 
and $\mathcal{S}^{-1}$ giving an identity as a result. 

\subsection{Basic Dualities moves}\label{subsec: newBasic Duality Moves}

The two dualities schematically depicted in \eqref{n=2_planar_fusion_to_identity} and \eqref{n=2_chiral_fusion_to_identity} can be interpreted as the field theory equivalent of the multiplication between two operators $\mathcal{S}$ 
and $\mathcal{S}^{-1}$ giving an identity as a result. This statement is analogous to that for the $\mathcal{S}_\mathcal{N}=4$ theory. To extend further the idea that the $\mathcal{S}$-wall theory is an oporator we need to define on which object it acts and how. The objects will be $\mathcal{N}=2$ QFT blocks, that are simple WZ models that can be glued together to form generic $\mathcal{N}=2$ theories. As we will show in a moment, it is possible to define an action of the $\mathcal{S}$-wall theory on QFT-blocks. We refer to these identities as \textit{basic duality moves}.

To derive the $\mathcal{N}=2$ duality moves we start from the $\mathcal{N}=4$ basic moves which we repeat below for convenience in their Lagrangian form:
\begin{equation}\label{fig:n4_basicmoves_lag}
    \includegraphics[width=\textwidth]{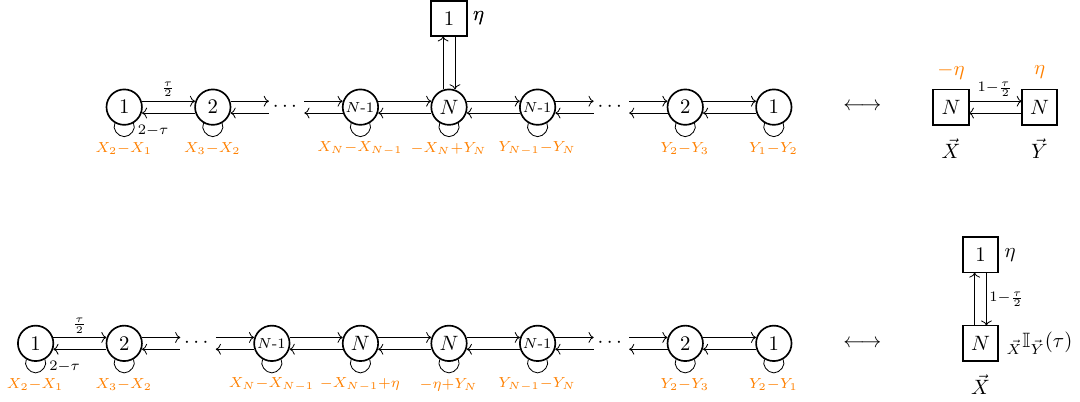}
\end{equation}

\subsubsection{Chiral Bifundamental $\mathcal{S}$-dualization}
\begin{figure}
    \centering
    \includegraphics[width=1\textwidth]{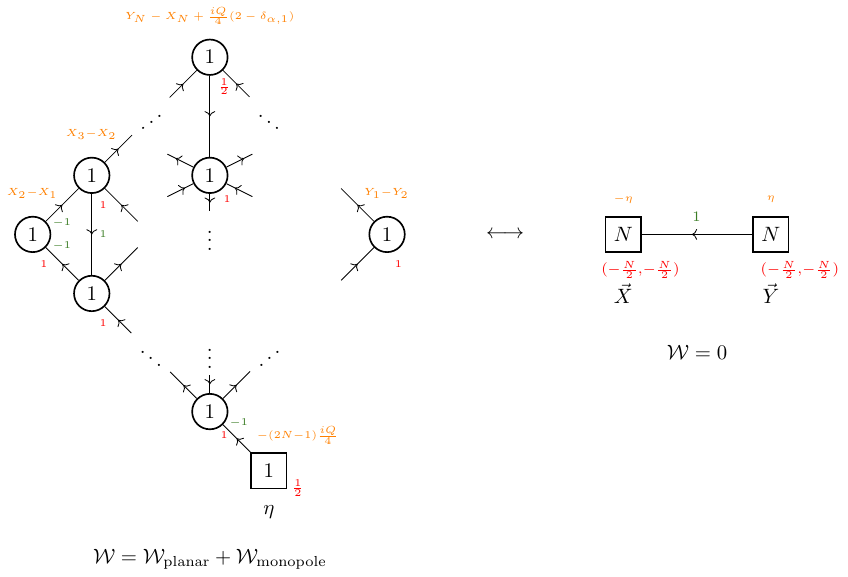}
    \caption{A chiral bifundamental on the r.h.s.~is $\mathcal{S}$-dualized into planar flavor sandwiched between two $\mathcal{S}$-walls. As usual, on the planar side the FIs in orange are those of all the nodes in the corresponding column below. The term $\delta_{\alpha,1}$ indicates that there is a shift valid only for the $1$-st node in the middle column, counting from below.}
    \label{quiv: planar_braid_n=2}
\end{figure}

Our starting point is the 
$\mathcal{N}=4$ basic move in the first line of \eqref{fig:n4_basicmoves_lag}.
We perform the real mass deformations specified by the following shifts of fugacities, on both sides:
\begin{equation}
    \begin{cases}
    \vec{X}\to \vec{X} + N \frac{\tau}{2} \\
    \vec{Y} \to \vec{Y} + (N+1) \frac{\tau}{2} \\
    u^{(I)}_{\alpha}\to u^{(I)}_{\alpha}+\frac{\tau}{2}(2\alpha-I-1), \quad &\text{for}\; I=1, \dots, N \\
    u^{(I)}_{\alpha}\to u^{(I)}_{\alpha}+\frac{\tau}{2}(2\alpha+I-2N-1), \quad &\text{for}\; I=N+1, \dots, 2N-1 \\
    \eta \to \eta - \frac{N}{2}\tau
    \end{cases}
\end{equation}
where $\vec{X}$ and $\vec{Y}$ are fugacities of the two $U(N)$ global symmetries
(which are preserved by the deformations),
$\vec{u}^{(I)}$ are gauge fugacities for the $I$-th gauge node in the quiver (where we start to count from the left), and $\eta$ is the fugacity for the  flavor. The flow generated by these deformations yields the duality in Figure \ref{quiv: planar_braid_n=2}. We interpret this duality as a \textit{basic move} relating a \textit{chiral} bifundamental to a \textit{planar} flavor  sandwiched between two $\mathcal{S}$-walls.

We define the \textit{left-pointing} chiral bifundamental block with trial R-charge $r$ as:
\begin{equation}
    \includegraphics[]{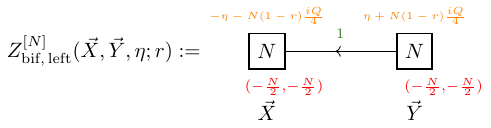}
\end{equation}
whose $\mathbf{S}_b^3$ partition function is given by:
\begin{equation}    
\begin{split}
    Z^{[N]}_{\text{bif},\, \text{left}}(\vec{X},\vec{Y},\eta;r) \; := \; &
    e^{2\pi i \sum_j X_j (-\eta - N(1-r) iQ)}\;
    e^{2\pi i \sum_j Y_j (\eta + N(1-r) iQ)} 
    \\ &
    e^{\frac{N\pi i}{2}\sum_j(X_j^2+Y_j^2) - \pi i \sum_{j,k}X_jY_k} \;
    \prod_{j,k=1}^N \;
    s_b\bigg(\frac{iQ}{2}(1 -r) + X_k-Y_j\bigg) \,.
\end{split}
\end{equation}

Then on the r.h.s.~of Figure \ref{quiv: planar_braid_n=2} we identify the chiral bifundamental block\footnote{Recall that in this Section we assign trial R-charge 0 to the diagonal bifundamentals of the $\mathcal{S}$-wall theory, which in turn fixed the trial R-charge of the bifundamental on the r.h.s.~of Figure \ref{quiv: planar_braid_n=2} to 1.} $Z^{(N)}_{\text{bif},\, \text{left}}(\vec{X},\vec{Y},\eta;1)$, and the duality in Figure \ref{quiv: planar_braid_n=2} corresponds to the following identity between partition functions:

\begin{align}    \label{eq:Z_bif_N_left}
         Z_{\text{bif},\, \text{left}}^{[N]}(\vec{X},\vec{Y},\eta;1) = &
         \int d\vec{Z} \DeltaOne[\vec{Z}] e^{-\frac{i \pi}{2} (\eta^2-2\eta Z_1+Z_1^2+iQ[(2N-1) \eta-Z_1-2\sum_{j=2}^{N}Z_j])} \nonumber \\
         & 
         Z_{\mathcal{S}}(\vec{X},\vec{Z}) s_b\left(\frac{iQ}{2}+Z_1-\eta \right) Z_{\mathcal{S}^-}(\vec{Y},\vec{Z}) \,.
\end{align}
On the r.h.s.~the two $\mathcal{S}$-walls are glued as in the case of the planar fusion to identity in \eqref{n=2_planar_fusion_to_identity} except for the CS level of the bottom node being $1$ instead of $1/2$ and different FI terms.

We write the move depicted in Figure \ref{quiv: planar_braid_n=2} in compact form as:
\begin{equation} \label{quiv: planar_n=2}
    \includegraphics[]{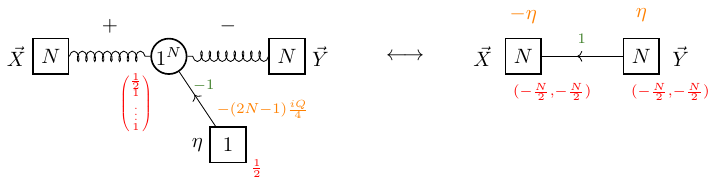}
\end{equation}
Where in the figure above we specify the shift of the CS levels of the $N$ $U(1)$ nodes in the middle column by a red column vector. 

Similarly, we can define a right-pointing bifundamental block, whose $\mathcal{S}^3_b$ partition function is given by:
\begin{equation}
Z^{[N]}_{\text{bif},\, \text{right}}(\vec{X},\vec{Y},\eta;r):=
Z^{[N]}_{\text{bif},\, \text{left}}(-\vec{X},-\vec{Y},-\eta;r) \,.
\end{equation}
Also, an equivalent definition is:
\begin{equation}
Z^{[N]}_{\text{bif},\, \text{right}}(\vec{X},\vec{Y},\eta;r):=
Z^{[N]}_{\text{bif},\, \text{left}}(\vec{Y},\vec{X},-\eta;r) \,.
\end{equation}
whose basic move in short notation is: 
\begin{equation} \label{quiv: planar_n=2}
    \includegraphics[]{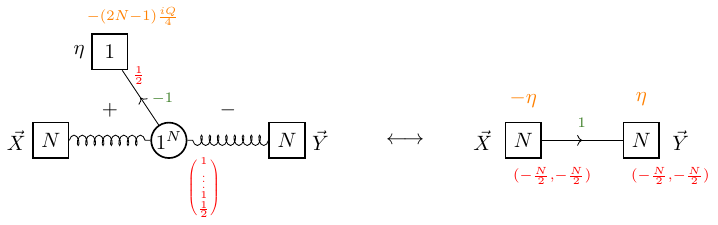}
\end{equation}
Where in this notation it is meant that the fundamental chiral is connected to the topmost node of the cental column of gauge nodes, instead of the bottom one as in Figure \ref{quiv: planar_braid_n=2}.
The corresponding partition function identity is:
\begin{align}    \label{eq:Z_bif_N_right}
         Z_{\text{bif},\, \text{right}}^{[N]}(\vec{X},\vec{Y},\eta;1) = & \int d\vec{Z} \DeltaOne[\vec{Z}] e^{-\frac{i \pi}{2} (\eta^2-2\eta Z_N+Z_N^2+iQ[-(2N-1) \eta+Z_1+2\sum_{j=2}^{N}Z_j])} \nonumber \\ 
         & Z_{\mathcal{S}}(\vec{X},\vec{Z}) s_b\left(\frac{iQ}{2}-Z_N+\eta \right) Z_{\mathcal{S}^-}(\vec{Y},\vec{Z}) 
\end{align}

\subsubsection{Planar Bifundamental $\mathcal{S}$-dualization}

\begin{figure}
    \centering
    \includegraphics[]{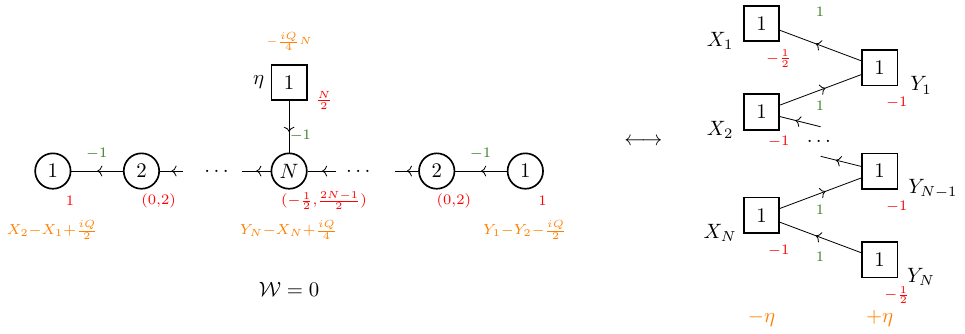}
    \caption{A planar bifundamental on the r.h.s.~is $\mathcal{S}$-dualized into a flavor sandwiched between two $\mathcal{S}$-walls. As usual, on the planar side the background FIs in orange are those of all the nodes in the column below.}
    \label{quiv: chiral_braid_n=2}
\end{figure}

Starting again from the first
$\mathcal{N}=4$ basic move  in \eqref{fig:n4_basicmoves_lag},
we perform  the real mass deformation
specified by the following shift of fugacities:
\begin{equation}
    \begin{cases}
    X_j \to X_j+(2j-N)\frac{\tau}{2}\\
    Y_j \to Y_j+\frac{\tau}{2}(2j-N+1)\\
    \vec{u}^{(I)}\to \vec{u}^{(I)}-\frac{\tau}{2}I, \quad \text{for} \; I=1,\dots, 2N-1\\
    \eta \to \eta - \frac{N+1}{2}\tau\,.
    \end{cases}
\end{equation}
The flow generated by these mass deformations
break the emergent $U(N)_{X,Y}$ symmetries and yelds the duality depicted in Figure \ref{quiv: chiral_braid_n=2}.
We define the left-pointing planar bifundamental block with trial R-charge $r$ as:
\begin{equation}
    \includegraphics[]{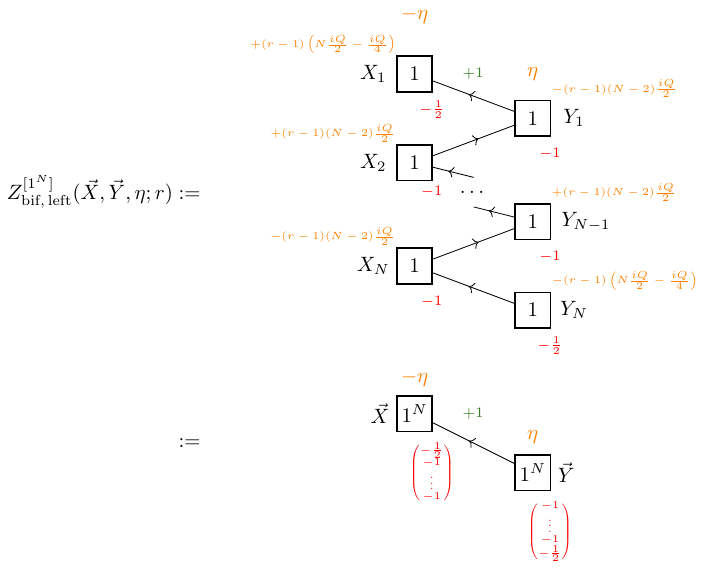}
\end{equation}
In the second line we introduced a short notation wherein we suppressed the specific shift in the background FI of each single node, leaving only the common $\eta$ dependence manifest. Also the background mixed-CS interaction in green is meant to be present for each bifundamental linking two $U(1)$ symmetries while the background CS levels are given as a colum in red.
The contribution to the $\mathbf{S}^3_b$ partition function is given below:
\begin{equation}
    \begin{aligned}
        Z^{[1^N]}_{\text{bif},\, \text{left}}(\vec{X},\vec{Y},\eta;r) &:= e^{\frac{\pi i}{2}(X_1^2+Y_N^2)}\;e^{\pi i (\sum_{j=2}^NX_j^2 + \sum_{j=1}^{N-1}Y_j^2)}\;e^{2\pi i \eta
        \sum_j(Y_j-X_j)}\;e^{-\pi i \sum_{j}Y_j(X_j+X_{j+1})}\\ & 
        e^{2\pi i \sum_J X_J (r-1) (N+2-2J) \frac{iQ}{2} }
        e^{-2\pi i \sum_J Y_J (r-1) (N-2J) \frac{iQ}{2} }
        e^{2\pi i (Y_N \frac{iQ}{4} - X_1 \frac{iQ}{4})}
        \\ &
        \prod_{j=1}^N\;
        s_b\bigg(\frac{iQ}{2} - r\frac{iQ}{2} +X_{j}-Y_{j}\bigg)\;
        \prod_{j=1}^{N-1}\;
        s_b\bigg(\frac{iQ}{2} - r\frac{iQ}{2} + Y_{j}-X_{j+1}\bigg) \,.
    \end{aligned}
\end{equation}
On the r.h.s.~of Figure \ref{quiv: chiral_braid_n=2} we recognize the left-pointing chiral bifundamental block\footnote{Recall that in this Section we assign trial R-charge 0 to the bifundamentals of the $\mathcal{S}$-wall theory, which in turn fixed the trial R-charge of the bifundamental on the r.h.s.~of Figure \ref{quiv: chiral_braid_n=2} to 1.}  $Z^{[1^N]}_{\text{bif},\, \text{left}}(\vec{X},\vec{Y},\eta;1)$.

The $\mathbf{S}^3_b$ partition function identity corrresponding to the duality in Figure \ref{quiv: chiral_braid_n=2} is:
\begin{align} \label{eq:Z_bif_1N_left}
    Z_{\text{bif},\, \text{left}}^{[1^N]}(\vec{X},\vec{Y},\eta;1) = & \int d\vec{Z} \DeltaN[\vec{Z}]e^{-\frac{i \pi}{2}[N\eta^2+iQ(N\eta-\sum_{j=1}^NZ_j)+           \sum_{j=1}^N(Z_j^2-2\eta Z_j)]} \nonumber \\
    & Z_{\mathcal{S}}(\vec{Z},\vec{X}) \prod_{j=1}^N s_b\left(\frac{iQ}{2}+Z_j-\eta \right) Z_{\mathcal{S}^-}(\vec{Z},\vec{Y}) \,.
\end{align}
On the r.h.s.~the two $\mathcal{S}$-walls are glued as in the case of the chiral fusion to identity in \eqref{eq:manchirfus}. There is only a slight modification for the CS level of the middle $U(N)$ node to which the flavor is attached, which has now CS level $(-1/2, (2N-1)/2)$ and a shifted FI.

We can write the move in Figure \ref{quiv: chiral_braid_n=2} in compact form as:
\begin{equation} \label{quiv: chiral_n=2}
    \includegraphics[]{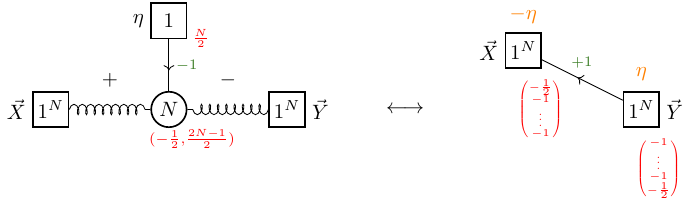}
\end{equation}

Similarly, we can define a right-pointing planar bifundamental block, whose $\mathcal{S}^3_b$ partition function is given by:
\begin{equation}
    Z^{[1^N]}_{\text{bif},\, \text{right}}(\vec{X},\vec{Y},\eta;r)
    :=
    Z^{[1^N]}_{\text{bif},\, \text{left}}(\vec{Y},\vec{X},-\eta;r) \,
\end{equation}
and whose basic move in short notation is: 
\begin{equation}
    \includegraphics[]{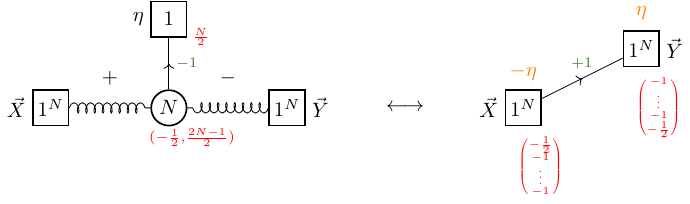}
\end{equation}
Notice that in short form the right-pointing planar bifundamental has opposite slope.
The corresponding partition function identity is:
\begin{align}    \label{eq:Z_bif_1N_right}
    Z_{\text{bif},\, \text{right}}^{[1^N]}(\vec{X},\vec{Y},\eta;1) = &
    \int d\vec{Z} \DeltaN[\vec{Z}] e^{-\frac{i \pi}{2}[N\eta^2+iQ(-N\eta+\sum_{j=1}^NZ_j)+           \sum_{j=1}^N(Z_j^2-2\eta Z_j)]} \nonumber \\
    & Z_{\mathcal{S}}(\vec{Z},\vec{X}) \prod_{j=1}^N s_b\left(\frac{iQ}{2}-Z_j+\eta \right) Z_{\mathcal{S}^-}(\vec{Z},\vec{Y}) \,.
\end{align}

\subsubsection{Chiral fundamental $\mathcal{S}$-dualization}
We now discuss the \textit{inverse} basic moves.  A possible way to obatain them is by ``multiplying" an $\mathcal{S}$ and $\mathcal{S}^{-1}$-wall respectively on the right and left sides of the basic moves in Figures \ref{quiv: planar_braid_n=2} and \ref{quiv: chiral_braid_n=2} and using the appropriate \textit{fusion to identity} properties.\\
However, we will follow a different route which is to start back from the inverse $\mathcal{N}=4$ basic move in the second line of \eqref{fig:n4_basicmoves_lag} and perform a suitable mass deformation.

\begin{figure}[ht]
    \centering
    \includegraphics[]{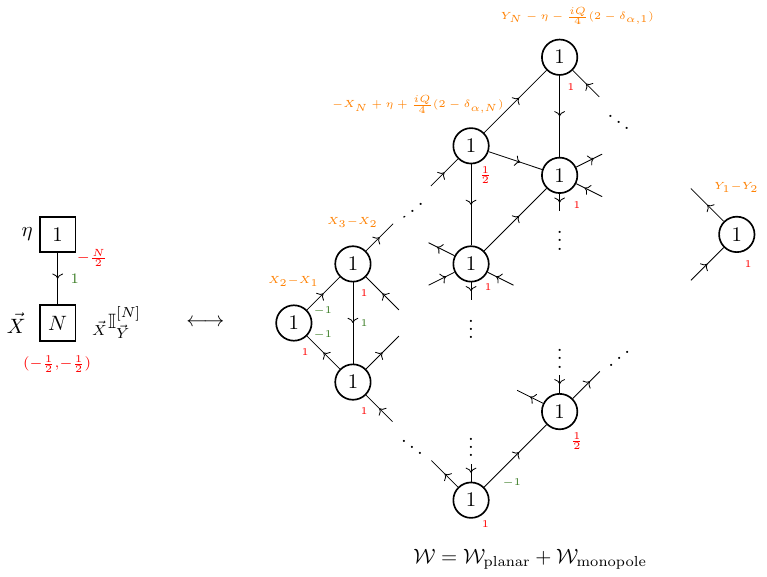}
    \caption{The planar mirror dual of the chiral anti-fundamental flavor block is shown here. Mixed CS interactions have been suppressed for brevity. Notice that w.r.t.~the fusion to identity property in Figure \ref{eq:manplanfus}, altough being very similar, in the basic move the planar quiver has an extra column of height $N$ in the middle. As usual, on the planar side we give the FI parameter of all the nodes in a colum on top of it. The term $\delta_{\alpha,i}$ implies a shift in the FI of the $i$-th gauge node in the column, counting from below.}
\label{pquiv:invbasmov_n=2}
\end{figure}

Let us start from the basic move for a chiral fundamental. Starting from \eqref{fig:n4_basicmoves_lag} we take the real mass deformation
\begin{equation}
    \begin{cases}
    X_j,Y_j \to X_j,Y_j \\
    u_{\alpha}^{(I)}\to u_{\alpha}^{(I)}+\frac{\tau}{2}(2\alpha-I), \quad \text{for}\; I=1,\dots, N\\
     u_{\alpha}^{(I)}\to u_{\alpha}^{(I)}+\frac{\tau}{2}(2\alpha+I-2N-2), \quad \text{for}\; I=N+1,\dots, 2N\\
    \eta \to \eta + \frac{\tau}{2}
    \end{cases}
\end{equation}
where again $u_{\alpha}^{(m)}$ represents the set of parameters for the $m$-th gauge node in the quiver (starting from the left), $X_j,Y_j$ are the parameters for the global $U(N)$ symmetries (unbroken by the deformation) and $\eta$ is the FI parameter. The resulting basic move is shown in Figure \ref{pquiv:invbasmov_n=2}. 

\noindent On the l.h.s.~we identify the \textbf{chiral Flavor Block} consisting  of a single chiral in the anti-fundamental of $U(N)$, equipped with a chiral identity wall. Together with this block we can also define an analogous fundamental chiral flavor block. The two blocks can be defined graphycally as:
\begin{equation}\label{eq:chiral_flav}
    \includegraphics[]{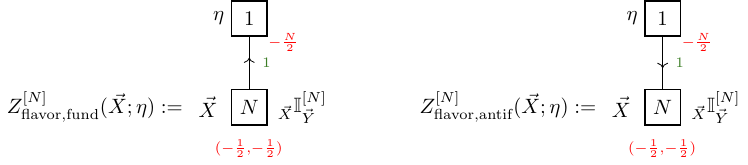}
\end{equation}

The contribution to the $\mathbf{S}^3_b$ partition function is given below\footnote{The CS contact term indicated for the global $U(N)$ symmetry in the $\mathbf{S}^3_b$ partition function has been calculated by including the contribution from the identity wall as well.}:
\begin{equation}    \label{eq:Z_flav_N_+}
        Z^{[N]}_{\text{flavor},\text{fund}}(\vec{X},\eta)=  e^{\frac{i \pi}{2} \left[N\eta^2 + \sum_{j=1}^N(X_j^2-2\eta X_j)\right] } \; \prod_{j=1}^N s_b\left( \eta - X_j \right) \identityN[\vec{X},\vec{Y}]
\end{equation}
\begin{equation}    \label{eq:Z_flav_N_+}
        Z^{[N]}_{\text{flavor},\text{antif}}(\vec{X},\eta)=  e^{\frac{i \pi}{2} \left[N\eta^2 + \sum_{j=1}^N(X_j^2-2\eta X_j)\right] } \; \prod_{j=1}^N s_b\left( -\eta + X_j \right) \identityN[\vec{X},\vec{Y}]
\end{equation}
We emphasize the CS level for the $U(N)_{\vec{X}}$ symmetry in \eqref{eq:chiral_flav} refers to the contribution coming from the single chiral that has been integrated out. The overall background CS level for this symmetry has to take into account also the shift $(-N,0)$ encoded in the identity wall, as previously mentioned so that the total CS level for the chiral block is $(-N-\frac{1}{2},-\frac{1}{2})$.

In the duality in Figure \ref{pquiv:invbasmov_n=2} we recognize a chiral flavor block on the l.h.s.~and the corresponding $\mathbf{S}^3_b$  partition function identity is:
\begin{equation}\label{eq:Z_flav_N_+}
   Z_{\text{flavor},\text{fund}}^{[N]}(\vec{X},\eta)= \int d \vec{Z}\; d \vec{W} \; Z_{\mathcal{S}}(\vec{X},\vec{Z}) \; Z^{[1^N]}_{\text{bif},\, \text{right}}(\vec{Z},\vec{W},-\eta;0)\; Z_{\mathcal{S}^{-}}(\vec{Y},\vec{W}) \,.
\end{equation}
Where on the r.h.s.~we recognize a planar bifundamental sandwiched between two $\mathcal{S}$-walls.
In short notation the duality in Figure \ref{pquiv:invbasmov_n=2} can be depicted as:
\begin{equation}\label{quiv: flavor_n=2_planar}
    \includegraphics[]{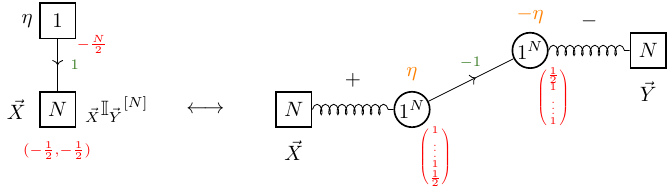}
\end{equation}

There is also a variant of this move,
corresponding to the $\mathcal{S}$-dualization of a chiral anti-fundamental. 
\begin{equation}\label{quiv: flavor_n=2_planar}
    \includegraphics[]{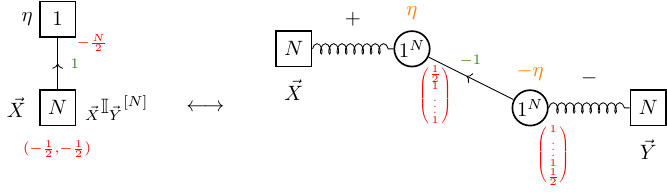}
\end{equation}
Which corresponds to the $\mathbf{S}^3_b$  partition function identity:
\begin{equation}    \label{eq:Z_flav_N_-}
   Z_{\text{flavor},\text{antif}}^{[N]}(\vec{X},\eta)= \int d \vec{Z}\; d \vec{W} \; Z_{\mathcal{S}}(\vec{X},\vec{Z}) \; Z^{[1^N]}_{\text{bif},\, \text{left}}(\vec{Z},\vec{W},-\eta;0)\; Z_{\mathcal{S}^{-}}(\vec{Y},\vec{W}) \,.
\end{equation}

\subsubsection{Planar fundamental $\mathcal{S}$-dualization}

\begin{figure} 
    \centering
    \includegraphics[width=\textwidth]{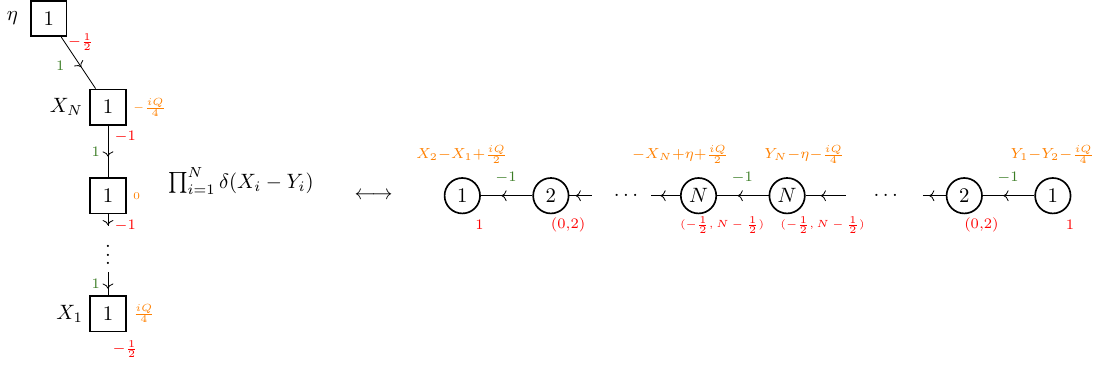}
    \caption{The planar mirror dual of the chiral anti-fundamental flavor block is shown here.}
    \label{quiv:manchirS}
\end{figure}

Starting again from the 
$\mathcal{N}=4$ basic move in the second line of \eqref{fig:n4_basicmoves_lag}, we perform the mass deformation defined by the following deformation breaking the $U(N)_{X,Y}$ symmetries:
\begin{equation}
    \begin{cases}
    X_j \to X_j - (2N-2j+1)\frac{\tau}{2} \\
    Y_j \to Y_j- (2N-2j+1)\frac{\tau}{2} \\
    \vec{u}^{(I)}\to \vec{u}^{(I)}+(2N-I+1)\frac{\tau}{2}, \quad \text{for}\; I=1, \dots, 2N \\
    \eta \to \eta 
    \end{cases}
\end{equation}
This limit yelds the duality in Figure \ref{quiv:manchirS}.
On the l.h.s.~we identify the anti-fundamental \text{Planar flavor block} for which we introduce the compact notation:
\begin{equation}
    \includegraphics[]{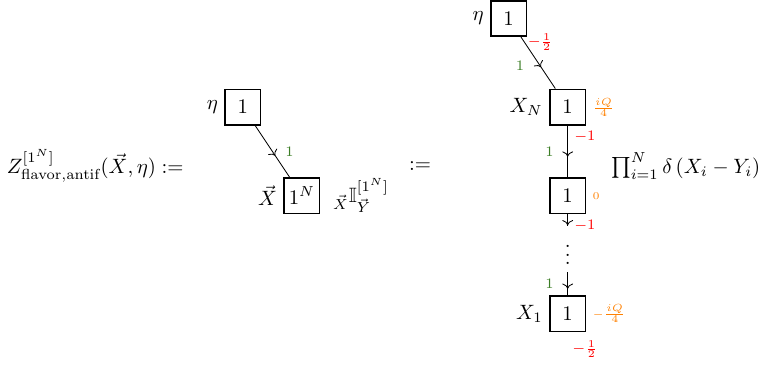}
\end{equation}
Notice that all the arrows point downward.
Similarly we define the planar fundamental block as:
\begin{equation}
    \includegraphics[]{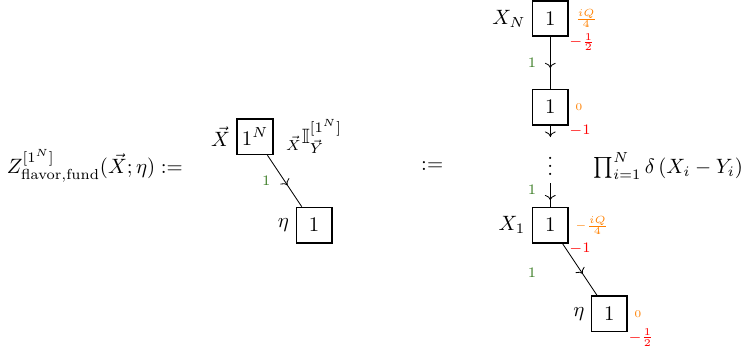}
\end{equation}
Notice that for the fundamental planar block the chiral is attached to the bottom of the string of chirals belonging to the Identity-wall.
The contribution to the $\mathbf{S}^3_b$ partition function is given below:
\begin{equation}
Z_{\text{flavor},\text{fund}}^{[1^N]}(\vec{X};\eta)=
e^{\frac{i \pi}{2} (\eta^2-2\eta X_1+X_1^2)} \;
s_b\left(\eta-X_1 \right )\identityOne[\vec{X},\vec{Y}] 
\end{equation}

\begin{equation}
Z_{\text{flavor},\text{antif}}^{[1^N]}(\vec{X};\eta)=
e^{\frac{i \pi}{2} (\eta^2-2\eta X_N+X_N^2)} \;
s_b\left(X_N-\eta\right )\identityOne[\vec{X},\vec{Y}] 
\end{equation}

The duality in Figure \ref{quiv:manchirS} corresponds to the $\mathbf{S}^3_b$ partition function identity:
\begin{equation}    \label{eq:Z_flav_1N_-}
    Z_{\text{flavor},\text{antif}}^{[1^N]}(\vec{X};\eta)= \int d \vec{Z}\; d \vec{W} \; Z_{\mathcal{S}}[\vec{Z};\vec{X}] \; Z^{[N]}_{\text{bif},\, \text{left}}(\vec{Z};\vec{W};-\eta;0)\; Z_{\mathcal{S}^{-}}[\vec{W};\vec{Y}]
\end{equation}
We represent the previous move  in compact form as:
\begin{equation}\label{plfvSd}
    \includegraphics[]{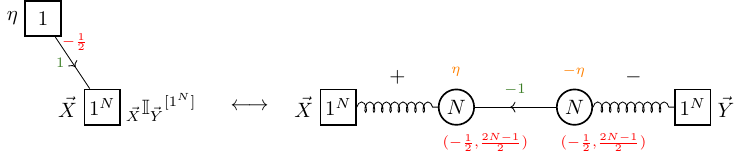}
\end{equation}

There is also a variant of this move that is the $\mathcal{S}$-dual of a planar fundamental block.
\begin{equation}
    \includegraphics[]{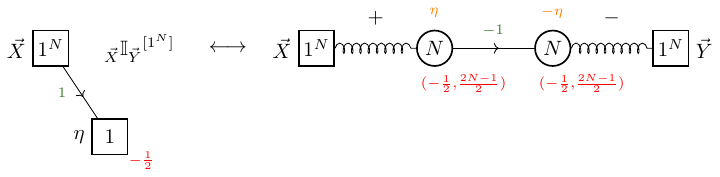}
\end{equation}
Which corresponds to the following identity between partition functions:
\begin{equation}    \label{eq:Z_flav_1N_+}
    Z_{\text{flavor},\text{fund}}^{[1^N]}(\vec{X};\eta)= \int d \vec{Z}\; d \vec{W} \; Z_{\mathcal{S}}[\vec{Z};\vec{X}] \; Z^{[N]}_{\text{bif},\, \text{right}}(\vec{Z};\vec{W};-\eta;0)\; Z_{\mathcal{S}^{-}}[\vec{W};\vec{Y}]\,.
\end{equation}

\section{The Chiral-Planar dualization Algorithm at work}\label{sec: Examples}
In this section we use the algorithm introduced in the previous section to produce new examples of the chiral-planar $\mathcal{N}=2$ mirror dualities. 

Along the lines of \cite{Comi:2022aqo}, we define the steps of the algorithmic dualization of a non-abelian chiral $\mathcal{N}=2$ quiver to the be the following:
\begin{itemize}
    \item \textbf{Step 1:} The quiver is chopped into its basic QFT blocks that can be either bifundamentals or (anti-)fundamental chiral fields. This step is performed by freezing the gauge interactions that will be reintroduced later.
    
    \item \textbf{Step 2:} Each QFT block is dualized via the basic duality moves described in Section \ref{subsec: newBasic Duality Moves}.
    These identities, or \textit{basic moves} are summarized for reference in Table \ref{tab:basicmoves}. 
    
    \item \textbf{Step 3:} The dualized blocks are glued back together by turning back on the gauge interactions frozen at the first step. The pairs $\mathcal{S}^{-1}\mathcal{S}$ fuse to
    Identity-walls  as described in Section \ref{subsec:N=2 S-wall}.
\end{itemize}
Each of these steps will be illustrated in more detail in the following sections with explicit examples. \\

We will use the algorithm to dualize chiral $\mathcal{N}=2$ quivers into abelian-planar quivers. Although the same algorithm can be applied in reverse — starting from a planar abelian quiver to obtain its chiral dual — we do not discuss this case in the present work.\\

\begin{table}
\centering
\begin{tabular}{m{3cm}|m{5cm}|m{3cm}}
    QFT block & $\mathcal{S}$-dualization & $\mathbf{S}^3_b$ p.f.~identity \\
    \hline & & \\ 
    \simpBifLNN[N] & \simpBifLNNMirror[N] & Fig.~\ref{quiv: planar_braid_n=2}, eq.~\eqref{eq:Z_bif_N_left}  \\
    \hline & & \\
    \simpBifRNN[N] & \simpBifRNNMirror[N] & Eq.~\eqref{eq:Z_bif_N_right} \\
    \hline & & \\
    \simpAFund[N] & \simpAFundNMirror[N] & Fig.~\ref{pquiv:invbasmov_n=2}, eq.\eqref{eq:Z_flav_N_+} \\
    \hline & & \\
    \simpFund[N] & \simpFundNMirror[N] & Eq.~\eqref{eq:Z_flav_N_-}  \\

\end{tabular}
\caption{Summary of the basic moves that are relevant to run the algorithm. In the first column are given the basic QFT blocks, in the second column their $\mathcal{S}$-duals. In the last column we give the reference to the $\mathbf{S}^3_b$ partition function identity and  to the extended quiver form. We avoided to give all the CS-interactions and FI terms, we only provided those that are dynamical either in the starting chiral theory or in the reulting planar theory. Notice that in order to keep track of all the details more information are needed, see Section \ref{sec: S_walls,QFT_blocks_Duality_Moves} for a complete discussion.}
\label{tab:basicmoves}
\end{table}

The current version of the algorithm has some limitations that we list below.
\begin{itemize}
    \item We can dualize chiral quivers with fixed CS levels. 
    A $U(N)$ node with $F$
    chirals will have CS level $(k,k+lN)$ with $k=-\tfrac{F}{2}+N$  and $l=-1$. Moreover, each pair of gauge nodes connected by a bifundamental chiral will be also coupled via a mixed-CS interaction. 

    \item We can dualize only chiral quivers with constant ranks $U(N)$. To algorithmically dualize quivers with non-constant ranks we would need to introduce asymmetric blocks and the chiral-planar version of the Hanany-Witten duality moves derived in \cite{Comi:2022aqo}.

    \item We also limit ourselves to cases where each gauge node sees 
    $[n_f,n_a]$  chirals, where $F=n_f+n_a$
    with
    \begin{equation}\label{eq:SQCD_cases}
    [n_f,n_a] = 
        \begin{cases}
        [N+F_1,\;N+F_2]\\
        [2N+F_1,\;F_2]\\
        [F_1,\;2N+F_2]\\
        \end{cases}
    \qquad F_1, F_2 \geq 0
    \end{equation}
\end{itemize}

The algorithm can be further extended to overcome the current limitations, but we leave this for future work.

Despite these limitations, we can extend the algorithm's reach by applying the strategies outlined in Sections \ref{subsec: SQCD_example} and \ref{sec: FTUN_mirror}, as we will demonstrate in Section \ref{furtherexamples}.

Additionally, starting from the dualities generated by the algorithm, one can overcome the limitations by performing suitable massive deformations; we leave a detailed study of this direction for future work.

It is important to emphasize that, even within its current scope, the algorithm already produces a rich variety of examples that capture essential features of the new type of mirror duality proposed in this paper. These will be discussed in detail in the following sections.

\subsection{Chern-Simons SQCD with fundamental and anti-fundamental Matter}\label{subsec: cssqcd_eg}

As a first example of the application of the algorithm, we consider the case of  SQCD with $[n_f,n_a]$ flavors, where $F=n_f+n_a$
satisfying the constraint in eq.~\eqref{eq:SQCD_cases}.
The second and third cases are analogous, therefore we explicitly present only the first and second case in eq.~\eqref{eq:SQCD_cases}.

\subsubsection{$U(N)$ CS-SQCD with $[F_1+N, F_2+N]$ Chiral Multiplets}\label{subsec: UN_F1+N_F2+N}
We start from the first case in \eqref{eq:SQCD_cases}, therefore we consider
$U(N)_{(-\frac{F_1+F_2}{2}, -\frac{F_1+F_2}{2}-N)}$ SQCD with $[F_1+N, F_2+N]$ chiral multiplets, depicted in short as:
\begin{equation}  \label{eq:SQCD_N_N}
    \includegraphics[]{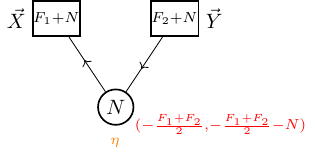}
\end{equation}
To apply our algorithm we begin
by chopping the theory into chiral QFT blocks
using the basic ingredients defined in Section \ref{sec: S_walls,QFT_blocks_Duality_Moves}.
We do so by considering $F_1$ chiral fundamental blocks, $F_2$ chiral anti-fundamental blocks and  two chiral bifundamental blocks,
in complete analogy 
with the  $\mathcal{N}=4$ SQCD case done in Figure \ref{fig:N=4Algorithm}, where the two bifundamentals account effetively for $2N$ flavors. 
The QFT block decomposition is schematically depicted below.
\begin{equation}\label{eq:SQCD1_decomp}
    \includegraphics[width=\textwidth]{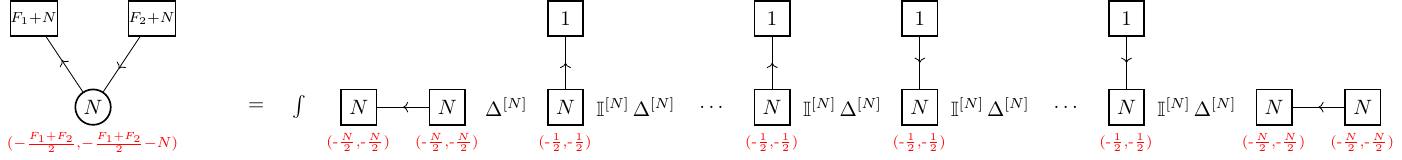}
\end{equation}
In this schematic picture the act of multiplication between two consecutive blocks is understood. The measure of integration between two blocks is given by
a $U(N)$ vector field
$\Delta^{[N]}$ which is precisely the measure defined in \eqref{eq:DeltaN}. To avoid cluttering we suppressed the labeling of global symmetry parameters.

Let's first check that gluing back the QFT blocks we recover the 
$U(N)_{(-\frac{F_1+F_2}{2}, -\frac{F_1+F_2}{2}-N)}$ SQCD.
The gluing  procedure consist in identifying and gauging two $U(N)$ global symmetries, one coming from each of the two blocks glued together. This means that there must be a total of $(\#\text{QFT blocks}-1)$ integrations to account for the gluing. Taking into account the presence of $F_1+F_2$ Identity-walls coming from flavor blocks, each freezing an integration, we are left with  a single dynamical, i.e.~non-frozen, $U(N)$ gauge symmetry.

We recall that each QFT block, Identity-wall and measure of integration comes together with CS interactions. In the schematic figure we give explictly only those coming from the QFT-blocks and leave the remaning implicit. The resulting CS level upon gauging and implementing the Identity-walls can be computed as follows. Each factor of $\Delta^{[N]}$ contributes as $(N,0)$ to the CS level. Each flavor block consists of a (anti-)fundamental chiral field, with a CS level $(-\tfrac{1}{2},-\tfrac{1}{2})$, and an identity, contributing as $(N,0)$, for a total contribution of $(-N-\tfrac{1}{2},-\tfrac{1}{2})$. Lastly, a bifundamental chiral contributes with a CS level of $(-\tfrac{N}{2},-\tfrac{N}{2})$.
Therefore the total CS level is:
\begin{equation}
\resizebox{.95\hsize}{!}{ $
    \underbrace{{2\color{CScolor}(-\frac{N}{2},-\frac{N}{2})}}_{\text{bifundamental blocks}} +
    \underbrace{(F_1+F_2){\color{CScolor}(-N-\frac{1}{2},-\frac{1}{2})}}_{\text{flavor blocks}} +
    \underbrace{{(F_1+F_2+1)\color{CScolor}(N,0)}}_{\text{chiral measure}} = {\color{CScolor}(-\frac{F_1+F_2}{2},-\frac{F_1+F_2}{2}-N)} $}
\end{equation}
which matches the CS level of the SQCD theory we are considering.\\

\begin{figure}
    \centering
    \includegraphics[width=\textwidth]{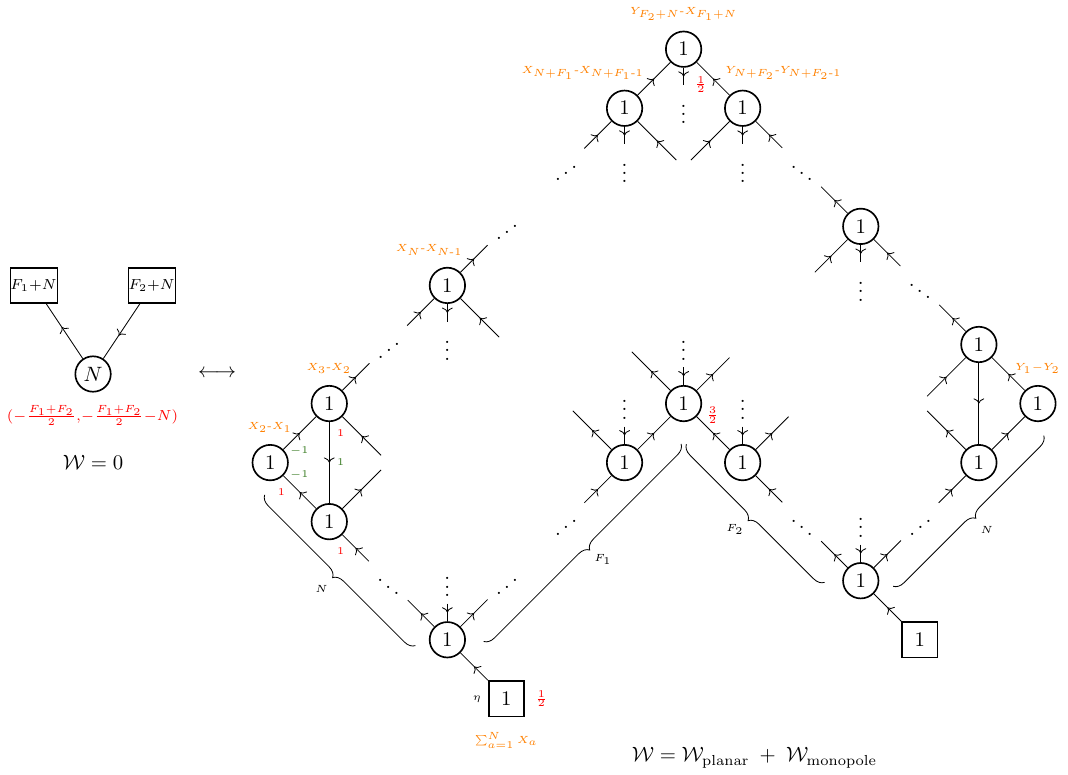}
    \caption{The $\mathcal{N}=2$ planar mirror dual of $U(N)_{(-\frac
    {F_1+F_2}{2}, -\frac{F_1+F_2}{2}-N)}$ SQCD with $[F_1+N, F_2+N]$ chiral multiplets. The superpotential terms and (mixed) Chern-Simons interactions are as described in Section \ref{subsec: SQCD_example}.}
    \label{eq:extended_mirror_SQCD_N_N}
\end{figure}

Now we dualize each QFT block in the decomposition in \eqref{eq:SQCD1_decomp} by exploiting the basic moves derived in Section \ref{sec: S_walls,QFT_blocks_Duality_Moves} and summarized in Table \ref{tab:basicmoves}, we obtain:
\begin{equation}
    \includegraphics[width=\textwidth]{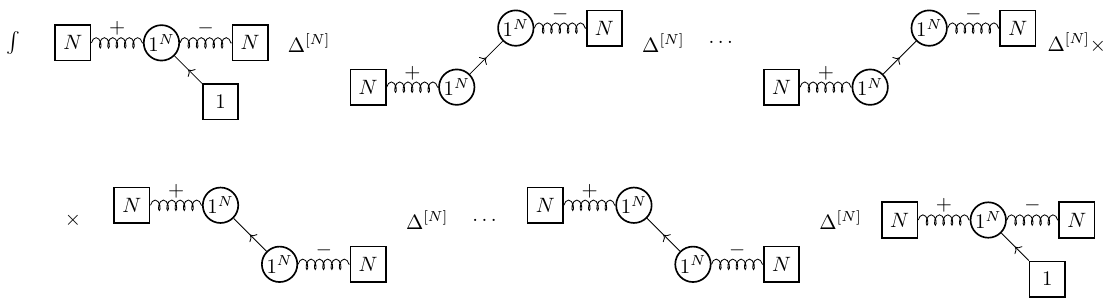}
\end{equation}
Now we can use the chiral fusion to identity \ref{eq:manchirfus} to remove pairs of $\mathcal{S}$-walls glued through their $U(N)$ symmetry. Schematically this operation can be represented as:

\begin{equation}
    \dots \;
	\begin{tikzpicture}[baseline=0]
		\node at (-1.5,0) (f1) [gauge,black] {$1^N$};
		\node at (0,0) (g) [gauge, black] {$N$};
		\node at (1.5,0) (f2) [gauge,black] {$1^N$};
		    \draw[decorate,decoration={coil,segment length=4pt}] (f1)--(g) node[midway,above] {$+$};
			\draw[decorate,decoration={coil,segment length=4pt}] (g)--(f2)node[midway,above] {$-$};
	\end{tikzpicture}
    \; \dots \qquad \rightarrow \qquad
    \dots \;
	\begin{tikzpicture}[baseline=0]
		\node at (0,0) (f1) [gauge,black] {$1^N$};
	\end{tikzpicture}
    \; \dots
\end{equation}
This allows us to remove all but the first and last $\mathcal{S}$-walls.
The resulting generalized quiver in short notation is: 
\begin{equation} \label{eq:mirror_SQCD_N_N}
    \includegraphics[]{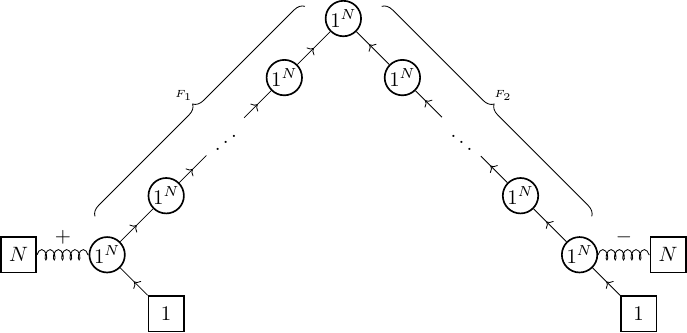}
\end{equation}
The generalized quiver in \eqref{eq:mirror_SQCD_N_N} can finally be written in the explicit Lagrangian form by using the planar UV completion of the $\mathcal{S}$-walls and the definition of the planar bifundamental and fundamental blocks. The result is depicted in Figure \ref{eq:extended_mirror_SQCD_N_N}.

Let us comment some of the details of the duality in Figure \ref{eq:extended_mirror_SQCD_N_N}.

The global symmetry of the SQCD theory is, in general, $S[U(F_1+N)\times U(F_2+N)] \times U(1)$, where the first factor is the flavor symmetry while the second one is the topological symmetry\footnote{For special cases, namely $F_1=F_2=0$ the topological symmetry enhances to $SU(2)$. This can also be understood by considering the Aharony-like dual for the theory \cite{Amariti:2021snj}, which is an abelian gauge theory, and then taking the mirror dual to that theory \cite{Tong:2000ky}. We will comment on this feature later in Section \ref{subsec: symmenhanc} and study other chiral theories exhibiting enhancement of the topological symmetry.}. In the mirror dual the $U(1)$ symmetry is the only flavor symmetry, taking into account gauge transformations and superpotential constraints. 

The $\mathcal{W}_{\text{monopole}}$ superpotential contains a term that is linear in the monopole with $-1$/$+1$ magnetic flux under two nodes connected by a vertical line,
which reduce to a single $U(1)$ the topological symmetry of each column.
So the UV topological symmetry is $U(1)^{F_1+F_2+N-1}$
which enhances in the IR to $S[U(F_1+N)\times U(F_2+N)]$, so that the global symmetry of the mirror quiver matches that of the SQCD.

In the SCQD the chiral ring is generated by mesonic operators constructed from the fundamentals $Q_i$ and antifundamentals $\tilde{Q}_j$
that form a bifundamental representation of $SU(F_1+N) \times SU(F_2+N)$. These are mapped to monopole operators that are gauge invariant in the planar theory.
For example the meson $Q_{F_1+N} \tilde{Q}_{F_2+N}$ is mapped to the monopole with flux $+1$ under the topmost gauge node of the abelian dual. The other mesons are mapped to monopoles with two strings of $+1$ fluxes starting at the topmost node and propagating along the two upper diagonals of the planar quiver.
There are $(F_1+N)(F_2+N)$ such monopoles which reproduce the bifundamental representation of the enhanced $SU(F_1+N)\times SU(F_2+N)$ global symmetry.

Notice that in this case, the planar dual never exhibits a mesonic chiral ring generator. This predicts that, for all values of $F_1$ and $F_2$, there are no dressed gauge-invariant monopoles in the SQCD theory that would parametrize a non-compact moduli space.

\subsubsection{$U(N)$ CS-SQCD with $[F_1+2N, F_2]$ Chiral Multiplets} We now consider the second case in \eqref{eq:SQCD_cases} (which is analogous to the third one), the 
$U(N)_{(-\frac{F_1+F_2}{2}, -\frac{F_1+F_2}{2}-N)}$ SQCD with $[F_1+2N, F_2]$ chiral multiplets:
\begin{equation}  \label{eq:SQCD_N_A}
    \includegraphics[]{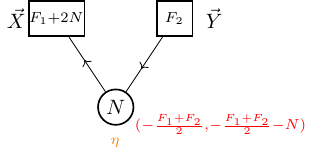}
\end{equation}
We decompose this theory into fundamental blocks according to the rules described in the previous section:
\begin{equation}
    \includegraphics[width=\textwidth]{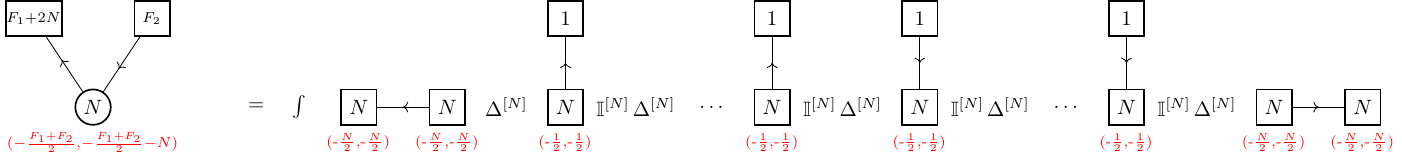}
\end{equation}
where the diagonal gauging of adjacent $U(N)$ fugacities is understood. The difference w.r.t.~\eqref{eq:SQCD1_decomp} is that the last bifundamental is right-pointing, so that it contributes as $N$ extra fundamentals instead of anti-fundamentals. We dualize each block using the basic moves in Table \ref{tab:basicmoves}:
\begin{equation}
    \includegraphics[width=\textwidth]{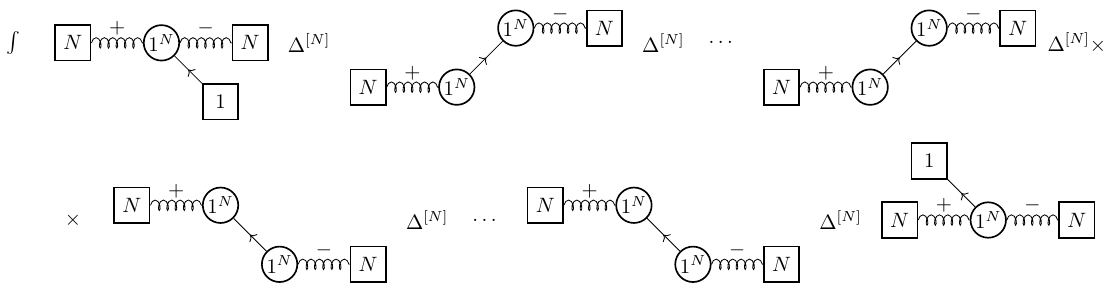}
\end{equation}

\begin{figure}
    \centering
    \includegraphics[width=\textwidth]{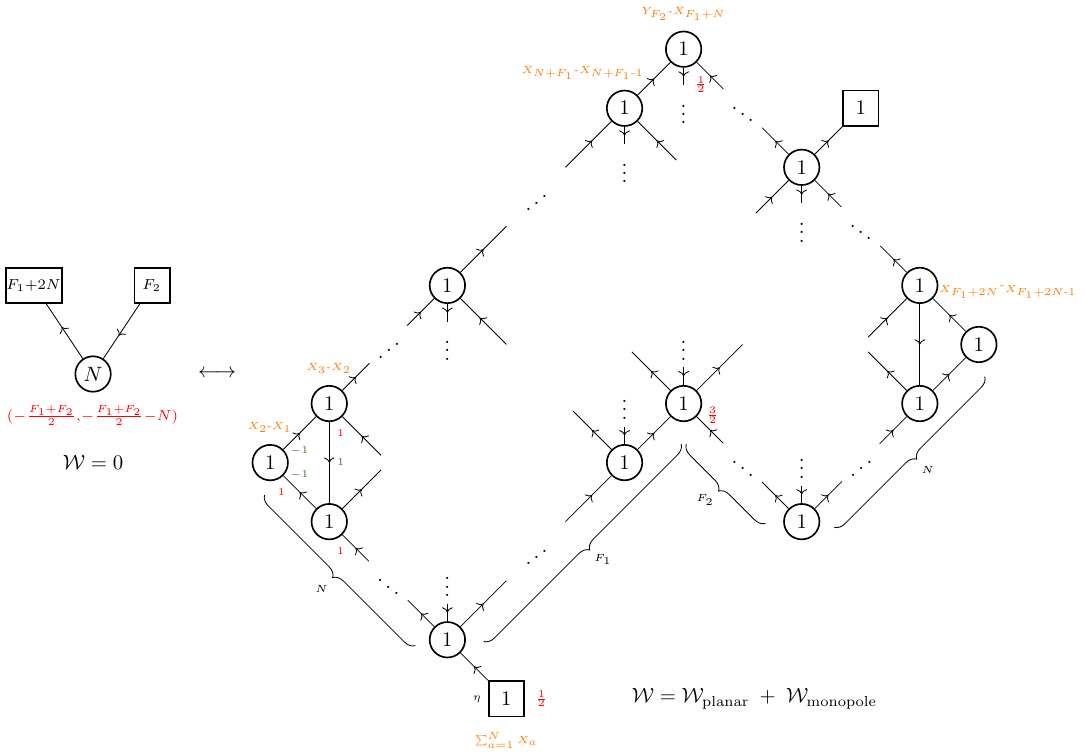}
    \caption{The $\mathcal{N}=2$ planar mirror dual of $U(N)_{(-\frac
    {F_1+F_2}{2}, -\frac{F_1+F_2}{2}-N)}$ SQCD with $[F_1+2N, F_2]$ chiral multiplets. The superpotential terms and (mixed) Chern-Simons interactions are as described in Section \ref{subsec: SQCD_example}.}
    \label{eq:extended_mirror_SQCD_N_A}
\end{figure}

The dualized QFT blocks can be glued back reintroducing gauge interactions with $\Delta^{[N]}$ measure, we also implement Identity-walls and obtain:
\begin{equation} \label{eq:mirror_SQCD_N_A}
    \includegraphics[]{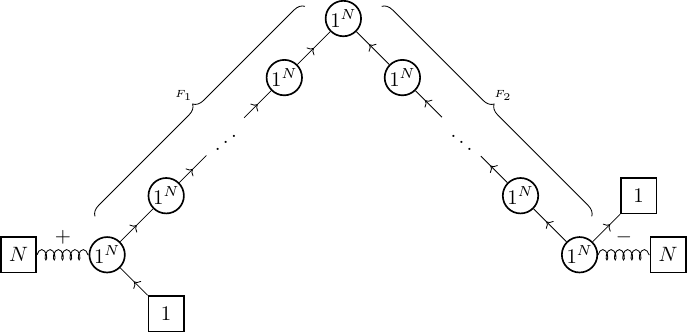}
\end{equation}
This quiver can be written explicitly in Lagrangian form using the definitions of the planar blocks, the result in depicted in Figure \ref{eq:extended_mirror_SQCD_N_A}.

The features of this duality are very similar w.r.t.~the duality in Figure \ref{eq:extended_mirror_SQCD_N_N}, so that the map between global symmetries and chiral ring generators works similarly as that explained at the end of Section \ref{subsec: UN_F1+N_F2+N}\footnote{In this case there can not be any enhancement of the topological symmetry for special values of $F_1$ and $F_2$ differently from the case of the SQCD with $[F_1+N,F_2+N]$ flavors.}. 

However, in this case there can be an extra chiral ring generator which is a mesonic operator from the point of view of the planar dual. It is possible to deduce that such operator exists only if $F_2 < N$. This operator maps to a gauge invariant (dressed) monopole operator with positive magnetic charge of the SQCD. It is possible to verify, for example through a Superconformal Index expansion, that in the SQCD this operator exists for the same range of $F_2$.

To summarize, the duals of the SQCD theories considered in this Section can be schematically depicted as:
\begin{equation}\label{eq:tangramSQCD1}
\begin{tikzpicture}[baseline=(current bounding box).center]
    \node at (-1,1.5) (m1) [manifest,black] {$_{F_1+N}$};
    \node at (1,1.5) (m2) [manifest,black] {$_{F_2+N}$};
    \node at (0,0) (g1) [gauge, black] {$N$};
    \draw[-<-] (g1)--(m2);
    \draw[-<-] (m1)--(g1);
    \draw[CScolor,right] (g1)++(.2,-0.3) node {$_{(-\frac{F_1+F_2}{2},-\frac{F_1+F_2}{2}-N)}$};
\end{tikzpicture}
\quad\leftrightarrow\quad
\raisebox{-.5\height}{\includegraphics[scale=1]{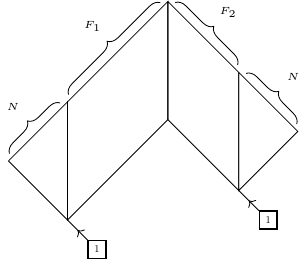} }
\end{equation}

\begin{equation}\label{eq:tangramSQCD2}
\begin{tikzpicture}[baseline=(current bounding box).center]
    \node at (-1,1.5) (m1) [manifest,black] {$_{F_1+2N}$};
    \node at (1,1.5) (m2) [manifest,black] {$_{F_2}$};
    \node at (0,0) (g1) [gauge, black] {$N$};
    \draw[-<-] (g1)--(m2);
    \draw[-<-] (m1)--(g1);
    \draw[CScolor,right] (g1)++(.2,-0.3) node {$_{(-\frac{F_1+F_2}{2},-\frac{F_1+F_2}{2}-N)}$};
\end{tikzpicture}
\quad\leftrightarrow\quad
\raisebox{-.5\height}{\includegraphics[scale=1]{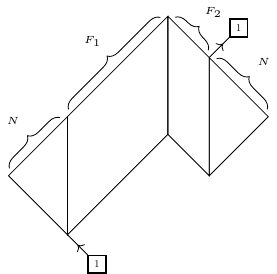} }
\end{equation}

\begin{equation}\label{eq:tangramSQCD3}
\begin{tikzpicture}[baseline=(current bounding box).center]
    \node at (-1,1.5) (m1) [manifest,black] {$_{F_1}$};
    \node at (1,1.5) (m2) [manifest,black] {$_{F_2+2N}$};
    \node at (0,0) (g1) [gauge, black] {$N$};
    \draw[-<-] (g1)--(m2);
    \draw[-<-] (m1)--(g1);
    \draw[CScolor,right] (g1)++(.2,-0.3) node {$_{(-\frac{F_1+F_2}{2},-\frac{F_1+F_2}{2}-N)}$};
\end{tikzpicture}
\quad\leftrightarrow\quad
\raisebox{-.5\height}{\includegraphics[scale=1]{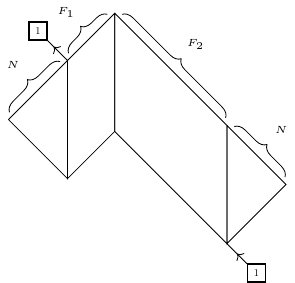} }
\end{equation}
Notice that for high enough number of fundamentals and antifundamentals an SQCD theory can admit multiple mirror duals among the ones presented above. 
As an example, $U(N)$ SQCD with $n_f \geq 2N$ fundamentals and $n_a\geq N$ antifundamentals admits two inequivalent abelian planar duals, schematically \eqref{eq:tangramSQCD1} and \eqref{eq:tangramSQCD2}.
It would be interesting to understand whether these two duals, in the range of $n_f$ and $n_a$ mentioned above, can be connected by performing local Aharony-like dualities, in the spirit of the analysis performed in Appendix \ref{App:B} for the case of the flip-flip duality, but we leave this to future work.

It is possible to derive planar abelian duals for SQCD theories with $SU(N)$ gauge group or with $U(N)$ gauge group and CS levels $(k,k + lN)$ for any $l$ by applying Witten's $SL(2,\mathbb{Z})$ action \cite{witten2003sl2zactionthreedimensionalconformal} on both sides of the dualities described above. 
This was described in detail for the case of SQCD with only fundamental chirals in Section \ref{ga-ung-section}, the generalization to SQCD with both fundamentals and antifundamentals is analogous and we do not carry it out explicitly.


\subsection{Topological symmetry enhancement in chiral quivers}

\subsubsection{A ``local" balancing condition}\label{subsec: symmenhanc}
We now move to an example of a quiver theory to show an interesting pattern of topological symmetry enhancement. 

We consider the following quiver:
\begin{equation}\label{eq:quivex_start}
    \includegraphics[width=.6\textwidth]{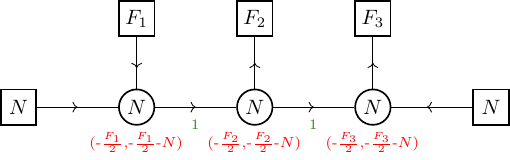}
\end{equation}
Notice that, as emphasized before, the CS levels for each gauge group are not arbitrary and are fixed by the amount of matter fields that each node sees following \eqref{eq:chiralCSrule}. 
The quiver presented here serves as a specific example, but it is straightforward to modify many of its properties. For instance, the directions of the bifundamental arrows can be flipped, and the flavor content can be generalized by considering different numbers of fundamentals and anti-fundamentals. However, for the sake of concreteness, we focus on this particular choice in the current discussion.

The mirror theory can be obtained by applying the algorithmic procedure described above, resulting in the following planar quiver. Here we only report the compact notation for the mirror quiver for ease of readability:
\begin{equation}\label{fig:gen_linear_quiver_mirror}
    \includegraphics[width=.6\textwidth]{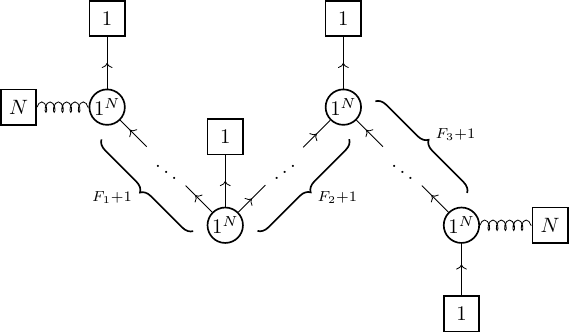}
\end{equation}

Therefore the planar quiver has the following structure:
\begin{equation}
    \includegraphics[width=.6\textwidth]{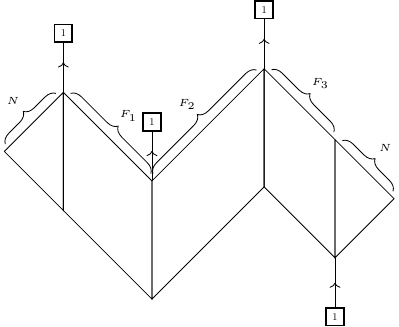}
\end{equation}

If we slightly modify the starting quiver in \eqref{eq:quivex_start} — for instance, by reversing the direction of a bifundamental arrow or swapping a fundamental flavor with an anti-fundamental — the resulting quiver in \eqref{fig:gen_linear_quiver_mirror} changes accordingly.  For example changing the arrow of a bifundamental changes a fundamental flavor attached to the top node of a column, with an anti-fundamental attached to the bottom node of the same column. Or also, changing a flavor from fundamental to anti-fundamental changes the slope of a planar bifundamental.

\begin{figure}
    \centering
    \includegraphics[]{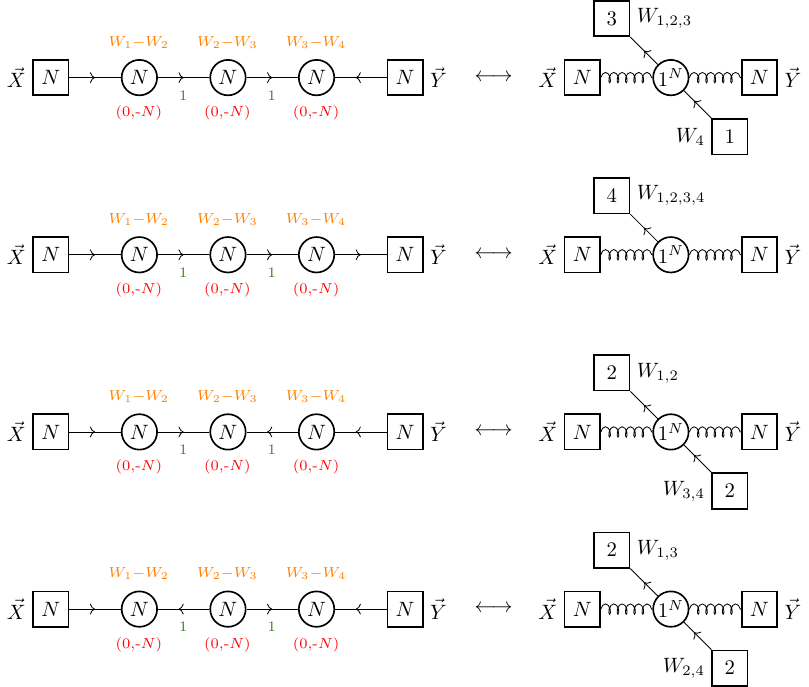}
    \caption{Chiral-Planar mirror dualities for a quiver with three gauge nodes.
The planar dual of each chiral quiver makes the enhancement of the topological symmetry manifest. In general, the topological symmetry enhances to $S[U(N_L) \times U(N_R)]$, where $N_L$ and $N_R$ are the numbers of left- and right-pointing bifundamentals, respectively. This enhancement occurs independently of whether bifundamentals with the same orientation are consecutive. Furthermore, examining the last two dualities, we observe that all theories with the same values of $N_L$ and $N_R$ are dual to each other. }
    \label{fig:quivex_enhancement}
\end{figure}

It is interesting to consider the special case where each gauge group has exactly $2N$ chiral (anti-)fundamentals, that is the case of $F_1=F_2=F_3=0$.
In this case the mirror planar quiver does not have sequences of planar bifundamentals, and all the flavor nodes are attached to the central column. The result is depicted in the first line of Figure \ref{fig:quivex_enhancement}. This leads to the flavor symmetry of the planar mirror theory to increse, in this case from $U(1)^4/U(1)$ to $SU(3)\times U(1)$. \\
Changing the orientation of the arrow produces different dualities whith possibly different enhancements of the global symmetry. Some other possibilities are depicted in Figure \ref{fig:quivex_enhancement}.
This, as far as we know, is a non-trivial prediction regarding the enhancement of topological symmetries in $\mathcal{N}=2$ quiver theories.

This suggests us that a rule to predict the enhancement of global symmetries. Whenever a $U(N)$ gauge node sees exactly $[N,N]$ flavors the topological symmetry enhances to at least $SU(2)$. When in a quiver more than one gauge node satisfied this condition then the enhancement is bigger. For example in Figure \ref{fig:quivex_enhancement}, it is shown a collection of linear quivers with $n=3$ nodes of rank $N$, with $N_R$ bifundamental chirals pointing to the right and $N_L$ pointing to the left, with $N_R + N_L = n+1$. Then the chiral-planar mirror duality predicts that the $U(1)^n$ topological symmetry enhances to $S[U(N_r)\times U(N_\ell)]$.
The limiting case with $n=1$ corresponds to SQCD with $U(N)_{(0,-N)}$ gauge group and $[N,N]$ flavors, where the topological symmetry enhances from $U(1)$ to $SU(2)$. This is a special case of the dualities already discussed in Subsubection \ref{subsec: UN_F1+N_F2+N}.
It is important to notice that the enhancement occurs for any quiver with the same values of $N_R$ and $N_L$, regardless of the ordering of the bifundamentals since the planar mirror dual is invariant under this choice. This fact implies a duality among all the quivers with the same value of $N_L$ and $N_R$, one such example is provided in Figure \ref{fig:quivex_enhancement}.

\subsubsection{A ``non-local" balancing condition} \label{subsec: overbalanced_mirror_example}

In Section \ref{subsec: SQCD_example}, we started from the $\mathcal{N}=4$ mirror pair of the $U(N)$ SQCD and considered the real mass deformation leading to the $\mathcal{N}=2$ mirror duality relating the chiral $U(N)$ SQCD to its planar quiver dual pair, depicted in Figure \ref{planar_mirror_general}. 

It is interesting to consider the alternative situation  where instead we take the chiral limit of the quiver on the dual side, which results in a planar limit for the SQCD. This can indeed be studied using the same strategies outlined in Section \ref{subsec: SQCD_example}; however, here we will instead use the algorithm to generate the example.
We begin with the  chiral limit of the mirror of SQCD, namely:
\begin{equation} \label{quiv: chiral_u_n_mirror}
    \includegraphics[]{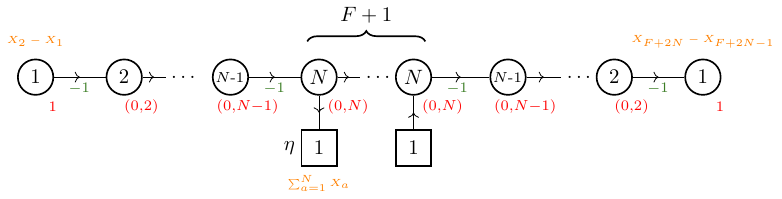}
\end{equation}
We decompose this theory into fundamental blocks according to the rules described in the previous section, dualize each of them and then glue-back the result implementing the Identity-walls. These steps are depicted in the following figure:
\begin{equation}
    \includegraphics[width=\textwidth]{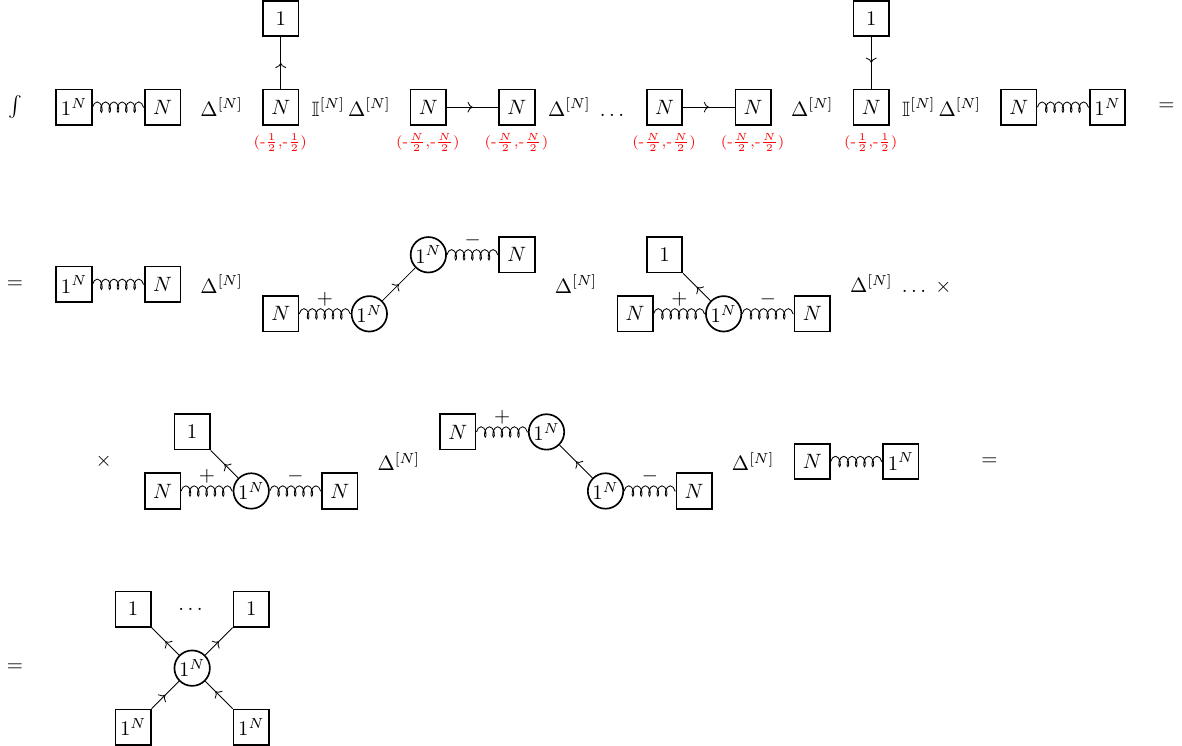}
\end{equation}
Notice that the quiver considered in \eqref{quiv: chiral_u_n_mirror} actually does not follow the rule prescribed at the beginning of the section since it contains two tails
built with asymmetric $U(N)\times U(M)$ bifundamentals. However
we can bypass the problem of dualizing asymmetric bifundamental blocks by noticing that 
the  two tails  can be thought simply as a $\mathcal{S}$ and $\mathcal{S}^{-1}$-walls.
These walls do not need to be dualized and  in the last step they fuse to Identity-walls with two other walls coming from the dualization of the flavor blocks.

\begin{figure}[ht]
    \centering
    \includegraphics[width=.6\textwidth]{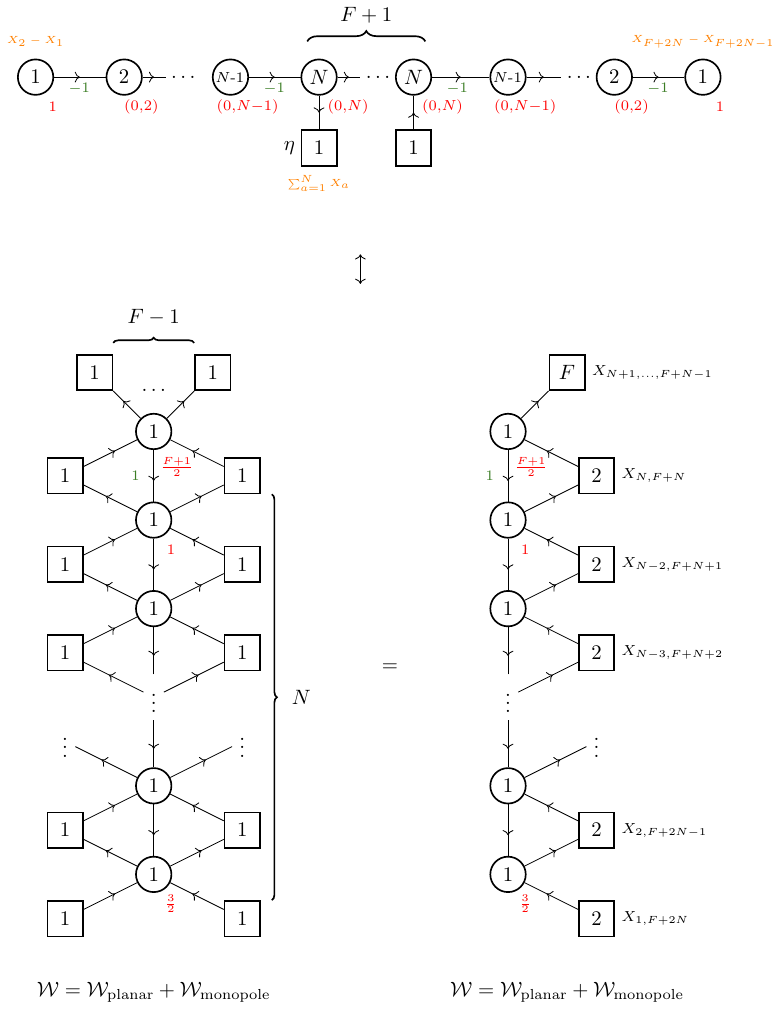}
    \caption{
    On the top, the mirror dual of the planar limit of the $\mathcal{N}=4$ SQCD, reported from \ref{quiv: chiral_u_n_mirror}.
    On the bottom, the planar limit of the $\mathcal{N}=4$ SQCD.
    We have suppressed the FI parameters and all the background terms for brevity. The FI terms for the gauge nodes are $\eta +\frac{iQ}{4}((F-1) \delta_{1,\alpha}-\delta_{N,\alpha})$. On the r.h.s.~it is depicted the same theory but after recognizing the manifest flavor symmetry $S[U(F-1) \times U(2)^N]$ and writing it explicitly.}
    \label{eq:overbalanced_mirror}
\end{figure}

The resulting quiver can be then written explicitly in Lagrangian form, providing the planar mirror description for the original linear quiver gauge theory (\ref{quiv: chiral_u_n_mirror}), as shown in Figure \ref{eq:overbalanced_mirror}. 

Interestingly, in the planar dual, the flavor symmetry is not the naive $U(1)^{F+2N}/U(1)$, but is instead enhanced to $S[U(F-1) \times U(2)^N]$, as shown in \eqref{quiv: chiral_u_n_mirror}.
This reveals an intriguing pattern of topological symmetry enhancement. The appearance of the $U(F-1)$ factor can be understood using the rule proposed in the previous section: the central part of the quiver consists of a sequence of $U(N)$ nodes, each with $[N, N]$ flavors and bifundamentals oriented in the same direction. However, the emergence of the multiple $U(2)$ factors is a new feature specific to this duality.
Notably, this latter enhancement involves topological symmetries associated with non-adjacent nodes — that is, it is non-local — in contrast to the local structure underlying the $U(F-1)$ enhancement. Nonetheless, both enhancements are manifest in the planar dual.

\section{Further examples and conclusions}\label{furtherexamples}
In Section \ref{sec: Examples} we used the dualization algorithm introduced in Section \ref{sec: S_walls,QFT_blocks_Duality_Moves} to construct a series of examples to illustrate interesting features of the chiral-planar $\mathcal{N}=2$ mirror duality.

However, as discussed in detail at the beginning of Section \ref{sec: Examples}, the algorithm does have a few limitations. Nevertheless, with the present work, we aim to argue that a much larger class of — if not all — $\mathcal{N}=2$ quiver theories admit a planar abelian dual. Although we do not offer a formal proof of this statement, our claim is supported by the observation that the strategies outlined in Section \ref{subsec: SQCD_example} can be applied to any pair of $\mathcal{N}=4$ mirror theories, and we expect the outcome to consistently take the form of a $\mathcal{N}=2$ chiral-planar mirror duality.
\footnote{Indeed this argument assumes that the 3d $\mathcal{N}=4$ mirror theories are Lagrangian. This is not the case for many interesting $\mathcal{N}=4$ theories and in this case the strategy of Section \ref{subsec: SQCD_example} can not be applied since the partition function and the EOM might not be known.}
Therefore we expect that the assumptions needed to apply the algorithm, for example in the cases analyzed in Section \ref{sec: Examples}, are mostly technical and can be overcome 
by either performing directly a real mass limit, similar to the anlysis in Section \ref{subsec: SQCD_example} or by developing a more sophisticated algorithm. Here we focus on the former possibility, discussing how the limitations of the analysis described in Section \ref{sec: Examples} can be overcome:

\begin{itemize}
    \item So far we considered  $U(N)$ gauge nodes with $[n_f,n_a]$ flavors satisfying the condition in eq.~\eqref{eq:SQCD_cases}. In Subsection \ref{subsec: ex5_SQCD} we provide an example of a $U(N)$ SQCD with flavors outside the range \eqref{eq:SQCD_cases}.
    \item For quiver theories we always assume that the gauge nodes have the same rank. We already relaxed this condition by considering an example of a quiver with sequences of gauge nodes $U(1)-U(2)-\dots -U(N)$. In Subsection \ref{subsec: ex5_quiver} we provide an example of a more generic quiver.
    \item Although we developed the algorithm for linear quivers, we do not expect the chiral-planar duality to be restricted to $\mathcal{N}=2$ theories with linear topology. In Subsection \ref{subsec: ex5_circular}, we present  examples involving a circular quiver.
    
    \item We developed the  algorithm  for quivers with $U(N)$ gauge groups, however $\mathcal{N}=4$ mirror dualities are  known for theories with different gauge group. In Subsection \ref{subsec: ex5_usp} we provide the planar mirror of a $USp(2N)$ SQCD as an example of a non unitary $\mathcal{N}=2$ theory.

    \item So far, we have considered quivers with $U(N)$ gauge groups and CS levels $(k, k + lN)$, where $k = -\tfrac{F}{2} + N$ and $F$ denotes the total number of flavors coupled to the gauge group. The current version of the  algorithm applies to quivers  with  $l=-1$, but we already argued that it is possible to generalize $l$ using Witten's $SL(2,\mathbb{Z})$ action \cite{witten2003sl2zactionthreedimensionalconformal}. 
    We can clearly find planar duals also for quivers with more general values of $k$.
    In Subsection \ref{subsec: ex5_gencs} we provide an example of an SQCD with $k \neq - \tfrac{F}{2} + N$.
\end{itemize}
All the examples provided are computed starting from known Lagrangian $\mathcal{N}=4$ mirror dualities and using the techniques of Section \ref{subsec: SQCD_example}. Indeed, it is possible to think of many more generalizations. We leave those possibilities fo a future work.

To conclude, we point out that some of these limitations of the algorithm could be overcome by extending it with new moves that are already known in the $\mathcal{N}=4$ case, we would only need to perform the SUSY-breaking deformation of these moves to generalize further the algorithm.

\subsection{$U(2)$ SQCD with $[3,1]$ flavors}
\label{subsec: ex5_SQCD}

Let us consider the $U(2)$ SQCD.
There is only one combination not included in eq.~\eqref{eq:SQCD_cases} that is $n_f=3$ and $n_a=1$ (and equivalently $n_f=1$ and $n_a=3$). Using the methods outlined in Section \ref{subsec: SQCD_example} we find the following chiral-planar mirror duality:
\begin{equation}
    \includegraphics[]{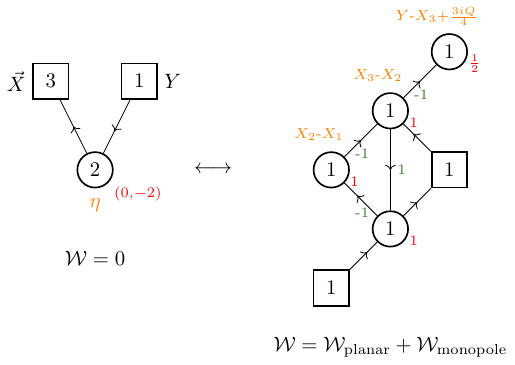}
\end{equation}
As usual, on the planar side we write the FI of all the gauge nodes in a column on top of it. 

Notice that on the planar side there is a mesonic chiral ring generator given by the shortest path connecting the the flavors nodes (modulo F-terms). This predicts that in the SQCD there is a gauge invariant monopole operator that maps to it, which is indeed the case.

\subsection{Quiver with non-constant ranks}\label{subsec: ex5_quiver}
Let's now consider a chiral-planar duality involving a chiral  quiver  with non-constant ranks.

For example we start from the following 3d $\mathcal{N}=4$ mirror duality among two linear quivers:
\begin{equation}
    \includegraphics[]{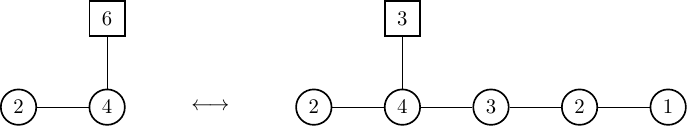}
\end{equation}
Here we are drawing $\mathcal{N}=4$ quivers in  $\mathcal{N}=4$ notation  where each line represents a hypermultiplet and circles include the full $\mathcal{N}=4$ vector multiplet.
We then apply the strategies outlined in Section \ref{subsec: SQCD_example} to derive the following chiral-planar mirror duality:
\begin{equation}
    \includegraphics[]{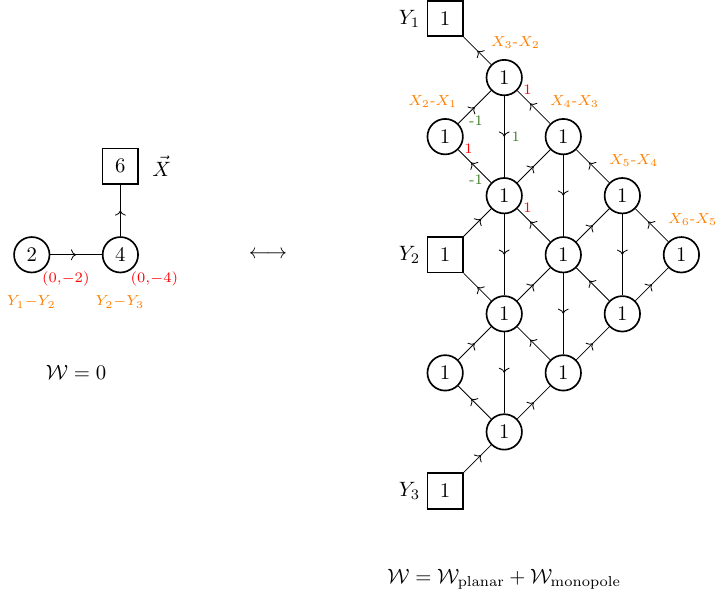}
\end{equation}
The label on top of the column indicates that all the gauge node in the same column have the same FI.
Also, the superpotential $\mathcal{W}_{\text{monopole}}$ 
contains, in addition to the usual monopoles following the rules described in Section \ref{subsec: SQCD_example}, the additional monopole:
\begin{equation}
\mathcal{W}_{\text{monopole}} \supset \mon^{\left(\:\hspace{-2pt}
        \resizebox{50pt}{!}{%
        \begin{tabular}{ccccc} 
        	&0&&&\\
	- &&0&&\\
	&-&&0&\\
	&&0&&0\\
	&+&&0&\\
	+ &&0&&\\
        	&0&&&
        \end{tabular}}\right)
        }
\end{equation}
Which is compatible with being in the superpotential since it has R-charge 2, is gauge invariant and compatible with the global symmetry of the resulting theory.

\subsection{Circular quivers}\label{subsec: ex5_circular}
We now consider an example of $\mathcal{N}=4$ mirror circular quivers
 (see for example \cite{Assel:2012cj} and references therein)
and perform the real mass deformation to  find a chiral-planar circular  mirror pair.

To perform these steps following the strategies of Section \ref{subsec: SQCD_example}, it is obviously necessary to have an exact (fully refined) $S^3_b$ partition function identity for $\mathcal{N}=2^*$ mirror dualities between circular quivers. This will be provided in the upcoming work \cite{CircularAlgorithm}, authored by one of us, where the $\mathcal{N}=4$ dualization algorithm will be extended to circular quivers. Relying on those results, we are then able to obtain the circular chiral-planar mirror pairs.

As an example we consider the following $\mathcal{N}=4$ mirror duality:
\begin{equation}
    \includegraphics[width=\textwidth]{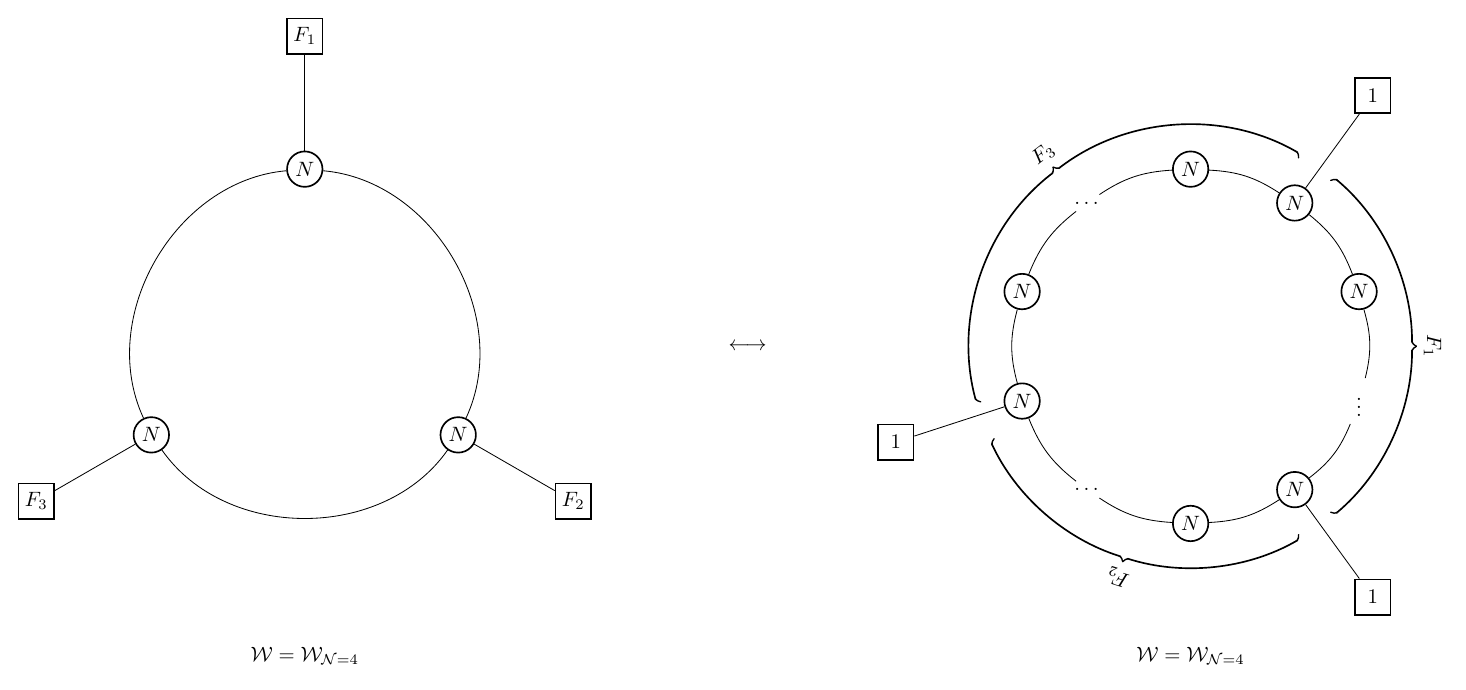}
\end{equation}
Where the braces labelled by $F_i$ indicate that the gauge nodes are connected by $F_i$ bifundamental hypermultiplets.

Performing a real mass deformation leads to the following chiral-planar mirror duality:
\begin{equation}
    \includegraphics[width=\textwidth]{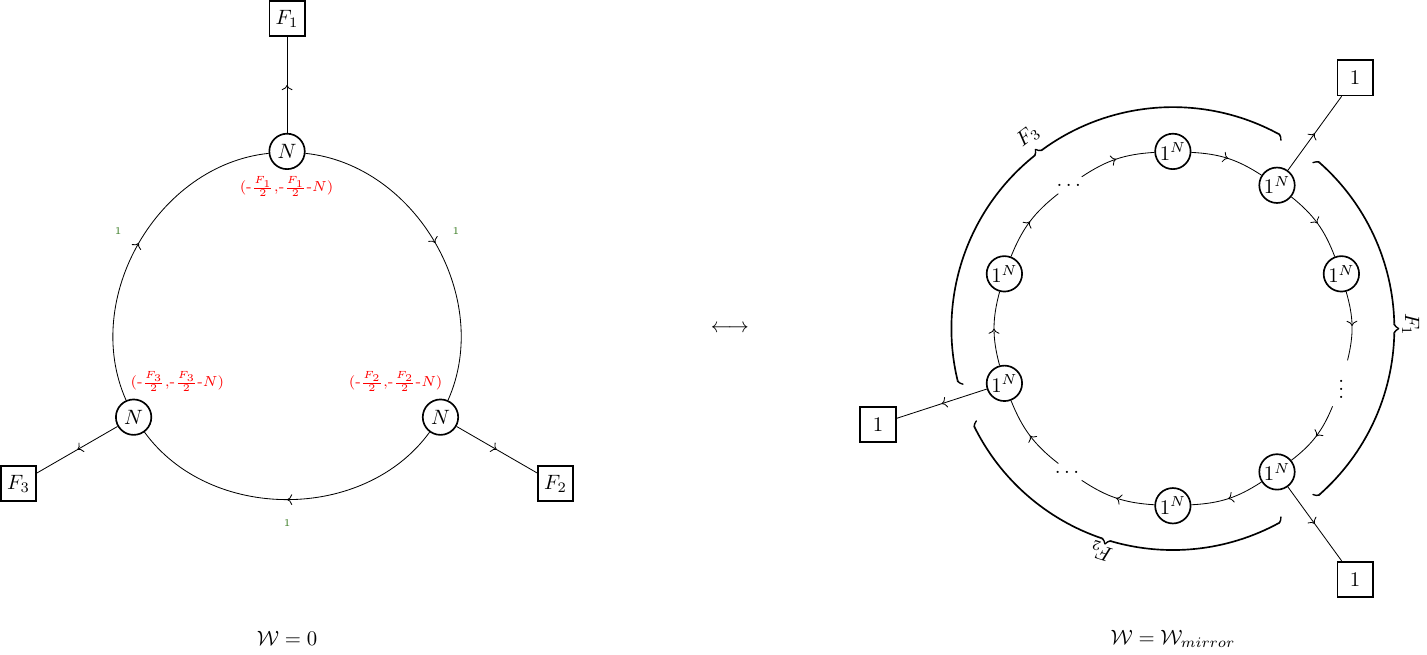}
\end{equation}
For the planar dual we provide only the short notation for brevity. This can be expanded in a complete Lagrangian description of the theory using the definition of the short notation for the planar blocks.

\subsubsection{Adjoint $U(N)$ SQCD}
As a limiting case of circular $\mathcal{N}=4$ mirror dualities we consider the duality for the adjoint $U(N)$ SQCD which is:
\begin{equation}
    \includegraphics[]{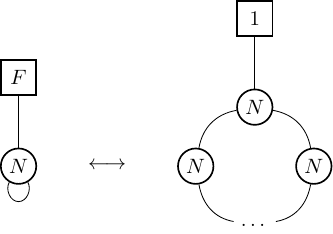}
\end{equation}
Where the arch denotes the presence of an extra adjoint hypermultiplet. If we perform the SUSY-breaking deformation we obtain the following chiral-planar mirror duality:
\begin{equation}
    \includegraphics[]{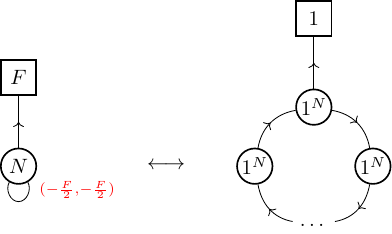}
\end{equation}
Note that in the $\mathcal{N}=4$ adjoint SQCD there is an $SU(2)$ global symmetry rotating the two adjoint chirals in the hypermultiplet. The real mass parameter associated to this symmetry partecipates in the SUSY-breaking real mass deformation and it is in fact tuned so that one of the two adjoint chiral multiplets remain massless, while the other acquires a real mass.
Due to the integration of the adjoint chiral we obtain an extra shift in the resulting CS-level of $(-N,0)$. therefore we land on the adjoint $U(N)$ SQCD with CS level $(-\tfrac{F}{2},-\tfrac{F}{2})$, thus with equal $SU(N)$ and $U(1)$ levels.

On the r.h.s.~we have the planar dual of the $\mathcal{N}=2$ adjoint SQCD, showing that the planar mirror duality is a feature that might be extended also to theories with tensor matter.

\subsection{$USp(2N)$ SQCD}\label{subsec: ex5_usp}
Throughout the paper we only focused on gauge theories with $U(N)$ gauge group. However, we can try to perform the same analysis on theories with different gauge group, whenever a suitable $\mathcal{N}=4$ mirror duality is accessible.

As an example, we consider the $\mathcal{N}=4$  mirror dualilty for the $USp(2N)$ SQCD
\cite{Kapustin:1998fa,Porrati:1996xi,Hanany:1999sj}:\footnote{The $\mathcal{N}=4$ $USp(2N)$ SQCD also enjoys a mirror dual which has orto-symplectic gauge groups \cite{Feng:2000eq} on top of the unitary mirror dual that we consider here. It would be interesting to perform the same analysis in this case to understand how the planarization limit might affect an orto-symplectic theory. However, we leave this analysis for a future work.}:
\begin{equation}
    \includegraphics[]{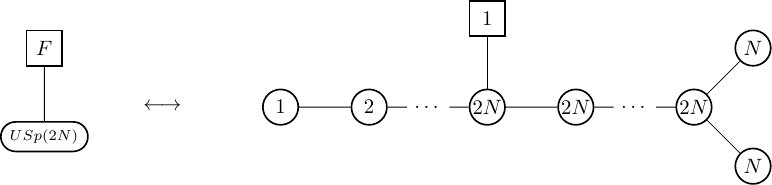}
\end{equation}
Also in this case to implement the strategies of Section \ref{subsec: SQCD_example} we need the exact
$S^3_b$ partition function identity
and this will be provided in the upcoming work \cite{Uspalgorithm}, authored by two of us, where the $\mathcal{N}=4$ dualization algorithm will be extended to $USp$ theories.
Building on these results,
we perform a real mass deformation leading to a chiral-planar mirror duality for the $\mathcal{N}=2$ $USp(2N)$ SQCD. For clarity we now report the result in the special case $N=2$ and $F=9$.
\begin{equation}\label{fig:ex_uspsqcd}
    \includegraphics[]{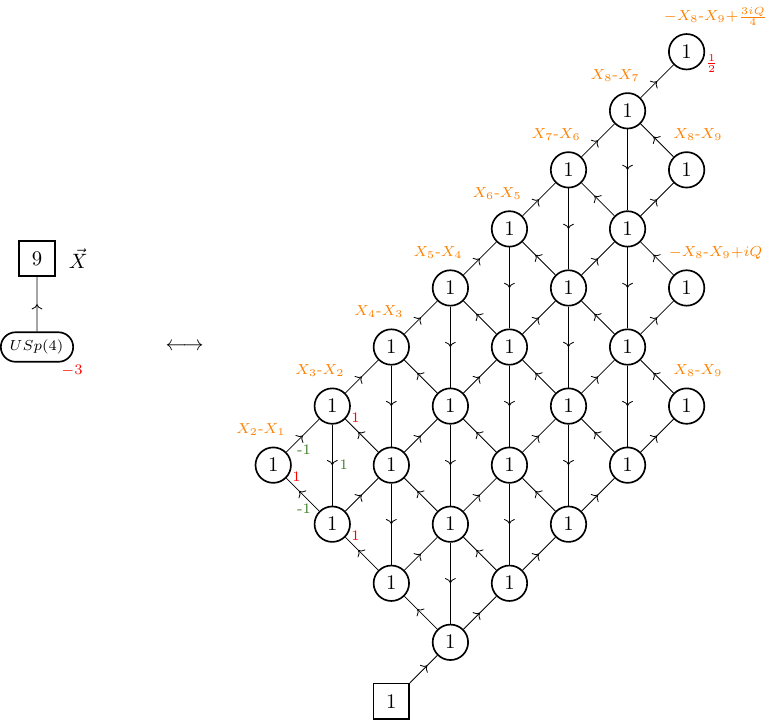}
\end{equation}
In general, the real mass has the effect of breaking the flavor $SO(2F)$ global symmetry down to $U(F)$ and also generates a CS level of $(2N+2-F)$, where the integration of the fundamentals provide a contribution of $-F$ and the adjoint chiral of $2+2N$.

The superpotential $\mathcal{W}_{\text{monopole}}$ 
contains, in addition to the usual monopoles following the rules described in Section \ref{subsec: SQCD_example}, the two additional monopoles:
\begin{equation}
\mathcal{W}_{\text{monopole}} \supset \mon^{\left(\:\hspace{-2pt}
        \resizebox{50pt}{!}{%
        \begin{tabular}{cccc} 
        	&&&-\\
	&&-&\\
	&0&&0\\
	&&+&\\
	&0&&+\\
	$\cdots$&&0&\\	
	&0&&0\\
	&&0&\\
	&0&&\\
        \end{tabular}}\right)
        }
+
\mon^{\left(\:\hspace{-2pt}
        \resizebox{50pt}{!}{%
        \begin{tabular}{cccc} 
        	&&&0\\
	&&0&\\
	&0&&-\\
	&&-&\\
	&0&&0\\
	$\cdots$&&+&\\	
	&0&&+\\
	&&0&\\
	&0&&\\
        \end{tabular}}\right)
        }
\end{equation}
\\
Taking into account the monopole superpotential, each column contributes as a single $U(1)$ UV topological symmetry, except for the last column which contributes as two $U(1)$ topological symmetries. Together all the $U(1)$ topological symmetries enhance in the IR to $U(F)$.

From the planar mirror dual we observe that there is no mesonic chiral ring generator and then we can conclude that also in the $USp$ SQCD there is no gauge invariant monopole generating a chiral ring.

In conclusion, let us mention that the mirror dual depicted in \eqref{fig:ex_uspsqcd} can be generalized to any $N$ and for $F > 2N+1$. It is interesting to notice that in the case $F=2N+1$ the $\mathcal{N}=4$ theory exibit a global symmetry enhancement where a $U(1)$ magnetic symmetry emerges in the IR. We do not attept this analysis which is made more complicated by the fact that now we have an emergent symmetry for which we can perform a real mass deformation, whose Cartan's subgroup is not even visible in the UV SQCD decription.

\subsection{Further real mass deformations}\label{subsec: ex5_gencs}
Starting from any of the duality presented in this paper one may try to perform further real mass deformations to flow to theories with more generic CS level. As an example let us consider the chiral-planar mirror dual for the $U(2)$ SQCD with 5 fundamental flavors:
\begin{equation}
    \includegraphics[]{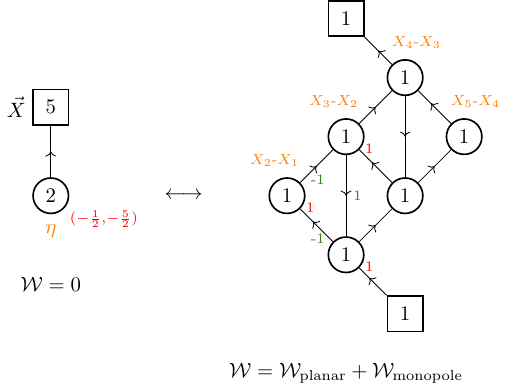}
\end{equation} 
If we give a negative mass to a fundamental chiral on the l.h.s.~we flow to a $U(2)$ SQCD with 4 fundamental flavors and with CS level $(-1,-3)$ outside the range of values considered so far: 
$(k,k+lN)$ with $k=-\tfrac{F}{2}+N$ and $l=-1$.
\footnote{If instead we perform a positive real mass deformation, we obtain a $U(2)$ SQCD with four fundamental flavors and Chern-Simons levels $(0, -2)$, which lies within the range. It is therefore a non-trivial consistency check to verify that the planar dual correctly reduces to the $F=4$, $N=2$ case of the general result depicted in Figure \ref{planar_mirror_general}. This turns out to be the case, although we do not provide a complete account of the analysis here.}. 
By using the same strategies as in Section \ref{subsec: SQCD_example} we can study the effect of the real mass deformation in the planar mirror dual. We discover that the resulting duality is:
\begin{equation}
    \includegraphics[]{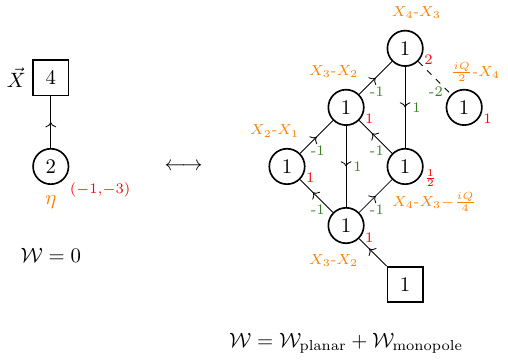}
\end{equation}
On the planar side the effect of the real mass is to give a positive/negative mass to the pair of chirals connecting the third column to the rightmost node, and also a positive mass to the top flavor. This has the effect of shifting the level of some self and mixed-CS interactions
and some of the FI parameters.

We can further simplify the quiver theory by noticing that the $U(1)_1$ 
gauge node on the r.h.s.~can be confined causing only a shift in the CS-level of the top node which is 
coupled to it by a mixed-CS interaction (see \eqref{eq:tftintegration}). The resulting duality is then:
\begin{equation}
   \includegraphics[]{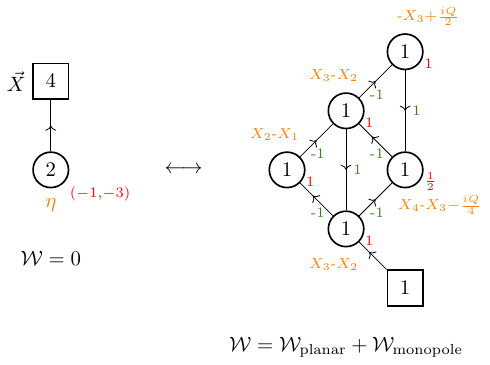}
\end{equation}

From the  planar mirror dual we see that there is no chiral ring in the two theories since no mesonic operator is present in the planar theory.

Generalizations of this strategy give rise to new chiral–planar dualities beyond those presented in this paper, including cases with more general Chern-Simons levels. This direction will be explored in the upcoming work \cite{RealMasses}.

\acknowledgments

We thank Amihay Hanany, Antonio Amariti, Cyril Closset, Noppadol Mekareeya, Pierluigi Niro and Simone Giacomelli for useful discussions. 

SP and SB are Partially supported by the MUR-PRIN grant No. 2022NY2MXY (Finanziato dall’Unione europea- Next Generation EU, Missione 4 Componente 1 CUP H53D23001080006).
SR is supported by the MUR-PRIN grant No. 2022NY2MXY.

\appendix

\nocite{*}
\section{Charges of monopoles}\label{app: monopole}
In this appendix we rapidly review how to compute the R-charge, the charges under global symmetries and the representation under gauge symmetries.

We first consider the case of a generic $U(N)_{(k,k+lN)}$ gauge theory with $n_f$ fundamental and $n_a$ anti-fundamental chiral fields:
\begin{center}
\tikzstyle{flavor}=[rectangle,draw=red!50,thick,inner sep = 0pt, minimum size = 6mm]
\tikzstyle{manifest}=[rectangle,draw=blue!50,thick,inner sep = 0pt, minimum size = 6mm]
\tikzstyle{gauge}=[circle,draw=black!50,thick,inner sep = 0pt, minimum size = 6mm]
\tikzset{->-/.style={decoration={
  markings,
  mark=at position .5 with {\arrow{<}}},postaction={decorate}}}
  \begin{equation} 
\begin{tikzpicture}  
    \node at (0,0) (g1) [gauge, black] {$N$};
    \node at (2,0) (f1) [flavor,black] {$A$};
    \node at (-2,0) (f1) [flavor,black] {$F$};
    \draw[->-] (0.3,0) -- (1.7,0);
    \draw[->-] (-1.7,0) -- (-0.3,0);
    \draw[red](0.5,-0.5) node {$_{(k,k+lN)}$};
    \draw [](-2,-0.7) node {$\vec{Y}$};
    \draw [](2,-0.7) node {$\vec{X}$};
    \draw (0,0.6) node {$\vec{u}$};
    \end{tikzpicture} \end{equation}
\end{center}
where $\vec{X}$ and $\vec{Y}$ are fugacities related to the $S[U(A) \times U(F)]$ global symmetry, and $\vec{u}$ are the gauge fugacities. 
The R-charge of a monopole operator with generic magnetic flux $\vec{m}$ is \cite{Aharony:1997bx,Borokhov:2002cg,Benini_2011}:
\begin{equation}
    R[\mon^{\vec{m}}] = \Lambda \sum_{i=1}^N m_i - \frac{1}{2} \bigg[ \sum_{i \neq j} |m_i - m_j| + n_f(r_f -1)\sum_{i=1}^N |m_i| + n_a(r_a -1)\sum_{i=1}^N |m_i|
    \bigg]
    \label{monfor}
\end{equation}
where $\Lambda$ is the value of the mixing between the R-symmetry and the topological symmetry. Also $r_{f/a}$ is the R-charge of the fundamentals and anti-fundamentals. The representation of the monopole under the gauge symmetry can be obtained instead computing the following quantity:
\begin{equation}\label{eq: gauge_charge_monopole}
    Q_{\text{gauge}}[\mon^{\vec{m}}] = -\sum_{i=1}^N (k Z_i + l \sum_{j=1}^N Z_j)m_i + \sum_{i=1}^N (n_f-n_a)Z_i |m_i| 
\end{equation}
which gives the highest weight of the representation of the bare monopole. Notice that one has also to take into account that the magnetic flux $\vec{m}$ generically breaks the $U(N)$ gauge symmetry. The unbroken gauge symmetry is in general $\prod_{k=1}^L U(M_k)$, with $\sum_{k=1}^L M_L = N$ generate by a magnetic flux where the first $M_1$ entries are equal, then the following $M_2$ are equal and so on. In this analogy the parameters associated to the Cartans are the $Z_{1,\ldots,M_1}$ for $U(M_1)$, then $Z_{M_1+1,\ldots,M_1+M_2}$ and so on. 

In the present paper, a crucial role is also played by quiver abelian gauge theories with mixed CS interactions and matter in bifudnamental representation of pairs of gauge groups. We are thus interested in the charges of monopoles in such quiver theories. We will consider monopoles with non-vanishing magnetic fluxes for multiple gauge groups and vanishing flux under global symmetries.
In general, given an abelian quiver theory with the following data:
\begin{itemize}
\item $n_g$ abelian gauge nodes with gauge parameters $u_i$, $i=1,\dots, n_g$
\item $n_f$ flavor nodes
\item Global symmetry $U(1)^r$, not including topological symmetries (this may be the Cartan of a larger non-abelian symmetry). We denote as $y_a, a=1,\dots,r$ the corresponding parameters.
\item CS levels $k_i$,  BF levels $k_{ij}$ and  FI parameters $\lambda_i$,
as described in Section \ref{subsec: SQCD_example}.
\item Bifundamental chirals $\alpha_{ij}$, which is in the antifundamental representation (charge $-1$) under the $i$-th node and fundamental representation (charge $+1$) under the $j$-th node
\item Antifundamentals $\tilde{\beta}_{ia}$ and fundamentals $\beta_{ai}$
\end{itemize}
we consider a monopole with gauge fluxes $\vec{m} = (m_1, \dots, m_{n_g})$. 
The charges of the monopole under the gauge and global symmetries can be encoded in a polynomial of the fugacities given by:
\begin{equation}	\label{eq:monopole_charge}
\begin{split}
\mathcal{M}(\vec{m}) =& -\frac{1}{2} \Bigg(
		\sum_{\alpha_{ij}} |m_i - m_j| \left((R[\alpha_{ij}]-1) \frac{iQ}{2} - u_i + u_j + \sum_{m=1}^{r} Q_m [\alpha_{ij}] y_m \right)
		\\&\qquad+
		\sum_{\tilde{\beta}_{ia}} |m_i| \left( (R[\tilde{\beta}_{ia}]-1) \frac{iQ}{2}-u_i + \sum_{m=1}^{r} Q_m [\tilde{\beta}_{ia}] y_m \right)
		\\&\qquad+
		\sum_{\beta_{ia}} |m_i| \left((R[\beta_{ia}]-1) \frac{iQ}{2} + u_i + \sum_{m=1}^{r} Q_m [\beta_{ia}] y_m \right)
	\Bigg)
\\&
	-\sum_{i=1}^{n_g} k_i m_i u_i
	+\sum_{i=1}^{n_g} \lambda_i m_i
	-\frac{1}{2}\sum_{i<j} k_{ij} (m_i u_j + m_j u_i)
\end{split}
\end{equation}
where $Q_m$ is the charge under a global symmetry $U(1)_m \subset U(1)^r$ and $r[\alpha]$ is the R-charge of the fermion in the chiral multiplet $\alpha$.
The first line in \eqref{eq:monopole_charge} corresponds to the contribution coming from the bifundamentals, the second and third lines correspond to fundamentals and antifundamentals, respectively, and the last line corresponds to the contributions from CS terms, FI terms and BF terms. The term $\lambda_i$ is the mixing parameter between the topological symmetry and the R-symmetry.
The charges of the monopoles under the various symmetries are given by the corresponding parameters' coefficients, where the R-symmetry is instead the pure number obtained, i.e.~the one that does not multiply any parameter.


\section{Flip-Flip Duality from Local BCC-like Dualities}
\label{App:B}

In this appendix  we derive the flip-flip dualities for the $G[U(N)]$ theory by repeated application of local Aharony-like dualities \cite{Aharony:1997gp,Giveon:2008zn,Benini:2011mf,Aharony:2014uya,Closset:2023vos}, that we denote BCC-like dualities after \cite{Benini:2011mf}.
These dualities are well established and can be derived from Aharony duality via real mass deformations \cite{Benini:2011mf}.
As an example here we review the derivation of the following duality between $U(1)$ gauge theories with chiral matter:
\begin{equation}    \label{eq:U1_[1,2]}
\begin{gathered}
U(1)_{\frac{1}{2}} \text{ with } [2,1]
\\
\mathcal{W}=0, 
\end{gathered}
\qquad \leftrightarrow \qquad
\begin{gathered}
U(1)_{-\frac{1}{2}} \text{ with } [1,2]
\\
\mathcal{W}= \mathcal{W}_{Seiberg} + t^- \tilde{\mon}^-, 
\end{gathered}
\end{equation}
where on the r.h.s.~ there are two singlets flipping the mesons via $\mathcal{W}_{Seiberg}$ and one singlet $t^-$ flipping the monopole $\tilde{\mon}^-$, the monopole of the magnetic theory with GNO flux $-1$. On the r.h.s.~there are also background CS and mixed CS terms, discussed below.
It is convenient for our purposes to present the relevant dualities in quiver notation, in line with the main body of this paper.
We also introduce BF terms between the gauge group and the flavor groups, as well as background CS and mixed CS terms.
The duality \eqref{eq:U1_[1,2]} can be written as:
\begin{equation}	\label{quiv:U1_[1,2]}
\begin{tikzpicture}[baseline=(current bounding box.center)]
	\nodeCS(0,0)(g,$1$, $\frac{1}{2}$)
	\flavorCS(1,1)(ne,$1$, $0$)
	\flavorCS(-1,-1)(sw,$1$, $0$)
	\flavorCS(1,-1)(se,$1$, $0$)
	\arrowBFlr(g,ne)($-1$,right)
	\arrowBF(g,sw)($1$)
	\arrowBFlr(se,g)($-1$,left)
	\dottedBF(ne,se)($1$)
	\dottedBF(se,sw)($0$)
    \draw[FIcolor]  (g)++(0,0.5) node[anchor=south] {\tiny$\eta$};
	\node at (0,-1.7) (W) {$\mathcal{W}=0$};
\end{tikzpicture}
\qquad
\longleftrightarrow
\qquad 
\begin{tikzpicture}[baseline=(current bounding box.center)]
	\nodeCS(0,0)(g,$1$, $-\frac{1}{2}$)
	\flavorCS(1,1)(ne,$1$, $-\frac{1}{2}$)
	\flavorCS(-1,-1)(sw,$1$, $1$)
	\flavorCS(1,-1)(se,$1$, $-\frac{1}{2}$)
	\arrowBFlr(ne,g)($1$,right)
	\arrowBFlr(sw,g)($-1$,right)
	\arrowBFlr(g,se)($1$,right)
	\arrowBF(se,ne)($0$)
	\arrowBFlr(se,sw)($0$,right)
    \path[draw] (sw) edge[->-,bend right=15] node[midway,anchor=north] {\tiny$t_-$} (se);
    \draw[FIcolor]  (g)++(-0.2,0.5) node[anchor=south] {\tiny$-\eta-\frac{iQ}{4}$};
	\node at (0,-1.9) (W)
    {$\mathcal{W}=\mathcal{W}_{Seiberg} + t^- \tilde{\mon}^-$};
\end{tikzpicture}
\times e^{\pi i \frac{\eta^2}{2}}
\end{equation}
where $\mathcal{W}_{Seiberg}$ is the usual Seiberg-like cubic superpotential involving the dual quarks and the mesons. 
Additionally, on the r.h.s.~ there is a background CS term for the topological symmetry at level $-\frac{1}{2}$ in the magnetic phase as well as background FIs for flavor symmetries discussed below.

The duality \eqref{quiv:U1_[1,2]} can be derived from Aharony duality for $U(1)$ gauge group with 2 flavors by giving a large real mass to one fundamental chiral multiplet.
This procedure is well established in the literature, but we report it here for the sake of completeness. Aharony duality relates a $U(1)_0$ gauge theory with two flavors and vanishing superpotential to another $U(1)_0$ gauge theory with two flavors, four singlets flipping the mesons and two singlets flipping the monopoles. The corresponding $\mathbf{S}_b^3$ partition function identity is the following:
\begin{equation}	\label{eq:Aharony_U1_2,2}
\begin{split}
\mathcal{Z}_{U(1)_0}^{[2,2]} (-\vec{X}, \vec{Y}, \eta) &= 
\mathcal{Z}_{U(1)_0}^{[2,2]} (\frac{iQ}{2} - \vec{Y},\frac{iQ}{2} + \vec{X}, -\eta)
\\&\times \sum_{a,b=2}^{2} s_b \left( \frac{iQ}{2} -Y_a + X_b \right)
s_b \left( \frac{iQ}{2} +\frac{1}{2} (Y_1 + Y_2 - X_1 - X_2) \pm \eta \right)
\\& \times e^{-\pi i \eta ( X_1+X_2+Y_1+Y_2 )}
\end{split}
\end{equation}
where  $-\vec{X}$ and $\vec{Y}$ are the fugacities associated to the fundamentals and fundamentals of the electric theory, $\eta$ is the FI term and the $\mathbf{S}_b^3$ partition function of $U(1)_0$ with two flavors is:
\begin{equation}
\mathcal{Z}_{U(1)_0}^{[2,2]} (-\vec{X}, \vec{Y}, \eta) =
\int du\; e^{2\pi i u \eta} \sum_{a=1}^2 s_b \left(\frac{iQ}{2}- u + X_a \right)s_b \left(\frac{iQ}{2} +u - Y_a \right)
\end{equation}

The identity \eqref{eq:Aharony_U1_2,2} is equivalent to Theorem 5.1.11 of \cite{vandebult}.
In order to introduce mixed CS terms $k_{G,X_a}$, $k_{G,Y_a}$ between the gauge group and the flavor nodes we shift the FI parameter as follows:
\begin{equation}
\eta \to \tilde{\eta} = \eta - \frac{1}{2} (k_{G,X_1} X_1 + k_{G,X_2} X_2 + k_{G,Y_1} Y_1 + k_{G,Y_2} Y_2)
+\frac{ iQ}{4}
\end{equation}
In particular we have $k_{G,X_1}=k_{G,Y_1}=-1$, $k_{G,X_2}=1$ and $k_{G,Y_2}=0$.
We now introduce a large real mass in order to integrate out one of the fundamentals. At the level of the flavor fugacities this corresponds to the following limit:
\begin{equation}
\left\{
\begin{array}{l}
	X_1 \to X_1 -s
	\\
	X_2 \to X_2 -s
	\\
	Y_1 \to Y_1 - s
	\\
    Y_2 \to s
	\\
	\tilde{\eta} \to \tilde{\eta} - s
\end{array}
\right.
\end{equation}
Which is paired with the shift of gauge fugacities:
\begin{equation}
\left\{
\begin{array}{l}
	u \to u -s
	\\
	w \to w - s
\end{array}
\right.
\end{equation}
where $u$ and $w$ are the gauge fugacities for the electric and magnetic gauge groups, respectively. We now take the $s\to +\infty$ limit. One can check that the divergent phases cancel between the electric and magnetic sides and one obtains an identity corresponding to chiral Giveon-Kutasov duality for $U(1)_{\frac{1}{2}}$ with 2 fundamentals and 1 antifundamental \eqref{quiv:U1_[1,2]}.
The identity between $\mathbf{S}_b^3$ partition functions is given by:
\begin{equation}	\label{eq:Aharony_U1_1,2}
\begin{split}
\mathcal{Z}_{U(1)_{-\frac{1}{2}}}^{[2,1]} (-X_1, \vec{Y}, \eta) &= 
\mathcal{Z}_{U(1)_{\frac{1}{2}}}^{[1,2]} (\frac{iQ}{2} - Y_1,\frac{iQ}{2} + \vec{X}, -\eta-\frac{iQ}{4})
\\&\times \sum_{a=2}^{2} s_b \left( \frac{iQ}{2} -Y_1 + X_a \right)
s_b \left( \frac{iQ}{2} +\frac{1}{2} (Y_1  - X_1 -X_2 ) + \tilde{\eta} \right)
\\& \times e^{\frac{\pi i}{2} \phi}
\end{split}
\end{equation}
The phase $\phi$ encodes shifts in the background CS and BF levels for the flavor symmetries, as well as background FIs for the flavor symmetries:
\begin{equation}
\begin{split}
\phi = &
X_1^2-2 X_2^2+Y_1^2+ 2 X_1 Y_1+\lambda ^2
+2 \lambda  Y_1+4 \lambda  X_1+2 \lambda  X_2
\\&+
\frac{3 i \lambda  Q}{2}+i Q X_1-\frac{1}{2} i Q X_2+\frac{1}{2} i Q Y_1
+ f(iQ)
\end{split}
\end{equation}
where $f(iQ)$ includes terms independent on the flavor fugacities which are not relevant for our discussion.
Notice in particular that the mopole flipper $t^-$, contributing as:
\begin{equation}
s_b \left( \frac{iQ}{2} +\frac{1}{2} (Y_1  - X_1 - X_2 ) + \tilde{\eta} \right)
=
s_b \left(
-\frac{i Q}{4} -X_2+Y_1+\eta
\right)
\end{equation}
is charged with charge $\pm 1$ under two of the flavor groups, and it is therefore depicted as a bifundamental field connecting two flavor nodes in the quiver representation \eqref{quiv:U1_[1,2]}.

\subsection{Planar-Planar flip-flip dual via local dualization}\label{appff}
In this section we show that 
as in the  $\mathcal{N}=4$ case
the planar-planar \textit{flip-flip} duality 
can be demonstrated by iterative applications of local Aharony-like dualities.
However the local dualization now involves dualizing a \textit{column} of $U(1)$ gauge nodes.
Therefore, we consider the following duality for a $U(1)^N$ gauge theory: 

\begin{equation}	\label{quiv:planar_Aharony}
\begin{tikzpicture}[baseline=(current bounding box.center)]
	\nodeCS(0,0)(g1,$1$,$\frac{1}{2}$)
	\nodeCS(0,-2)(g2,$1$,$0$)
	\nodeCS(0,-4)(g3,$1$,$0$)
	
	\nodeCS(0,-8)(g4,$1$,$\frac{1}{2}$)
	\flavorCS(-0.85,-1)(l1,$1$,$-\frac{1}{2}$)
	\flavorCS(-0.85,-3)(l2,$1$,$0$)
	\flavorCS(-0.85,-7)(l3,$1$,$-\frac{1}{2}$)
	\flavorCS(0.85,1)(r1,$1$,$1$)
	\flavorCS(0.85,-1)(r2,$1$,$1$)
	\flavorCS(0.85,-3)(r3,$1$,$1$)
	\flavorCS(0.85,-7)(r4,$1$,$1$)
	\flavorCS(0.85,-9)(r5,$1$,$1$)
    
	\foreach \to/\from in {r1/g1, g1/r2, r2/g2, g2/r3, r3/g3, r4/g4, g4/r5}
	{ \arrowBFlr(\from,\to)($-1$,left) }
	
	\foreach \to/\from in {g4/l3, g3/l2, l2/g2, g2/l1, l1/g1}
	{ \arrowBF(\from,\to)($1$) }
	\foreach \from/\to in {r1/r2, r2/r3, r4/r5}
	{ \arrowBFlr(\from,\to)($1$,left) }
	
	\foreach \from/\to in {l2/l1}
	{ \dottedBF(\from,\to)($0$) }
	
	\node at (0,-6) () {$\vdots$};
	\node at (0,-10) () {$\mathcal{W} = \mathcal{W}_{planar} + 
    \mathcal{W}_{monopole}$};
\end{tikzpicture}
\qquad 
\longleftrightarrow
\qquad
\begin{tikzpicture}[baseline=(current bounding box.center)]
	\nodeCS(0,0)(g1,$1$,$-\frac{1}{2}$)
	\nodeCS(0,-2)(g2,$1$,$0$)
	\nodeCS(0,-4)(g3,$1$,$0$)
	
	\nodeCS(0,-8)(g4,$1$,$-\frac{1}{2}$)
	\flavorCS(-0.85,-1)(l1,$1$,$1$)
	\flavorCS(-0.85,-3)(l2,$1$,$1$)
	
	\flavorCS(-0.85,-7)(l3,$1$,$1$)
	\flavorCS(0.85,1)(r1,$1$,$\frac{1}{2}$)
	\flavorCS(0.85,-1)(r2,$1$,$0$)
	\flavorCS(0.85,-3)(r3,$1$,$0$)
	
	\flavorCS(0.85,-7)(r4,$1$,$0$)
	\flavorCS(0.85,-9)(r5,$1$,$\frac{1}{2}$)
	\foreach \to/\from in {r1/g1, g1/r2, r2/g2, g2/r3, r3/g3, r4/g4, g4/r5}
	{ \arrowBF(\to,\from)($1$) }
	
	\foreach \to/\from in {g4/l3, g3/l2, l2/g2, g2/l1, l1/g1}
	{ \arrowBFlr(\to,\from)($-1$,left) }
	\foreach \to/\from in {r1/r2, r2/r3, r4/r5}
	{ \dottedBF(\to,\from)($0$) }
	
	\foreach \to/\from in {l1/l2}
	{ \arrowBF(\to,\from)($1$) }
	
	\node at (0,-6) () {$\vdots$};
	\node at (0,-10) () {$\mathcal{W} = \mathcal{W}_{planar} + \mathcal{W}_{monopole}$};
\end{tikzpicture}
\end{equation}
on the r.h.s.~ there is a background CS term for the topological symmetry at level $1$. The superpotential $\mathcal{W}_{planar}$ includes cubic and quartic terms associated with the corresponding quiver's triangles and squares. 
The superpotential $\mathcal{W}_{monopole}$ is:
\begin{equation}    \label{app:mon_sup}
	\mathcal{W}_{monopole} = \mon_1^- \mon_2^+ + \mon_2^- \mon_3^+ + \dots +\mon_{N-1}^- \mon_N^+
\end{equation}
where $\mon_j^\pm$ is the monopole operators with GNO flux $\pm1$ under the $j$-th gauge group $U(1)$. The monopole operators $\mon_1^-$, $\mon_n^+$ and $\mon_j^\pm$, $j=2,\dots, N-1$ are gauge invariant since there are no BF couplings between adjacent gauge nodes. The monopole superpotential \eqref{app:mon_sup} is equivalent to the monopole superpotentials considered in the main body of the paper, because the monopoles with GNO flux $-1$ under the $j$-th node an $+1$ and the $(j+1)$-th node factorize into the product $\mon_{j}^- \mon_{j+1}^+$.
The monopole superpotential breaks all the topological symmetries to a diagonal $U(1)$.

We dualize the top node now using
the duality \ref{quiv:U1_[1,2]}. 
Then we can dualize the second node from the top using Aharony duality for $U(1)$ gauge group with 2 flavors. In the presence of BF terms involving the gauge and flavor symmetries this duality is given by:
\begin{equation}	\label{quiv:U1_[1,1]}
\begin{tikzpicture}[baseline=(current bounding box.center)]
	\nodeCS(0,0)(g,$1$, $0$)
	\flavorCS(-1,1)(nw,$1$, $0$)
	\flavorCS(1,1)(ne,$1$, $0$)
	\flavorCS(-1,-1)(sw,$1$, $0$)
	\flavorCS(1,-1)(se,$1$, $0$)
	\arrowBF(g,ne)($-1$)
	\arrowBF(g,sw)($1$)
	\arrowBF(nw,g)($1$)
	\arrowBFlr(se,g)($-1$,right)
	\dottedBF(nw,ne)($0$)
	\dottedBF(ne,se)($1$)
	\dottedBF(se,sw)($0$)
	\dottedBF(sw,nw)($0$)
    \draw[FIcolor] (g)++(0,0.3) node[anchor=south] {\tiny$\eta$};
	\node at (0,-1.7) (W) {$\mathcal{W}=0$};
\end{tikzpicture}
\qquad
\longleftrightarrow
\qquad 
\begin{tikzpicture}[baseline=(current bounding box.center)]
	\nodeCS(0,0)(g,$1$, $0$)
	\flavorCS(-1,1)(nw,$1$, $\frac{1}{2}$)
	\flavorCS(1,1)(ne,$1$, $-\frac{1}{2}$)
	\flavorCS(-1,-1)(sw,$1$, $\frac{1}{2}$)
	\flavorCS(1,-1)(se,$1$, $-\frac{1}{2}$)
	\arrowBFlr(ne,g)($1$,left)
	\arrowBFlr(sw,g)($-1$,right)
	\arrowBFlr(g,nw)($-1$,left)
	\arrowBFlr(g,se)($1$,right)
	\arrowBFlr(nw,ne)($0$,right)
	\arrowBFlr(se,ne)($0$,right)
	\arrowBFlr(se,sw)($0$,right)
	\arrowBFlr(nw,sw)($1$,right)
    \path[draw] (ne) edge[->-,bend right=15] node[midway,anchor=south] {\tiny$t_+$} (nw);
    \path[draw] (sw) edge[->-,bend right=15] node[midway,anchor=north] {\tiny$t_-$} (se);
    \draw[FIcolor] (g)++(0,0.3) node[anchor=south] {\tiny$-\eta$};
	\node at (0,-1.9) (W) {$\mathcal{W}=\mathcal{W}_{Seiberg} + t^- \tilde{\mon}^- + t^+ \tilde{\mon}^+$};
\end{tikzpicture}
\end{equation}

After the local dualization of the second node, two of the meson flippers and two of the monopole flippers take masses pairwise due to the planar and monopole superpotentials of the original theory. After integrating out these fields, we obtain the theory:
\begin{center}
\begin{equation}	\label{quiv:planar_Aharony_2steps}
\begin{tikzpicture}[baseline=(current bounding box.center)]
	\nodeCS(0,0)(g1,$1$,$-\frac{1}{2}$)
	\nodeCS(0,-2)(g2,$1$,$0$)
	\nodeCS(0,-4)(g3,$1$,$0$)
	
	\nodeCS(0,-8)(g4,$1$,$\frac{1}{2}$)
	\flavorCS(-0.85,-1)(l1,$1$,$1$)
	\flavorCS(-0.85,-3)(l2,$1$,$\frac{1}{2}$)
	
	\flavorCS(-0.85,-7)(l3,$1$,$-\frac{1}{2}$)
	\flavorCS(0.85,1)(r1,$1$,$\frac{1}{2}$)
	\flavorCS(0.85,-1)(r2,$1$,$0$)
	\flavorCS(0.85,-3)(r3,$1$,$\frac{1}{2}$)
	
	\flavorCS(0.85,-7)(r4,$1$,$1$)
	\flavorCS(0.85,-9)(r5,$1$,$1$)
	\foreach \to/\from in {r4/g4, g4/r5,  g1/l1, l1/g2, g2/l2}
	{ \arrowBFlr(\from,\to)($-1$,right) }
	
	\foreach \to/\from in {g4/l3, g3/l2, g1/r1, r2/g1, g2/r2, r3/g2}
	{ \arrowBFlr(\from,\to)($1$,right) }
	\dottedBF(r1,r2)($0$)
	\dottedBF(r2,r3)($0$)
	\arrowBFlr(l1,l2)($1$,left)

    \arrowBFlr(g3,r3)($-1$,left)
	
	\arrowBFlr(r4,r5)($1$,right)
	\arrowBFlr(r3,l2)($0$,right)
    \path[draw] (l2) edge[->-,bend right=15] node[midway,anchor=north] {\tiny$t_2^-$} (r3);

	
	\node at (0,-6) () {$\vdots$};
 \node at (0,-11) {$ \mathcal{W} = \mathcal{W}_{planar}+ \mathcal{W}_{monopole}$}; 
\end{tikzpicture}
\end{equation}
\end{center}
where:
\begin{equation}
 \mathcal{M}_{monopole} = \tilde{\mon}_1^{+}\tilde{\mon}_2^{-}  +  t_2^{-} \tilde{\mon}_2^{-}
 					+ t_2^{-} \mon_3^{+} + \mon_3^{-}\mon_4^{+} + \dots+ \mon_{n-1}^{-}\mon_n^{+}
\end{equation}

We continue to dualize the other nodes from top to bottom using the duality \eqref{quiv:U1_[1,1]}, producing additional singlets. Some mesons and monopole flippers acquire a mass at each step and can be integrated out. The bottom node can be dualized by a duality analogous to \eqref{quiv:U1_[1,2]}, producing an additional CS term for the topological symmetry at level $\frac{1}{2}$. 
The result of dualizing every gauge node is the theory on the r.h.s.~ of \eqref{quiv:planar_Aharony}, therefore the sequence of dualities described above provides a derivation of the duality \eqref{quiv:planar_Aharony} in terms of local Aharony-like dualities.\\

\begin{figure}[ht]
    \centering
    \includegraphics[width=.45\textwidth]{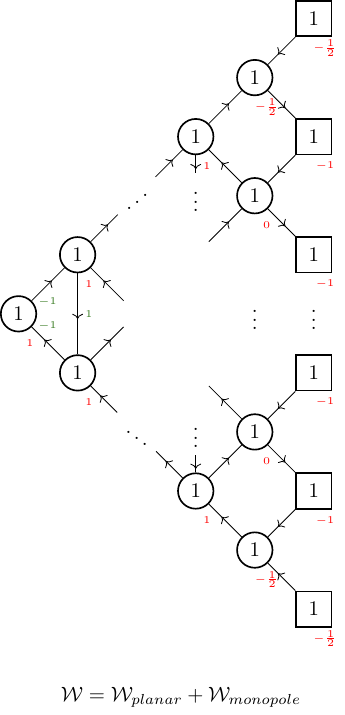}
    \caption{The resulting dual frame obtained by locally dualizing each column of the planar UV completion of the $G[U(N)]$ theory once. Notice in particular that there are no vertical bifundamentals in the last two columns and the CS levels change accordingly.}
    \label{quive:flip_1pass}
\end{figure}

Similarly, we can use the duality \eqref{quiv:planar_Aharony} to derive the flip-flip duality of the $G[U(N)]$ theory. Starting from the leftmost node we locally apply the duality \eqref{quiv:planar_Aharony} to the columns of the planar UV completion of the $G[U(N)]$ theory \eqref{planar_mirror_ftu}. One can verify at each step the relevant column has the correct CS and BF levels so that \eqref{quiv:planar_Aharony} can be applied. 
After dualizing every column the resulting theory is shown in Figure \ref{quive:flip_1pass}.

We now repeat the same procedure, dualizing the columns of \eqref{quive:flip_1pass} from left to right, stopping at the second to last column. We repeat this procedure, each time dualizing one less column than the last time. In total we apply the duality \eqref{quiv:planar_Aharony} $N(N-1)/2$ times. 
The resulting theory corresponds to the \textit{flip-flip} dual of the $G[U(N)]$ theory, depicted on the r.h.s.~ of \eqref{quiv:flip-flip_ghat}. 
Therefore this procedure provides a derivation of the \textit{flip-flip} \eqref{quiv:flip-flip_ghat} in terms of local Aharony-like dualities.

\subsection{Chiral-Chiral flip-flip dual via local dualization}\label{accff}
Similarly, it is possible to prove
the chiral \textit{flip-flip} duality 
via local dualization. 
The local dualities that we employ are well-establish Aharony-like dualities for chiral SQCD theories \cite{Benini:2011mf,Aharony:2014uya,Closset:2023vos}, which can be written in quiver notation as:
\begin{equation}	\label{eq:local_duality_chiral_1}
\begin{tikzpicture}[baseline=(current bounding box.center)]
	\node[gauge,black] (g1) at (0,0)  {$1$};
		\draw[red] (g1.south east)++(3pt,-3pt) node {\tiny$1$};
	\node[flavor,black] (f1) at (1.5,0)  {$2$};
		\draw[red] (f1.south east)++(3pt,-5pt) node {\tiny$(0,2)$};
	\arrowBF(g1,f1)($-1$)
	\draw (0.75,-0.75) node {$\mathcal{W} = 0$};
\end{tikzpicture}
\qquad\leftrightarrow\qquad
\begin{tikzpicture}[baseline=(current bounding box.center)]
	\node[gauge,black] (g1) at (0,0)  {$1$};
		\draw[red] (g1.south east)++(3pt,-3pt) node {\tiny$-1$};
	\node[flavor,black] (f1) at (1.5,0)  {$2$};
		\draw[red] (f1.south east)++(3pt,-5pt) node {\tiny$(1,1)$};
	\arrowBFlr(f1,g1)($1$,right)
	\draw (0.75,-0.75) node {$\mathcal{W} = t \mon^-$};
\end{tikzpicture}
\end{equation}
and:
\begin{equation}	\label{eq:local_duality_chiral_2}
\begin{tikzpicture}[baseline=(current bounding box.center)]
	\node[gauge,black] (g1) at (0,0) {$N$};
		\draw[red] (g1.south east)++(-4pt,2pt) node[anchor=north west] {\tiny (1,1)};
	\node[flavor,black] (f1) at (-1.5,0) {$N\!-\!1$};
		\draw[red] (f1.south east)++(-4pt,2pt) node[anchor=north west] {\tiny (-1,-1)};
	\node[flavor,black] (f2) at (1.5,0) {$N\!+\!1$};
		\draw[red] (f2.south east)++(-4pt,2pt) node[anchor=north west] {\tiny (0,$N$+1)};
	\arrowBFlr(g1,f1)($1$,right)
	\arrowBF(g1,f2)($-1$)
	\draw (0,-1) node {$\mathcal{W} = 0$};
\end{tikzpicture}
\qquad\leftrightarrow\qquad
\begin{tikzpicture}[baseline=(current bounding box.center)]
	\node[gauge,black] (g1) at (0,0) {$N$};
		\draw[red] (g1.south east)++(-4pt,2pt) node[anchor=north west] {\tiny (-1,-1)};
	\node[flavor,black] (f1) at (-1.5,0) {$N\!-\!1$};
		\draw[red] (f1.south east)++(-4pt,2pt) node[anchor=north west] {\tiny (0,$N$-1)};
	\node[flavor,black] (f2) at (1.5,0) {$N\!+\!1$};
		\draw[red] (f2.south east)++(-4pt,2pt) node[anchor=north west] {\tiny (1,1)};
	\arrowBFlr(f1,g1)($-1$,left)
	\arrowBFlr(f2,g1)($1$,right)
	\draw (0,-1) node {$\mathcal{W} = 0$};
\end{tikzpicture}
\end{equation}
on the r.h.s.~ there are also background CS and FI terms for the flavor symmetries, in particular on the r.h.s.~ of \eqref{eq:local_duality_chiral_1} there is a background CS term at level $-\frac{1}{2}$ for the topological symmetry.

We start from the chiral UV completion of the $G[U(N)]$ theory \eqref{g_n_chiral}:
\begin{equation} 
\begin{tikzpicture}[baseline=(current bounding box.center)]
    \node at (0,0) (g1) [gauge, black] {$1$};
    \node at (2,0) (g2) [gauge, black] {$2$};
    
    \begin{scope}[xshift=1cm]
    \node at (4,0) (g3) [gauge, black] {$_{N-2}$};
    \path[scale=1] (g2) -- node[midway] (d1) {$\ldots$} (g3);
    \node at (6,0) (g4) [gauge, black] {$_{N-1}$};
    \node at (8,0) (f) [flavor, black] {$N$};
    \draw[red] (4.8,-.4) node {\tiny${(0,N-2)}$}; \draw[red] (6.8,-.4) node {\tiny${(0,N-1)}$}; \draw[red] (9.3,-.4) node {\tiny${(-\frac{N+1}{2}, \frac{N-1}{2})}$};
    \end{scope}

 \arrowBFlr(g1,g2)($-1$,left);
 \arrowBFlr(g3,g4)($-1$,left);
 \arrowBFlr(g4,f)($-1$,left);
 \arrowBFlr(g2,d1)($-1$,left)
  \arrowBFlr(d1,g3)($-1$,left)
  
   \draw[red] (0.4,-0.4) node {\tiny$1$}; 
   \draw[red] (2.5,-0.4) node {\tiny${(0,2)}$};

    \draw (4.5,-1.5) node {$\mathcal{W}=0$};
 \end{tikzpicture} \end{equation}

We locally dualize the first node with the Aharony-like duality \eqref{eq:local_duality_chiral_1}. In the dual frame the gauge invariant monopole $\mon^{(-,0\dots)}$ is flipped by a gauge singlet denoted as $t_{N-1}$. Furthermore, the second node has CS level $(1,1)$ and we may dualize it with the basic duality \eqref{eq:local_duality_chiral_2}. Doing so does not produce additional singlets but has the effect of “lengthening" the monopole of the first gauge node to $\mon^{(-,-,0,\dots)}$, which must be dressed appropriately \cite{Benvenuti:2020wpc}. Therefore the singlet  $t_{N-1}$ now flips $\mon^{(-,-,0,\dots)}$. The third gauge node now has CS level $(1,1)$ and we may dualize it with \eqref{eq:local_duality_chiral_2}. From left to right, we continue this dualization procedure for all gauge nodes. Each dualization step further lengthens the monopole flipped by $t_{N-1}$, resulting in:
\begin{equation} 
\begin{tikzpicture}[baseline=(current bounding box.center)]
    \node at (0,0) (g1) [gauge, black] {$1$};
    \node at (2,0) (g2) [gauge, black] {$2$};
    
    \begin{scope}[xshift=1cm]
    \node at (4,0) (g3) [gauge, black] {$_{N-2}$};
    \path[scale=1] (g2) -- node[midway] (d1) {$\ldots$} (g3);
    \node at (6,0) (g4) [gauge, black] {$_{N-1}$};
    \node at (8,0) (f) [flavor, black] {$N$};
    \draw[red] (4.8,-.4) node {\tiny${(0,N-2)}$}; \draw[red] (6.8,-.4) node {\tiny${\text{(0,-(N-1))}}$}; 
    \draw[red] (9,-.6) node {\tiny${-\left(\frac{N-1}{2},\frac{N-1}{2}\right)}$};
    \end{scope}

 \arrowBFlr(g1,g2)($-1$,left);
 \arrowBFlr(g3,g4)($-1$,left);
 \arrowBFlr(f,g4)($1$,right);
 \arrowBFlr(g2,d1)($-1$,left)
  \arrowBFlr(d1,g3)($-1$,left)

   \draw[red] (0.4,-0.4) node {$_1$}; 
   \draw[red] (2.5,-0.4) node {$_{(0,2)}$};

    \draw (4.5,-1.5) node {$\mathcal{W}=t_{N-1} \mon^{(-,-,\dots,-)}$};
 \end{tikzpicture} \end{equation}

We may now dualize the first node again using \eqref{eq:local_duality_chiral_1}, producing an additional singlet $t_{N-2}$ flipping the monopole of the first node. Furthermore the long monopole $\mon^{(-,-,\dots,-)}$ is shortened to $\mon^{(0,-,\dots,-)}$. We continue to dualize the other nodes using \eqref{eq:local_duality_chiral_2}, from left to right, stopping at the second to last gauge node. Each duality shortens the monopole flipped by $t_{N-1}$ and lengthens the monopole flipped by $t_{N-2}$.
This second set of dualizations results in:
\begin{equation} 
\begin{tikzpicture}[baseline=(current bounding box.center)]
    \node at (0,0) (g1) [gauge, black] {$1$};
    \node at (2,0) (g2) [gauge, black] {$2$};
    
    \begin{scope}[xshift=1cm]
    \node at (4,0) (g3) [gauge, black] {$_{N-2}$};
    \path[scale=1] (g2) -- node[midway] (d1) {$\ldots$} (g3);
    \node at (6,0) (g4) [gauge, black] {$_{N-1}$};
    \node at (8,0) (f) [flavor, black] {$N$};
    \draw[red] (4.8,-.4) node {\tiny${\text{(0, -(N-2))}}$}; 
    \draw[red] (6.8,-.4) node {\tiny${\text{(0,-(N-1))}}$}; 
    \draw[red] (9,-.6) node {\tiny${-\left(\frac{N-1}{2},\frac{N-1}{2}\right)}$};
    \end{scope}

 \arrowBFlr(g1,g2)($-1$,left);
 \arrowBFlr(g4,g3)($1$,right);
 \arrowBFlr(f,g4)($1$,right);
 \arrowBFlr(g2,d1)($-1$,left)
  \arrowBFlr(d1,g3)($-1$,left)

   \draw[red] (0.4,-0.4) node {\tiny$1$}; 
   \draw[red] (2.5,-0.4) node {\tiny${(0,2)}$};

    \draw (4.5,-1.5) node {$\mathcal{W}=t_{N-2} \mon^{(-,\dots,-,0)}+t_{N-1} \mon_d^{(0,\dots,0,-)}$};
 \end{tikzpicture} \end{equation}

We continue this procedure, each time dualizing one less node that the last time. Each time we dualize the first gauge node we generate an additional singlet. At the end of this sequence of dualizations we obtain the \textit{flip-flip} dual of \eqref{g_n_chiral}, namely:
\begin{equation} 
\begin{tikzpicture}[baseline=(current bounding box.center)]
    \node at (0,0) (g1) [gauge, black] {$1$};
    \node at (2,0) (g2) [gauge, black] {$2$};
    
    \begin{scope}[xshift=1cm]
    \node at (4,0) (g3) [gauge, black] {$_{N-2}$};
    \path[scale=1] (g2) -- node[midway] (d1) {$\ldots$} (g3);
    \node at (6,0) (g4) [gauge, black] {$_{N-1}$};
    \node at (8,0) (f) [flavor, black] {$N$};
    \draw[red] (4.8,-.4) node {\tiny${\text{(0,-(N-2))}}$}; 
    \draw[red] (6.8,-.4) node {\tiny${\text{(0,-(N-1))}}$}; 
    \draw[red] (9,-.6) node {\tiny${-\left(\frac{N-1}{2},\frac{N-1}{2}\right)}$};
    \end{scope}

 \arrowBFlr(g2,g1)($1$,right);
 \arrowBFlr(g4,g3)($1$,right);
 \arrowBFlr(f,g4)($1$,right);
 \arrowBFlr(d1,g2)($1$,right)
  \arrowBFlr(g3,d1)($1$,right)

   \draw[red] (0.4,-0.4) node {\tiny${-1}$}; 
   \draw[red] (2.5,-0.4) node {\tiny${(0,-2)}$};

    \draw (4.5,-1.5) node {$\mathcal{W}=t_{1} \mon^{(-,0,\dots,0)}+t_{2} \mon_d^{(0,-,\dots,0)}+ \dots + t_{N-1} \mon_d^{(0,\dots,0,-)}$};
 \end{tikzpicture} \end{equation}

\bibliographystyle{JHEP}
\bibliography{References}

\end{document}